\documentclass[journal]{rmaa}

\usepackage{graphicx}
\usepackage{natbib}
\usepackage{hyperref}
\usepackage{amsmath}
\usepackage{adjustbox}
\usepackage{lastpage}

\usepackage{float}
\usepackage{txfonts}
\usepackage{booktabs}
\usepackage{amssymb,amsmath,graphicx,natbib,hyperref, color}
\usepackage{lipsum}

\providecommand{\aap}{A\&A}

\providecommand{\apjl}{ApJL}

\DeclareRobustCommand{\ION}[2]{%
\relax\ifmmode
\ifx\testbx\f@series
{\mathbf{#1\,\mathsc{#2}}}\else
{\mathrm{#1\,\mathsc{#2}}}\fi
\else\textup{#1\,{\mdseries\textsc{#2}}}%
\fi}

\newcommand{\ha}          {\mbox{H$\alpha$}}
\newcommand{\hb}          {\mbox{H$\beta$}}
\newcommand{\pyHII}       {{\sc pyHIIextractor}}
\newcommand{\hii}{\ION{H}{ii}}
\newcommand{\nii}{[\ION{N}{ii}]}
\newcommand{\oiii}{[\ION{O}{iii}]}
\newcommand{\IC}          {\mbox{IC~342}}

\newcommand{\AM}          {\mbox{AMUSING++}}
\newcommand{\Av}          {\mbox{$\mathrm{A_{V}}$}}
\newcommand{\fluxunits}   {\mbox{$10^{-17} \mathrm{erg\, cm^{-2}\, s^{-1}}$}}
\newcommand{\sigmaHa}{\mbox{$\mathrm{\sigma_{H\alpha}}$}}
\newcommand{\DeltaL}{\mbox{$\mathrm{\Delta\,L_{H\alpha}}$}}
\newcommand{\DeltaAv}{\mbox{$\mathrm{\Delta\,A_{v}}$}}
\newcommand{\DeltaOH}{\mbox{$\mathrm{\Delta\,O/H}$}}
\newcommand{\haLF}          {\mbox{H$\alpha$LF}}

\title{Physical properties of \hii\ regions at sub-kpc scales using Integral Field Spectroscopy on IC~342}

\author{
 J.~K.~Barrera-Ballesteros\altaffilmark{1},
 S.~F.~S\'anchez\altaffilmark{2},
 K.~Kreckel\altaffilmark{3},
 A.~Lugo-Aranda\altaffilmark{1},
 H.~Ibarra-Medel\altaffilmark{1}, 
 L.~Carigi\altaffilmark{1},
 N.~Drory\altaffilmark{4},
 D. Bizyaev\altaffilmark{5,6}, 
 J.~E.~Méndez~Delgado\altaffilmark{3}, 
 Guillermo Blanc\altaffilmark{7,8}
 }
\altaffiltext{1}{Universidad Nacional Aut\'onoma de M\'exico, Instituto de Astronom\'ia, AP 70-264, CDMX  04510, M\'exico}
\altaffiltext{2}{Universidad Nacional Aut\'onoma de M\'exico, Instituto de Astronom\'ia, AP 70-264, CDMX  04510, M\'exico}
\altaffiltext{3}{Astronomisches Rechen-Institut, Zentrum für Astronomie der Universität Heidelberg, Mönchhofstraße 12-14, 69120 Heidelberg, Germany}
\altaffiltext{4}{University of Texas at Austin, McDonald Observatory, 1 University Station, Austin, TX 78712, USA}
\altaffiltext{5}{Apache Point Observatory and New Mexico State University, P.O. Box 59, Sunspot, NM 88349-0059, USA}
\altaffiltext{6}{Sternberg Astronomical Institute, Moscow State University, Moscow, Russia}
\altaffiltext{7}{The Observatories of the Carnegie Institution for Science, 813 Santa Barbara St., Pasadena 91101, CA, USA}
\altaffiltext{8}{Departamento de Astronomía, Universidad de Chile, Camino del Observatorio 1515, Las Condes, Santiago, Chile}

\shortauthor{Barrera-Ballesteros J.K., et al.}
\shorttitle{Physical properties of the ionized gas for \IC}

\abstract{

In this study we use Integral Field Spectroscopic (IFS) observations for one of the closest galaxy to us, the grand design spiral \IC, to derive physical properties of \hii\ regions at sub-kpc scales. This IFS data represents, to our knowledge, the most comprehensive observational effort in the optical for this galaxy. The final IFS datacube consists of 349 individual pointings using the IFS instrumentation from the SDSS-IV MaNGA survey. Using a prototype of the data analysis pipeline that will be devoted to the SDSS-V Local Volume Mapper (LVM) survey, we measure different observables from the emission line in  the optical. In particular, using the flux map of the \ha\ emission line, we derive the location and sizes of \hii\ region candidates for \IC. Using the integrated flux for different emission lines within each region, we derived the radial distribution of different physical properties from the ionized gas (e.g., optical extinction, \ha\ luminosity, oxygen abundance, etc). Comparing with larger samples of galaxies with IFS data, our results suggest that physical properties of the ionized gas of \IC\ are similar to galaxies with similar stellar mass in the nearby universe.

}

\resumen{En este trabajo presentamos el análisis de las propiedades físicas de regiones \hii\ a escalas de sub-kpc en una de las galaxias mas cercanas a nosotros, \IC, usando datos de espectroscopia de campo integral (IFS por sus siglas en ingles). Estos datos representan las observaciones más completas en el óptico de esta galaxia. El cubo de datos esta compuesto por 349 observaciones individuales del cartografiado SDSS-IV MaNGA. Para el análisis de este cubo de datos usamos un prototipo del dataducto de análisis del cartografiado SDSS-V LVM, con el cuál estimamos las propiedades de las lineas de emisión. En particular, usando el mapa de la línea de emisión de \ha\ determinamos la ubicación y el tamaño de candidatos a regiones \hii. Usando las propiedades integradas para cada candidato estimamos la distribución radial para un gran numero de propiedades físicas (e.g., extinción, luminosidad de \ha, abundancia de oxigeno, etc). Comparando con galaxias de masa estelar similar observadas con IFS, nuestros resultados muestran que las propiedades del gas ionizado para \IC\ son similares con dichas galaxias.}

\begin{document}
\maketitle


\keywords{TBD}
\section{Introduction}
\label{sec:intro}

A theoretical \hii\ region is a sphere of gas that surrounds young massive stars (mostly O and B type). The UV emission from these stars ionizes this gas which cools down through emission lines, some of the brightest in the optical. Since OB stars are short lived ($<$ 15 Myr), the observed emission from these regions has been often used to trace recent star-formation. Furthermore thanks to the emission by the ionized gas contained in these regions we are able to gain deep knowledge of, for instance, the luminosity function of the \ha\ emission line (which in turn traces the massive end of the initial mass function), the chemical content of the interstellar medium (ISM) or their dynamical stage. The size of an \hii\ region is variable, from a few to hundreds of parsecs \citep[e.g., the Orion Nebula and NGC~5471, from $\sim$8 pc to $\sim$1 kpc, respectively,][]{Anderson_2014, Garcia-Benito_2011}. Aggregations of those large \hii\ regions are what is considered as giant \hii\ regions, usually observed in star-forming spirals \citep[e.g., ][]{Hodge-Kennicutt_1983, Dottori-Copetti_1989, Knapen_1998}  

Observationaly, those giant \hii\ regions are usually traced as clumpy \ha\ emission across the optical extension of late-type galaxies \citep[e.g.,][and references therein]{Kennicutt_2012}. The galactocentric distribution of these regions has been usually quantified by a radial gradient suggesting that, in most spirals, at their center the \ha\ flux of these regions is larger in comparison to those regions located at the outskirts \citep[e.g.,][]{Knapen_2004, Bigiel_2008}. Using \ha\ narrow-band imaging it is possible to characterize the distribution of \hii\ regions in a galaxy via the \ha\ luminosity function \citep[e.g.,][]{Gonzalez-Delgado_1997,Bradley_2006}. Using long-slit spectroscopy it has been possible to trace other physical properties of these regions in reduced samples of spiral galaxies such as chemical abundances, electronic density or the ionization parameter \citep[e.g.,][]{Pilyugin_2016}. 

In recent years, extragalactic astronomy has witnessed a revolution in optical observations thanks to the Integral Field Spectroscopic (IFS) technique applied in thousand of galaxies in the nearby Universe. Large IFS surveys such as CALIFA \citep{Sanchez_2012}, MaNGA \citep{Bundy_2015}, SAMI \citep{Croom_2012} or AMUSING++ \citep{Lopez-Coba_2017} have shown the spatially-resolved properties of the demographics of galaxies in the nearby universe. Of particular interest is the radial distribution of physical properties of the ionized gas derived from the emission lines parameter in the optical \citep[e.g., ][]{Sanchez-Menguiano_2018, Lopez-Coba_2017,Espinosa-Ponce_2022, Barrera-Ballesteros_2023}. Given the success of these large IFS surveys, different research groups have employed different IFS techniques to map the properties of the ionized gas in very nearby galaxies, reaching superb spatial resolutions (of the order of pc of tens of pc). Among these different IFS surveys it is worth to mention the TYPHOON, SIGNALS and PHANGS surveys \citep{DAgostino_2018,Rousseau-Nepton_2019, Emsellem_2022}. 

Although there are dedicated IFS studies to explore the properties of \hii\ regions in galaxies in the local universe ($D~<~$~5~Mpc) these are rather scarce. Using the MUSE instrument, it has been possible to explore the physical properties of \hii\ regions at sub kpc scales in the LMC  \citep{McLeod_2019}, NGC~300 \citep{McLeod_2020, McLeod_2021},  NGC~7793 \citep{Della_Bruna_2020, Della_Bruna_2021}, and M83 \citep{Della_Bruna_2022}. These studies are mainly focused in explore how the stellar feedback from recent star formation in \hii\ regions affects the kinematics of the interstellar medium. Such studies are of great importance in order to bridge the gap between the observations at sub-kpc scales with those scaling relations between the star formation and the dynamical pressure at kpc scales for hundreds of galaxies \citep{Barrera-Ballesteros_2023}. However, they lack of a deep exploration on the physical properties of the ISM for the \hii\ regions at sub-kpc scales. With this work we aim to shed some light in the radial distribution of the physical parameters at these physical scales derived from the emission lines in the optical.

In this study we focus on \IC. This galaxy is the closest grand design spiral to the Milky Way \citep[$\sim$3.3~Mpc][]{Saha_2002}. Thus, it has been mapped at different wavelengths. From {\it WISE} data \citet{Jarrett_2013} derived its total stellar mass \citep[$\log(\mathrm{M_{\star}/M_\odot)} \sim 10.8 $, but see ][for other estimations]{Zibetti_2009,Zhu_2010}, and total star formation rate \citep[$\sim 2.4 \mathrm{M_\odot\,\,yr^{-1}}$, see also][]{Kennicutt_2011}. Its molecular and neutral gas content has also been extensively mapped either in its nucleus or in its arms \citep[e.g.,][]{Rickard_1981,Ishizuki_1990, Crosthwaite_2000, Kuno_2007}. However, despite those studies, there has been scarce studies regarding the properties of the ionized gas in \IC. The only reported measurements of the emission line fluxes in four \hii\ regions at different radii of this galaxy come from long-slit spectroscopy \citep{McCall_1985}. Using these observations \citet{Pilyugin_2004} derived the oxygen abundance gradient for this galaxy. In this work we present the largest integral field spectroscopic study of \IC\ in the optical. We took advantage of ancillary observations from the MaNGA survey \citep{Bundy_2015} devoted to mosaic this galaxy. In particular, we focus in the radial distribution of the physical parameters that could be derived from the observed properties from the different emission lines in the optical within a given \hii\ region. Due to its high angular resolution, the current dataset also gives us a unique opportunity to test the data-analysis pipeline that is been used to determine the physical properties in the Local Volume Mapper \citep{Drory_2024}, an on-going SDSS-V survey \citep{Kollmeier_2017} aimed to map the physical properties of the ionized gas within the Milky Way and galaxies included in the Local Volume \citep{Konidaris_2020}.


This article is organized as follows: In Sec.~\ref{sec:Data} we present the IFS observations of \IC\ as well as the data analysis pipeline used to derived the maps of the quantities from the observations; in Sec.~\ref{sec:analysis} we describe the method used to estimate the position and size of the \hii\ regions, the derived physical quantities for each region and their classification in a emission-line diagnostic diagram; in Sec.\ref{sec:Radial} we present the galactocentric distribution of the physical properties derived from the \hii\ regions; in Sec.~\ref{sec:LF_Ha} we present the \ha\ luminosity function for \IC\ derived from the \hii\ regions while in Sec.~\ref{sec:residuals} we explore the correlations between the radial  residuals of the derived physical properties; finally in Sec.~\ref{sec:Disc} we summarize and discuss our results in the framework of recent IFS surveys.

\section{Data}
\label{sec:Data}

\subsection{IFS observations as MaNGA ancillary program}
\label{sec:obs}
The observations to map the disk of \IC\ were part of an ancillary program included in the MaNGA survey (PI. K. Kreckel). The MaNGA \citep[Mapping Nearby Galaxies at Apache Point Observatory][]{Bundy_2015} survey was part of the forth generation of surveys included in the Sloan Digital Sky Survey \citep[SDSS-IV, ][]{Blanton_2017}. This survey observed more than 10000 galaxies in the nearby universe using the IFS technique. Here we briefly review the main features of this survey. Observations took place at the Apache Point Observatory using its 2.5-m telescope \citep{Gunn_2006}. It used two spectrographs from the BOSS survey \citep[Baryon Oscillation Spectroscopic Survey][]{Smee_2013}. BOSS spectrographs achieve a nominal spectral resolution of \mbox{$R \equiv \lambda/\Delta\lambda \sim 1900$} covering a large portion of the optical spectra (from 3000 to 10000 \AA). These spectrographs were fed by joined fibers in bundles distributed in a hexagon-like array (also known as the Integral Field Units, IFUs). The number of fibers in a given bundle, or IFU, varies from 19 to 127. Since the diameter of each fiber is $\sim$ 2\arcsec, the Field-of-View (FoV) for each IFU varies between 12\arcsec and 32\arcsec. Observations of MaNGA targets are done using a plugplate system, where the locations for each IFU (and corresponding sky fibers), 12 spectrophotometric standard stars, and 16 guide stars have been pre-drilled. This provides the necessary observations for the acquisition, guiding and dithering of targets, along with all calibration data needed in the data reduction pipeline. A detailed description of the instrumentation of the survey can be found in \citet{Drory_2015}. The reader is refereed to \citet{Law_2016} for a detailed explanation of the data strategy (acquisition, reduction, etc). The MaNGA reduction pipeline includes wavelength calibration, corrections from fiber-to-fiber transmission, subtraction of the sky spectrum and flux calibration \citep{Yan_2016}. The final product is a datacube with $x$ and $y$ coordinates corresponding to the sky coordinates and the $z$-axis corresponds to the wavelength. 

The observations of \IC\ were taken part of an ancillary program that was needed to provide additional MaNGA science targets when the Milky Way is overhead.  \IC\ has been neglected in optical studies as it suffers from high foreground Milky Way extinction (A$_V$ = 1.530; \citealt{Schlafly_2011}), but at a galactic latitude of $\sim$10~deg it was well situated for this ancillary program. 

The goal was to produce a uniform, contiguous mosaic covering the central $11^\prime \times 11^\prime$ (10 kpc $\times$ 10 kpc). This reaches to 0.25 R$_{25}$ $\approx$ 1.25 R$_e$ (R$_{25}$ = 22$'$), well matched to the primary MaNGA sample. Each of these pointings results in a datacube with a spaxel size of 0.5\arcsec \citep[the spatial resolution is given by average size of the PSF per fiber which is typically  2.54\arcsec,][]{Law_2016}. Given the distance of \IC\ this corresponds to a spatial resolution of $\sim$ 32 pc. Thus, these IFS data represent observations with significantly higher physical resolution compared to typical MaNGA targets ($\sim$3~kpc), surpassing even what is achieved with other nearby galaxy surveys like PHANGS-MUSE ($\sim$70~pc; \citealt{Emsellem_2022}).

The final mosaic of \IC\ uses 49 plates, and was observed between early 2017 and late 2021. Three pilot plates (with plate numbers: 9673-9675) were initially observed to test the feasibility of this survey, with IFUs placed on HII regions that provide good radial and azimuthal coverage. 46 plates (with plate numbers: 10141-10150, 10480-10491, 12027-12050) were used to uniformly mosaic the central 10 arcmin diameter disk of \IC. 
In some cases, these overlap with the positions of the IFUs in the pilot plates. Note that plate 10490 had one dropped IFU bundle, leading to a gap in the coverage.  An additional 15 plates were planned, but were not completed before the end of MaNGA survey operations. Each plate was observed in three dither positions, following the standard MaNGA procedure to fill the interstitial regions\citep{Law_2016}, and only the two largest IFUs are used (i.e., the bundles with 127 and 91 fibers). Individual IFU positions are tiled such that no gaps exist when combining all IFU fibers into a single data cube. As a result, seven large contiguous regions were observed (Figure \ref{fig:HIIregions}) with a total of over 10$^5$ spectra (including dither positions)

Each of these plates was observed using the same strategy as  a standard MaNGA plate, thus it was possible to use the same data reduction pipeline as was used for the rest of the survey. 
 The diameter of \IC\ is just small enough that sky fibers could successfully be placed for each IFU within the 14\arcmin patrol field.  
Exposure times in the pilot observations were limited to 1~hour total (two visits of 30 minutes, with 10 minutes per dither), but in the full survey consist of 1.5~hours total integration times (two visits of 45 minutes, with 15 minutes per dither). 


\subsection{Deriving emission line properties using the LVM-DAP}
\label{sec:pDAP}

As we mention in the previous section, given the close distance of \IC, the IFS dataset provided by the MaNGA observations have a superb angular resolution making it very suitable to use an analysis pipeline dedicated to IFS dataset from Local Volume galaxies. For this reason we use this IFS dataset from \IC\ as a test-bed for the data analysis pipeline (DAP). The DAP is been used to the analysis of the data that is been observed  by the Local Volume Mapper (LVM) survey which is part of fifth generation of the SDSS collaboration. The details of the LVM-DAP are described in \citep{Sanchez_2025}, and its preliminary results are  presented in \citep{Drory_2024}, here we provided a brief description of the main features of this DAP, in particular we focus on the extraction of physical properties from the ionized gas. Although with significant variations, the LVM-DAP follows a similar methodology as the \texttt{pyPipe3D} pipeline \citep{Lacerda_2022}. This DAP is an update of the pipeline \texttt{Pipe3D} \citep{Sanchez_2015, Sanchez_2016} which has been widely used to extract two dimentional distributions of physical properties from both the stellar and ionized gas components for heterogeneous data sets (e.g., CALIFA, MaNGA, AMUSING++).

In contrast to nearby galaxies, for those galaxies included in the Local Volume (including the Milky Way) it is not longer valid to assume that the stellar continuum of a portion of a galaxy is composed by the linear combination of single stellar population spectra (SSPs). Thus the LVM-DAP derives the stellar population using a technique called resolved stellar population (RSP) to estimate the best stellar continuum of the data observed at the LVM. A detailed description of this extraction and the physical properties derived for the stellar component of the spectra is going to be described somewhere else (Mejia  et al., in prep.).  In this case, the estimation of the best stellar continuum is no longer based on a linear combination of different SSPs, but on a library of individual stellar spectra covering a wide range of stellar properties (e.g., effective temperature, and gravity, chemical abundances). For each spectra in each reduced datacube. The DAP starts by deriving the systemic velocity, velocity dispersion, and optical extinction from the stellar continuum by fitting  only four stellar spectra. Once the best estimations for these parameters are derived, a stellar model is derived and subtracted from the observed spectra and a set of Gaussian profiles are fitted to a pre-defined set of emission lines. Those Gaussian profiles are then substracted from the observed spectra and a new fit of the stellar component is performed now using a large set of stellar spectra yielding the final continuum model. The remaining ``gas-only'' spectrum is then analyzed using a weighted-moment non-parametric procedure to estimate the integrated flux, velocity, velocity dispersion, and EW of a pre-defined set of 192 emission lines. Contrary to the DAP, where the data is store in the form of RSS tables, the data for \IC\ is stored in the usual format of the MaNGA survey for each pointing, this is as maps for each of the properties for each the analyzed emission lines. Furthermore, for each property derived by the DAP in \IC\ we perform a mosaic using the maps from individual pointings. Those maps have a scale of 1 arcsec per pixel. For the details of the algorithms and their implementation we refer the reader to \citet{sanchez24} and \citet{Lacerda_2022} \footnote{http://ifs.astroscu.unam.mx/pyPipe3D/}. In Fig.~\ref{fig:HIIregions} we show the map of the \ha\ line whereas in Appendix \ref{app:flux} we present the maps of other strong emission lines. The maps of the physical properties from all the pointings can be found in \texttt{https://ifs.astroscu.unam.mx/MaNGA/IC342/output\_dap/}.

\subsection{The AMUSING++ sample}

To provide a comparison of the trends derived for the \hii\ regions within \IC\ with respect to sample of \hii\ regions located in a larger sample of galaxies, we make use of the AMUSING++ sample \citep{Lopez-Coba_2020}. This sample is an extension of the All-weather MUse Supernova Integral field Nearby Galaxies survey \citep[AMUSING][]{Galbany_2016, Galbany_2020}. This compilation of galaxies observed with the MUSE instrument include observations from a wide variety of scientific goals \citep[see details in][]{Lopez-Coba_2020}. Although the initial compilation includes 635 galaxies \citep{Lopez-Coba_2020}, an updated sample comprises 678 galaxies \citep{Lugo-Aranda_2024}. The galaxies in the AMUSING++ sample are selected so the MUSE Field-of-view have a good spatial coverage  ($\sim$ 2 effective radius), and that they are nearby galaxies ($0.004<~z~<0.06$). This sample has a significant coverage in both, stellar masses and morphology \citep[see details in ][]{Lugo-Aranda_2024}. That the sample include nearby galaxies ensures a good spatial resolution ($\sim$ 100 to 1000 pc).  Using the latest compilation of AMUSING++ sample, \citet{Lugo-Aranda_2024} characterized the physical properties of more than 52~000 \hii\ regions. We use the data for those \hii\ regions to compare them with the trends we observe in the \hii\ regions located in \IC\ (see Sec.\ref{sec:BPT}).

\begin{figure*}
    \centering
    \includegraphics[width=\textwidth]{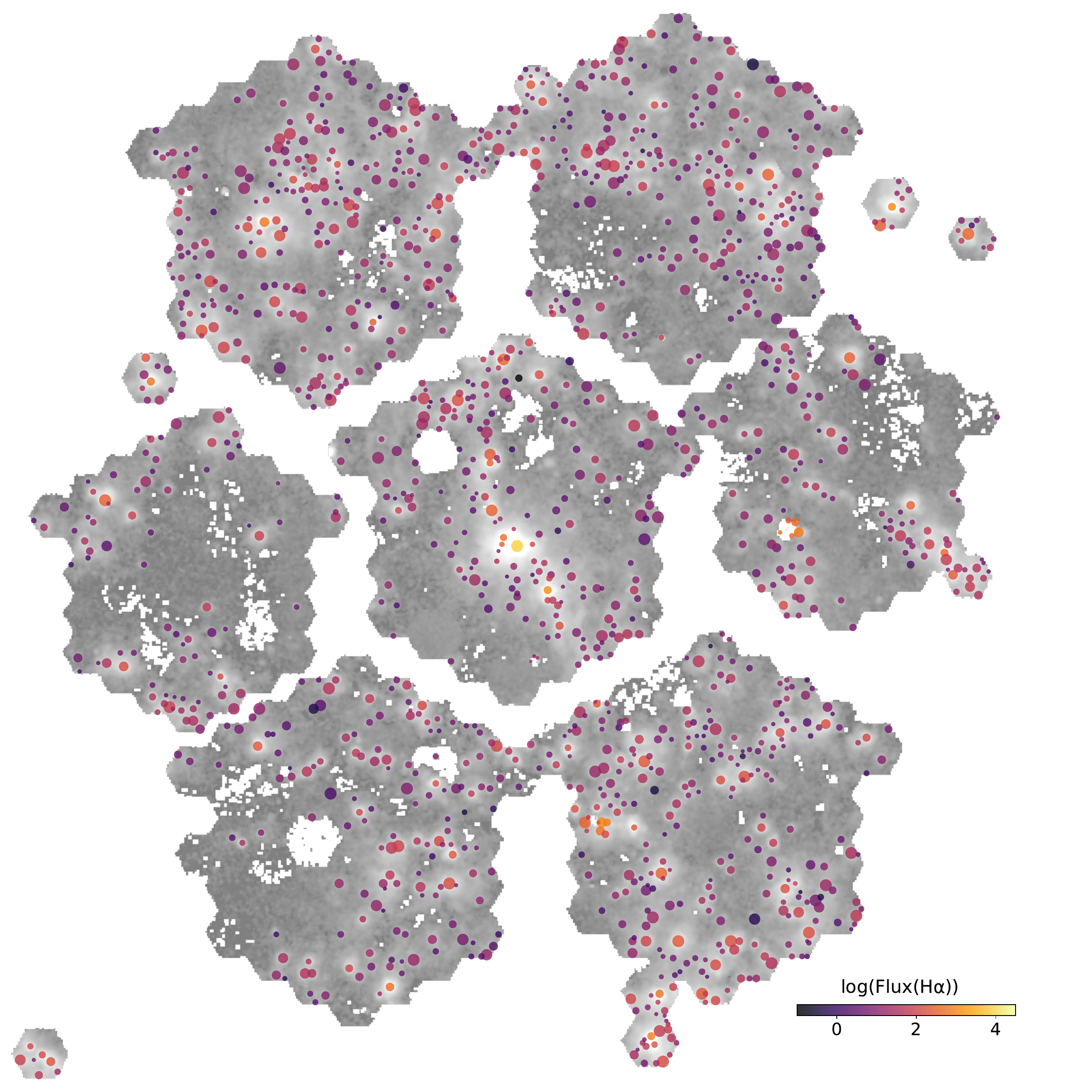}
    \caption{Flux of the \ha\ emission line from \hii\ candidates detected by \pyHII\ overplotted over the \ha\ emission line map derived using the DRP-LVM. }
    \label{fig:HIIregions}
\end{figure*}

\section{Analysis}
\label{sec:analysis}

\subsection{Detecting \hii\ regions using \textsc{PyHiiextractor}}

Having the maps of the physical properties derived from the different emission lines, we use the \ha\ emission mosaic to determine the locations and sizes of the \hii\ regions. In order to do this, we use the code \textsc{PyHiiextractor} \citep{LugoAranda_2022}. This code detects and extracts the properties of clumpy regions regions considered as \hii\ regions using an \ha\ emission-line image. If IFU data is available for that image, this code is also able to extract for each of the detected \hii\ regions the physical properties within that region (e.g., physical properties for each observed emission line, as well as, properties from the stellar continuum). Although \textsc{PyHiiextractor} also models the \ha\ diffuse emission (also known as diffuse ionized gas, DIG), in this study we focus only in the physical properties of the \hii regions. In a future study we will explore the properties of the DIG for \IC. In order to proceed to the estimation of the location and size of the \hii\ regions, the code requires a set of initial values. We refer the reader to \citet{LugoAranda_2022} for a detailed description of those parameter; here we describe the parameters we use in order to extract the physical information of the \hii\ regions. For this study, we systematically vary (i) the \ha\ flux threshold above the code detects clumpy regions, (ii) the maximum size of the detected \hii\ regions. We also set the code to search for \hii\ regions with 300 different sizes that vary from one spaxel to the maximum size. By visually inspecting the resulting sizes and locations of the \hii\ regions derived by \textsc{PyHiiextractor}, we find that appropriate initial values for the code are a value for (i) of \mbox{$\sim 5 \times 10^{-18}$ erg s$^{-1}\,\, \AA^{-1}$} and for (ii) the spatial resolution of the observations (set by the size of the fiber, $\sim$ 2.5 \arcsec). Since the range of sizes of the \hii\ begin with one spaxel (which does not have physical meaning), we discard regions with one spaxel of radius. We explore a wide range of maximum sizes for our \hii\ regions, however we find that having larger maxiuum sizes will yield spurious large \hii\ regions. In total we detect 4701 clumpy regions, however as we describe below to guarantee that we have reliable measurement of the physical properties of the \hii\ regions we apply further selection criteria. In Fig.~\ref{fig:HIIregions} we display as red circles the location and sizes of the \hii\ regions for \IC. As expected most of the \hii\ regions are located within the arms of the galaxy.   

To provide reliable estimations of the physical properties for each of the detected \hii\ regions, we use the following criteria: each region should have (i) all the emission line fluxes with positive values, and (ii) the signal-to-noise ratio of the \ha\ emission line has to be larger than three. In the next section we classify regions according to their location in their BPT diagram and their \ha\ equivalent width. From these criteria we have a final sample of 1155 \hii\ regions. 

\subsection{Derived Properties}

For each of the selected \hii\ regions we use the integrated/averaged properties derived from the LVM-DAP within its radius. Thus for each emission line we spatially integrated the flux, and average the other properties like the equivalent width, and the FWHM. Using this dataset we derive the physical properties that we are going to cover in this study. From the \ha/\hb\ ratio, known as the Balmer decrement, we derive the optical extinction, Av, following the prescription outlined in \citet{Catalan-Torrecilla_2015}. Using this estimation of Av, we derive the dust-corrected \ha\ luminosity for each region. On the other hand, to derived the other physical properties such as oxygen, nitrogen abundances, electronic density and ionization parameter we also make use of the integrated flux from other emission lines including the optical range of the observations. We refer the reader to consult the study from \citet{Espinosa-Ponce_2022}, to have a complete list of emission lines used in this study to derive the physical parameters mentioned above. We note that although in this study there is a large list oxygen abundance calibrators, we use in this study the one provided by \citet{Ho_2019}. Finally, regarding the kinematic properties of each region, we average the systemic velocity and velocity dispersion.

\subsection{The BPT diagnostic diagram and the clasification of \hii\ regions}
\label{sec:BPT}
\begin{figure*}
    \centering
    \includegraphics[width=\linewidth]{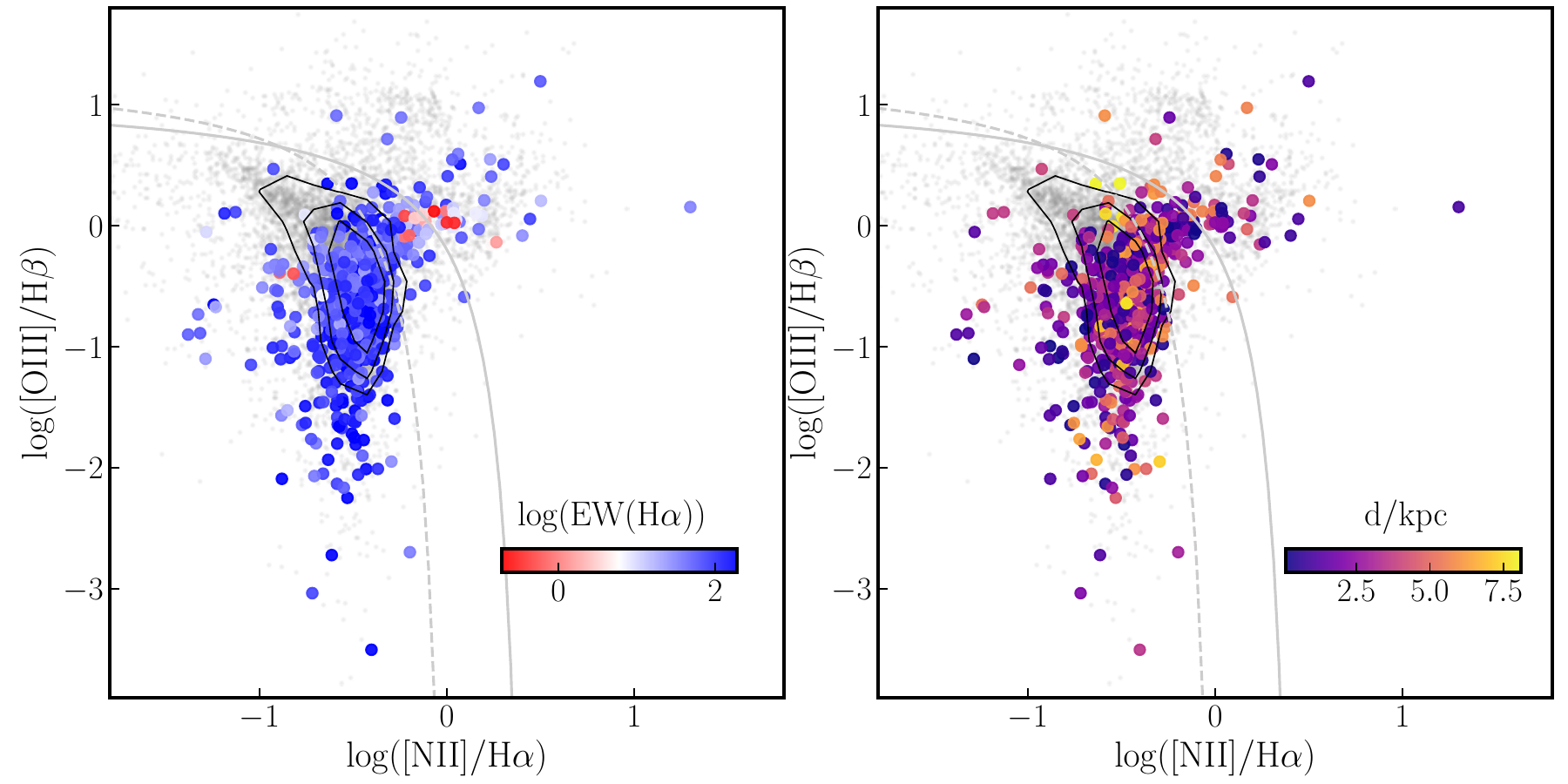}
    \caption{\hii\ candidates in the BPT diagram. {\it Left:} Candidates color codded by their \ha\ equivalent wdith. {\it Right:} Candidates color codded by their galactocentric distance. The dots represents the properties from the \AM\ sample. In both panels, the solid and dashed lines represent the Kauffman and Kewley demarcation lines. We use this demarcations lines and the \ha\ equivalent width to select  \mbox{\it bona fide} star-forming regions in \IC. }
    \label{fig:BPT}
\end{figure*}
Diagnostic diagrams have been widely used to classify galaxies/regions according to their ionizing source. This is the case of the BPT diagram \citep{BPT}. In this section we present the location of the candidates of \IC\ in the BPT diagram. Along with their EW(\ha), we use these line ratios to select the \mbox{\it bona fide} \hii regions. 

In Fig.\ref{fig:BPT} we plot the \hii\ candidates in the classical BPT diagram (i.e., \oiii/\hb\ vs \nii/\ha). As comparison, we also plot the line ratios from the \AM\ sample. Although most of the regions from both samples are located in the so-called 'star-forming branch', we note that the \AM\ sample does cover a wider dynamical range in the \nii/\ha\ axis than the \hii candidates from \IC. This is related to the differences we find previously in the radial distribution of the O3N2 ratio from these two samples. Furthermore, there are also \AM\ candidates with larger values of the \oiii/\hb\ ratio in comparison to those derived from the \IC\ sample. We suspect that this is due to regions with sources of ionization different than star-formation. However further exploration of the \AM\ sample of \hii\ candidates is beyond this study and will be addressed somewhere else (Aranda-Lugo et al. in prep.). 

A quantitative classification for the different ionization mechanisms is given by the demarcation lines proposed by \citet[][see dashed and solid lines in Fig.~\ref{fig:BPT}, respectively]{Kauffmann_2003, Kewley_2001}. Galaxies/regions below these demarcation lines are usually classified as star-forming while those above these demarcations could have different classifications (i.e., composite-like galaxy/regions  if the line ratios are between these demarcations or AGN-like galaxy/region if the ratios are above the Kewley demarcation line).  We find that the majority of regions lie below these demarcation lines in the BPT diagram ($\sim$80\% and $\sim$92\% for the Kauffmann and Kewley demarcation lines, respectively). 

The \ha\ equivalent width, along with the \nii/\ha\ ratio, has been also used to characterize the ionization source of galaxies/regions \citep[e.g.,][]{Cid-Fernandes2010, Lacerda_2018}. In the left panel of this figure we color code our candidates in the BPT diagram by their EW(\ha). We note that almost all the regions below those demarcation lines have large values of EW(\ha) with an average of $\sim$100~\AA. Conversely, most of the regions with low EW(\ha) are above the demarcation lines ($\sim$40~\AA). In comparison to the \AM\ sample, the IC\ candidates have large values of EW(\ha), both above and below the demarcation lines. In the right panel of Fig.\ref{sec:BPT} we color code the candidate according to their galactocentric distance. We do not find a clear trend between the distance and the location of the candidates in the BPT diagram. Regions above and below the demarcation line are distributed across the optical extension of \IC. Given the fact that flux ratios above the Kewley demarcation line have been typically associated with nuclear activity, this result suggest that a physical process other than nuclear activity could responsible for the ionization on those regions.

Depending of the study there are different thresholds to select star-forming galaxies/regions. \citet{Cid-Fernandes_2011} used as threshold EW(\ha) $>$ 6 \AA. In this study we use a more conservative value to select star-forming regions \citep[EW(\ha) $>$ 14 \AA][]{Lacerda_2020}. Thus, to obtain \mbox{\it bona fide} \hii regions from our sample of candidates we select those regions with flux ratios below the theoretical demarcation line proposed by \citet{Kewley_2001} as well as regions with the aforementioned EW(\ha). Using these criteria from the sample of 1155 candidates in \IC\ we select 960 as star-forming regions. Although it is beyond the scope of this study, we note that for those nebulae that were selected as non star-forming, a significant fraction are regions with low EW(\ha) below the Kewley demarcation line (121/195). This could indicate that the ionizing source is different as those expected from supernova remnants \citep[e.g.,][]{Vicens-Mouret_2023}. Further studies are require to assess the true nature of the ionization for these sources.

\section{Radial Distribution of Physical Parameters}
\label{sec:Radial}

\subsection{Properties derived from Balmer lines}
\label{sec:rad_Ha}
\begin{figure}
    \centering
    \includegraphics[width=\linewidth]{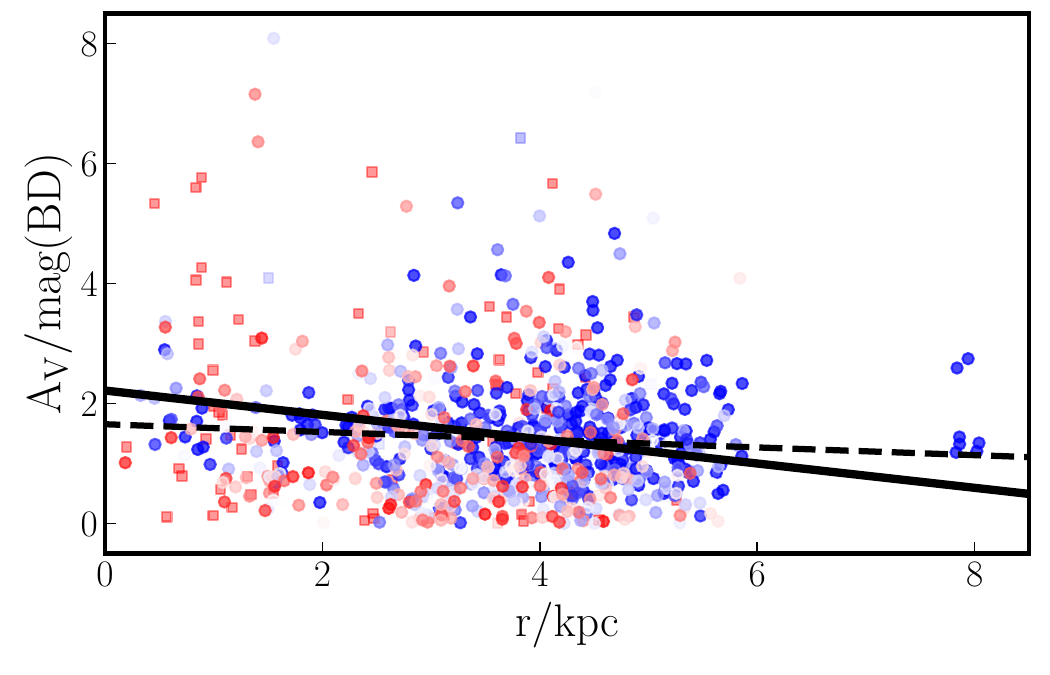}
    \caption{ Radial distribution of the optical extinction, Av, derived for the \hii\ regions in \IC. The symbols are color-coded by their EW(\ha). The solid line represent the best-fit radial gradient for the data. In average, we do not find a significant variation of Av for \IC. }
    \label{fig:rad_av}
\end{figure}
To obtain the \ha\ dust-corrected luminosity for our sample of \hii\ regions, it is necessary to estimate the extinction using the \ha/\hb\ ratio. In this section, we explore the radial distribution of this extinction. We also explore the radial distribution of both the \ha\ line equivalent width and the \ha\ luminosity.

Optical Extinction, $\mathrm{A_{V}}$: In Fig.~\ref{fig:rad_av} we plot the radial distribution of \Av\ for the sample of \hii\ candidates with a \ha/\hb\ ratio larger than 2.86. First, we note that the average extinction derived from the Balmer decrement for these candidates is close to one, $\Av \sim 1.3$ mag, with a large scatter measured from the standard deviation, $\Av \sim 1.4$ mag. Although the distribution of extinctions presents a large scatter, once the extinctions are represented against the galactocentric distance of the regions, it follows a rather decreasing trend. To further quantify this trend, we derive the radial gradient of the optical extinction by fitting a single-slope line to the dataset (black solid line in Fig.\ref{fig:rad_av}). We find that the best slope and zero-point are -0.18 mag / kpc and 2.21 mag, respectively. We note that there are regions with an extinction significantly larger than the median value, or in other words significantly above the best-fit gradient. These high-extinction regions are across the extension of the galaxy, with no clustering at a particular galactocentric distance. We also note that these regions cover a wide range of values of EW(\ha), as seen by the color code of the data points in Fig.~\ref{fig:rad_av}. In this figure we also segregate the candidates according to the selection criteria describe in Sec.\ref{sec:BPT}, this is between star and non-star forming regions (circles and squares, respectively). We note that in the central region (r$<$ 2~kpc) most of the regions with a high extinction are non-star forming candidates whereas at larger radii these  regions are mostly star-forming ones. To further understand the origin of these high-extincted regions, we map the residuals of the extinction with respect to its gradient in Fig.\ref{fig:AvRes_map}. We note that some of these regions appear to be clustered towards the arms of \IC\ (e.g., north-west arm at $\sim$ 4 kpc or the south-east pointing at $\sim$ 8 kpc). In the following sections we explore whether these atypical regions correlate with other properties of the regions. 

\begin{figure}
    \centering
    \includegraphics[width=\linewidth]{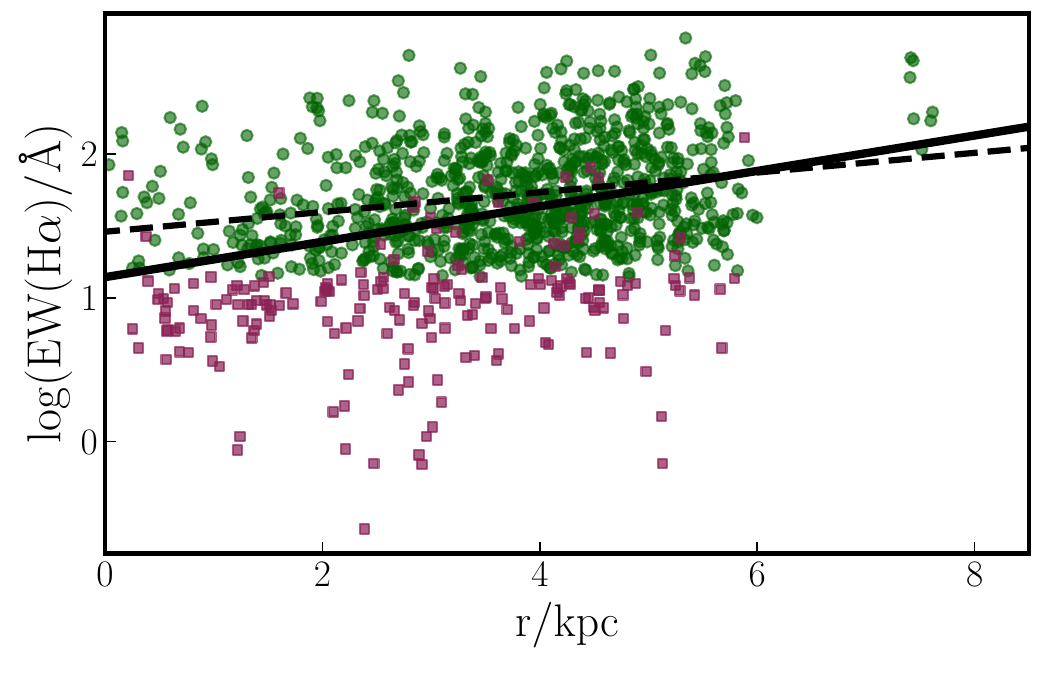}
    \caption{Radial distribution of the EW(\ha) for the \hii\ candidates in \IC.  Green, and pink symbols represent those regions classified as star and non-star forming, respectively. The solid and dashed lines represent the best-fit radial gradient for the entire data and using only \hii\ regions, respectively.}
    \label{fig:rad_EW}
\end{figure}

EW(\ha): Contrary to the the optical extinction, we find that in general the radial distribution of the EW(\ha) from the \hii\ candidates follows an incrasing trend (see Fig.~\ref{fig:rad_EW}). We plot this radial distribution color coded by the star formation status. We find that the star-forming regions follow a positive radial trend in contrast to non-star-forming regions with a large scatter and no evident trend (green circles and pink squares, respectively). On the other hand, when we derive the best single gradient for this radial distribution we find that selecting only star-forming regions lead to a shallower gradient in comparison to the one we derive using all \hii\ candidates (dashed and solid lines, respectively). In Fig.\ref{fig:EWRes_map} we color-coded for each position and radius of the \hii\ candidates the residual of the EW(\ha) with respect to the radial gradient we derive above. We find that the central star-forming regions are those with the larger enhancement of EW(\ha) whereas some of the external star-forming regions (e.g., at the outskirt north-west direction). On the other hand, we find that most of non star-forming regions have a deficit in their EW(\ha) in comparison to the derived gradient and are distributed in the central region of \IC. In the next section, we explore how the residuals from this radial gradient correlate or not with other physical properties derived from the emission lines.    

\begin{figure}
    \centering
    \includegraphics[width=\linewidth]{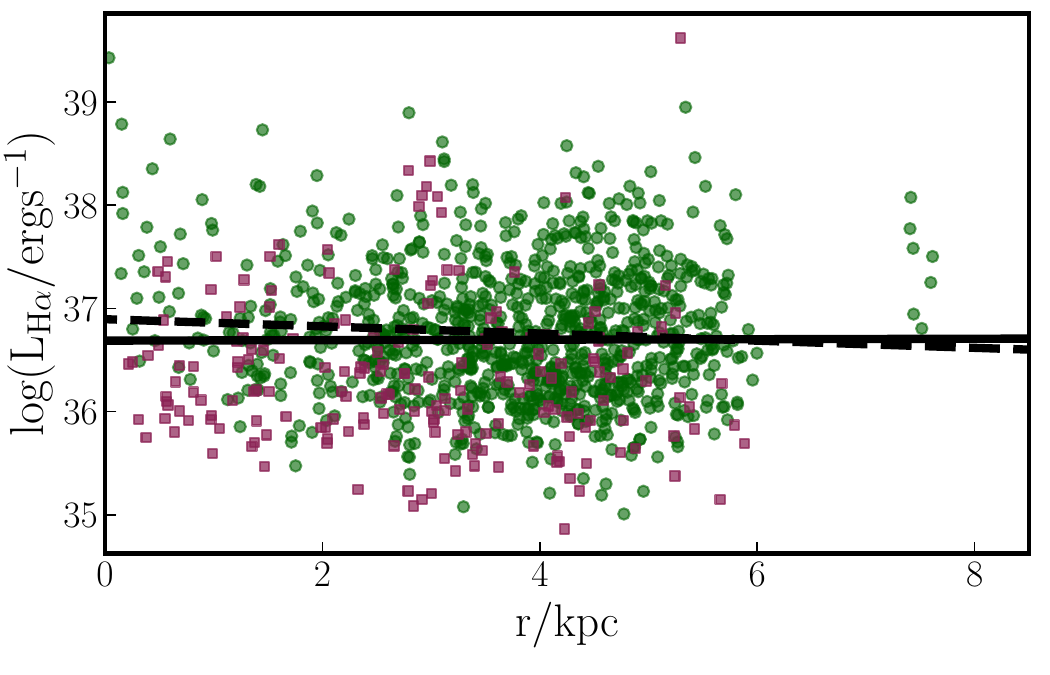}
    \caption{Radial distribution of the \ha\ luminosity. Colors of the symbols and lines are similar as those derived in Fig.\ref{fig:rad_EW}.}
    \label{fig:LHa_rad}
\end{figure}

\ha\ luminosity: In Fig.\ref{fig:LHa_rad} we plot the radial distribution of these luminosities. They follow a rather flat distribution with respect to their galactocentic distance. When we segregate our sample between star and non-star forming regions, we do not find a significant difference between the radial distribution of their \ha\ luminosities. For a given galactocentric distance both samples show a large scatter of their \ha\ luminosities. As for the previous parameters we derive the best radial gradient for these luminosities. We find a mild negative gradient, which gets even flatter when we derive it using only star-forming regions (solid and dashed lines, respectively).

\subsection{Oxygen Abundances}
\label{sec:rad_OH}
\begin{figure}
    \centering
    \includegraphics[width=\linewidth]{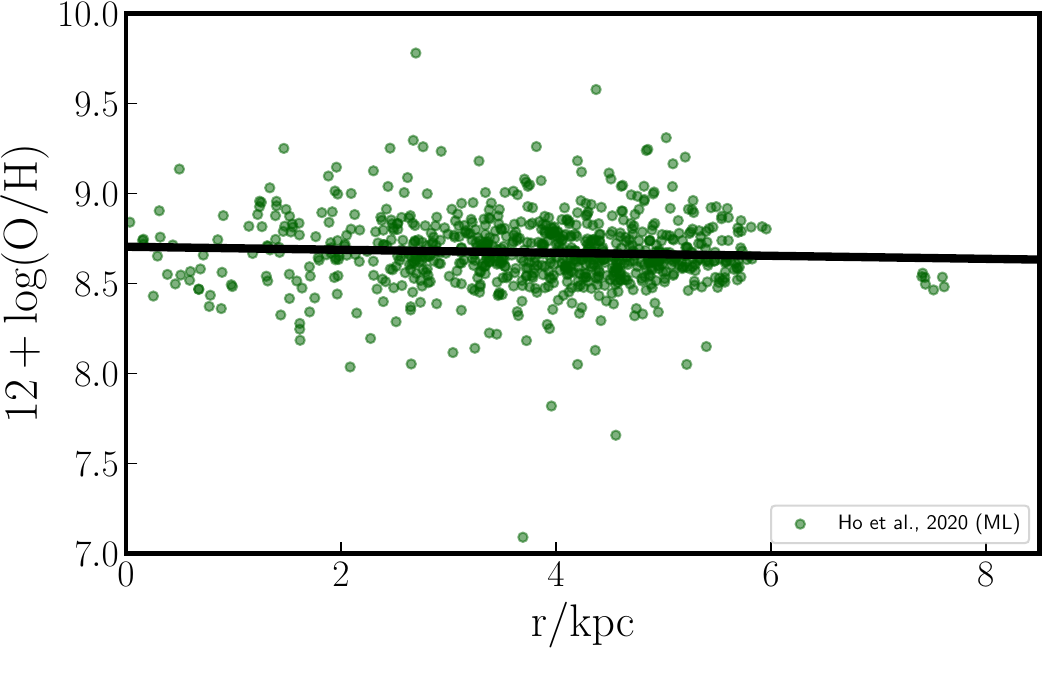}
    \caption{Radial distribution of the oxygen abundance using the calibrator from \citet{Ho_2019} for star-forming regions. The black solid line shows a best gradient fit representing the almost flat radial distribution of the oxygen abundance for the star-forming regions in \IC.}
    \label{fig:rad_OH}
\end{figure}
In Fig.~\ref{fig:rad_OH} we plot the radial distribution of the oxygen abundances for the sample of star-forming regions in \IC. As we mention in Sec.~\ref{sec:analysis}, from the large set of abundance calibrators, we use as fiducial calibrator the one derived from \citet{Ho_2019}. We note that the gradient provided from the best-fit is rather flat with a slightly negative slope. As in the other properties, we map in Fig.~\ref{fig:OHRes_map} the residuals of the oxygen abundance for each star-forming region where it is possible to derive the oxygen abundance. We do not find a cluster or specific location of regions with an excess or a deficit in their abundances. In Sec.~\ref{sec:residuals}, we further explore how the radial residuals of the different physical properties correlate with each other. 
\begin{figure}
    \centering
    \includegraphics[width=\linewidth]{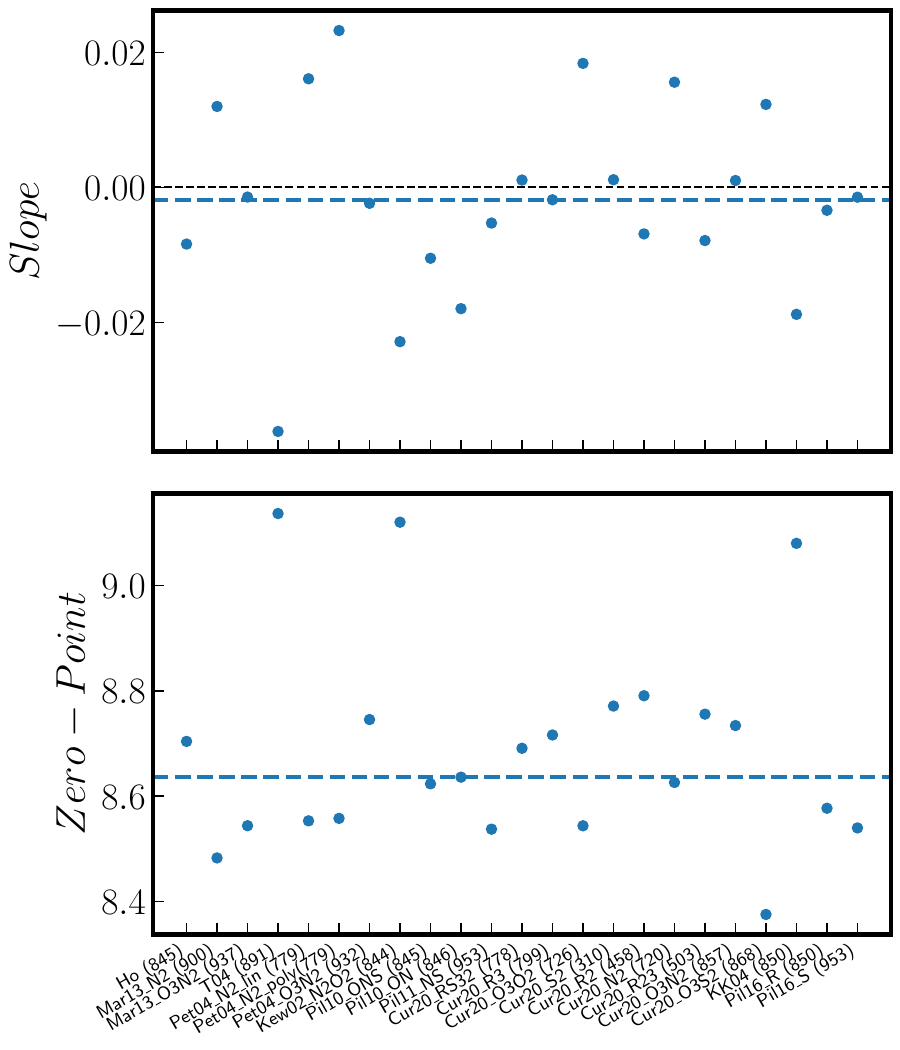}
    \caption{Best-fit parameters for the radial gradient of oxygen abundance using the set of calibrators described in Sec.~\ref{sec:analysis}: Slope and zero-point (top and bottom panels, respectively). In the top panel, the black-dashed line represents a flat gradient while he blue-dashed line represents the average of slopes from the different sets of calibrators. In the bottom panel the blue dashed line represents the average  of zero-points from the different calibrators. On average, the distribution of oxygen abundance in the star-forming galaxies of \IC\ is rather flat with a central oxygen abundance close to solar. We describe the number of star-forming regions used for each calibrators in parenthesis.}
    \label{fig:bf_OH}
\end{figure}

In order to quantify the impact of the abundance calibrators in the estimation of the best gradient, we use each of the calibrators presented in Sec.~\ref{sec:analysis} to derive the gradient of the oxygen abundance. In Fig.~\ref{fig:bf_OH} we plot the slope and zero-point from the best fit for the set of 23 calibrators. We note that although there are significant variations in both parameters, the average value of the slope is consistent with a flat radial distribution of the oxygen abundance. The central abundance given by the average zero-point is close to the solar abundance. Finally, we note that depending on the calibrator, the number of star-forming regions employed to estimate the best gradient varies. This difference in the sampling is due to (i) the emission lines required to derived a specific calibrator and (ii) the dynamical range imposed for each of them.   

\subsection{N/O ratio}
\label{sec:rad_NO}
\begin{figure}
    \centering
    \includegraphics[width=\linewidth]{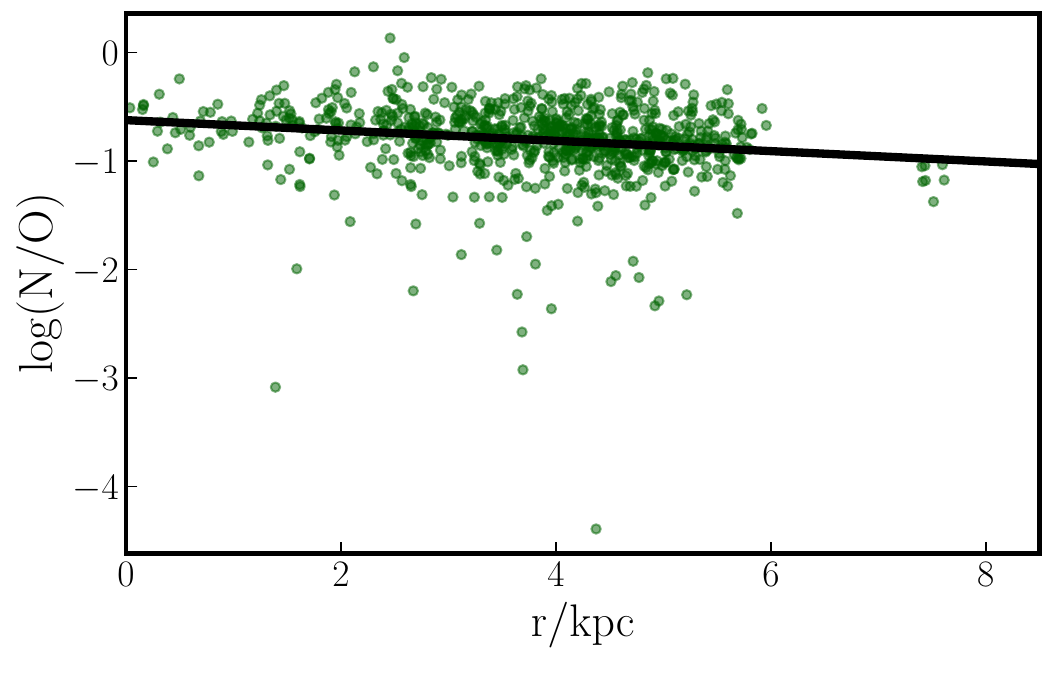}
    \caption{Radial distribution of the N/O ratio for star-forming regions.As for the oxygen abundance, the black line represent the best fit for this radial distribution.}
    \label{fig:rad_NO}
\end{figure}
\begin{figure}
    \centering
    \includegraphics[width=\linewidth]{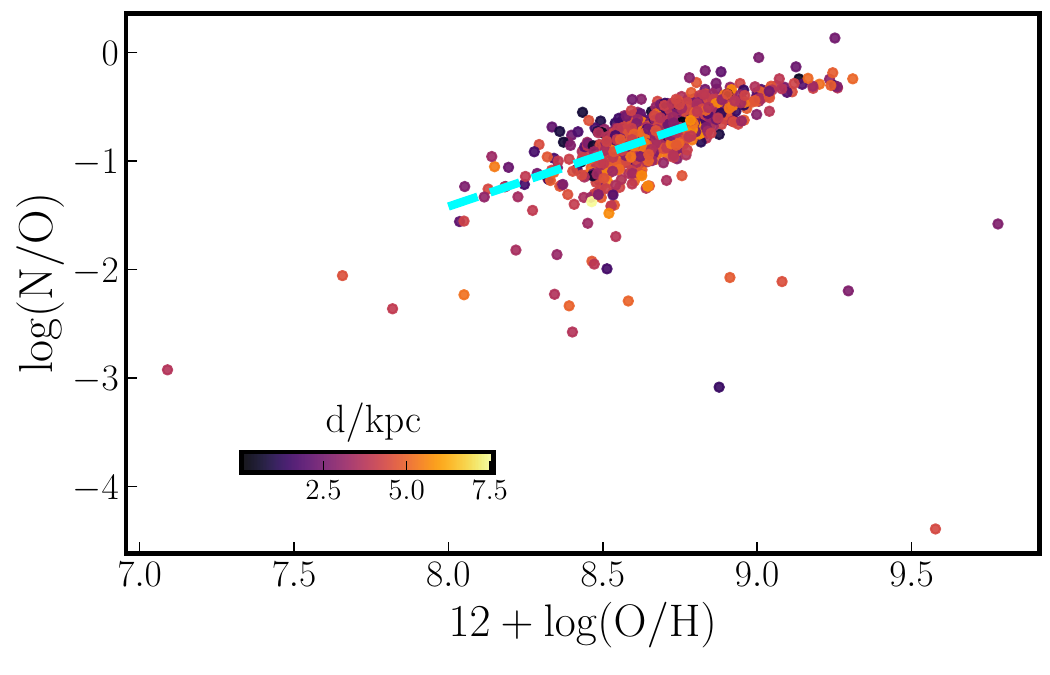}
    \caption{ Nitrogen-to-oxygen ratio against the oxygen abundance ratio. The cyan-dashed line represents the best fit for these two parameters derive from \hii\ regions observed within the CALIFA survey \citep{Espinosa-Ponce_2022}. We note that in comparison to a large sample of regions, those from \IC\ have large N/O ratios.}
    \label{fig:NOvsO}
\end{figure}
The measurement of chemical abundances from elements with different nucleosynthesis processes allows us to trace the scales of star formation in the evolution of \IC. Short-lived massive stars are traced by the oxygen abundance while nitrogen abundances traces longer lifetimes and less mass-star. As we mention in Sec.~\ref{sec:analysis}, the measurement of the nitrogen abundance is done using the calibrator presented in \citet{Pilyugin_2016}, whereas for the oxygen abundance we use our fiducial calibrator \citep{Ho_2019}.

In Fig.~\ref{fig:rad_NO}, we plot the radial distribution of the N/O ratio for the sample of \hii\ regions we detect in \IC. We find that a large fraction of the \hii\ regions show a decreasing value of this ratio as the galactocentric distance increases. For some few regions ($<$5\%) their N/O ratio is significantly smaller than regions at a similar galactocentric distance. The best fit of the radial distribution of the O/H ratio confirms a negative gradient (black line in Fig.~\ref{fig:rad_NO}). As for the oxygen abundance, in Fig.~\ref{fig:NORes_map} we color-coded the \hii\ regions with the residuals of the N/O ratio with respect to the derived radial gradient . We do not find a specific location where there is a larger deficit or enhancement of this ratio. We also do not find a possible correlation of the deficit  with respect to a morphological feature (e.g., an arm, inter-arm, etc).  

To further understand this ratio in the context of galaxy evolution, in Fig.~\ref{fig:NOvsO} we plot the N/O ratio against the oxygen abundance color-coded by their distance. We find that most of the regions, thus independent on the distance, are clustered in the well-studied linear relation between N/O and O/H \citep[e.g.,][]{Belfiore_2017, Espinosa-Ponce_2022}. We note that with respect to large samples of \hii\ regions included in a large number of galaxies \citep[e.g., the CALIFA survey][, see cyan-dashed line in Fig.~\ref{fig:NOvsO}]{Espinosa-Ponce_2022}, those from  \IC\ exhibit large metalicities and thus large N/O ratios. In relation to the regions with large deficit of the N/O ratio, we find that these \hii\ regions cover a wide range of metallicity. In conclusion, the bulk of \hii\ regions for \IC\ exhibit a similar trend in the N/O-O/H plane as \hii\ regions in other extragalactic sources. Furthermore, some \hii\ regions in this galaxy show larger values of the oxygen abundance and N/O ratio. Those regions appear to follow the linear trend described from large samples of \hii\ survey. Finally, it is not clear the underlying reason for having \hii\ regions with a significant deficit of the N/O ratio.  

\subsection{Electronic density and Ionization parameter}
\begin{figure}
    \centering
    \includegraphics[width=\linewidth]{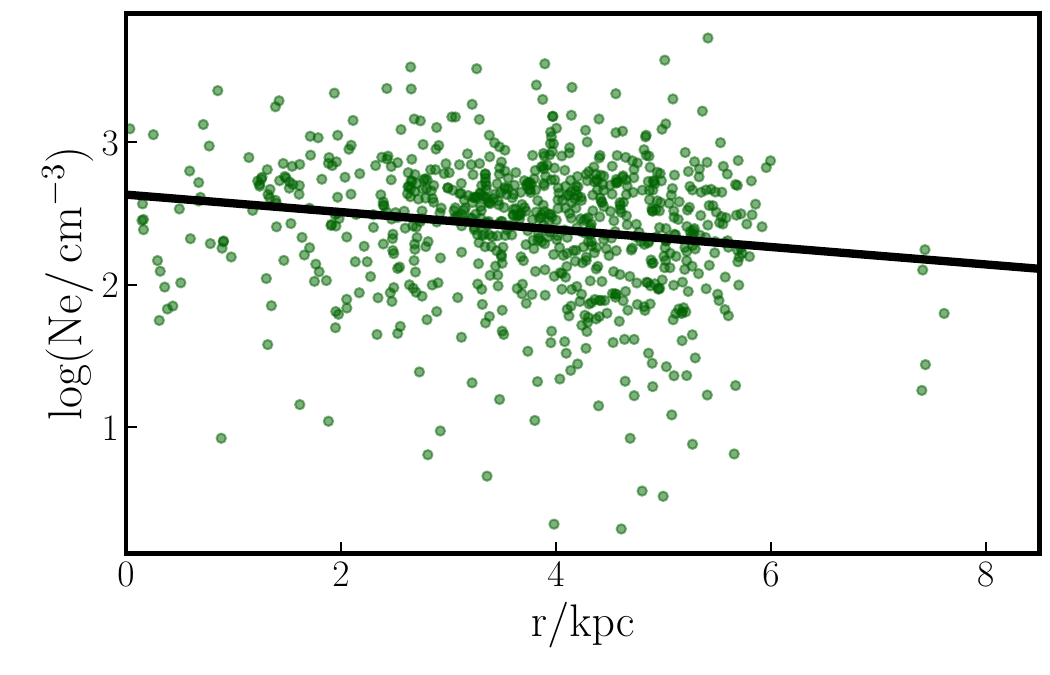}
    \caption{Similar to Fig.\ref{fig:rad_NO}, radial distribution of the electronic density for star-forming regions.}
    \label{fig:rad_Ne}
\end{figure}
\begin{figure}
    \centering
    \includegraphics[width=\linewidth]{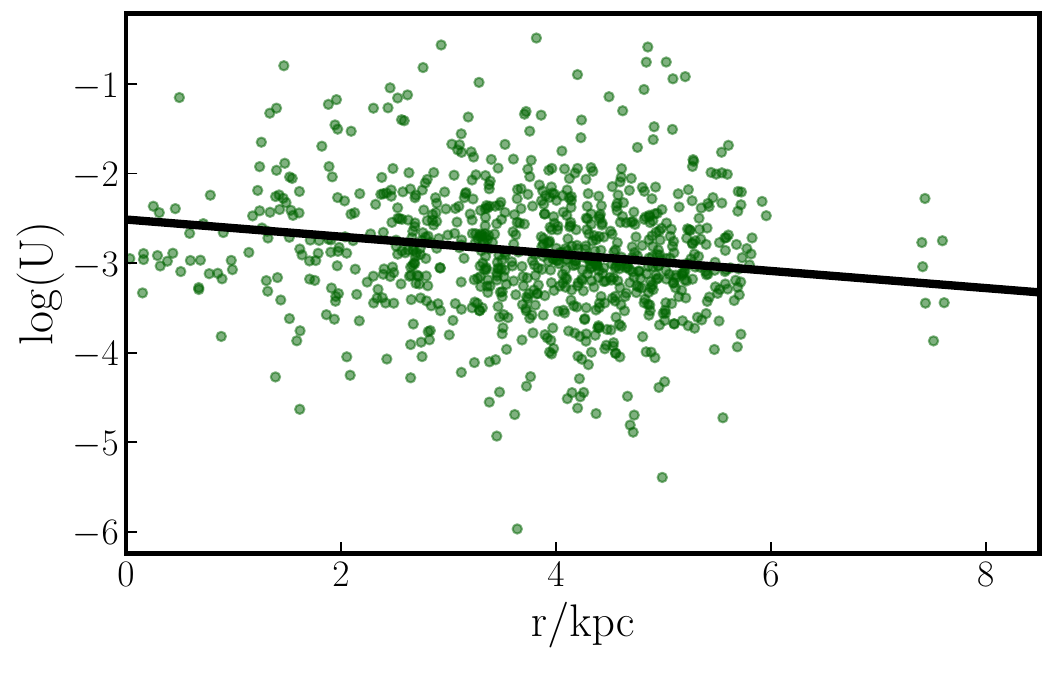}
    \caption{Similar to Fig.\ref{fig:rad_NO}, radial distribution of the ionization parameter for star-forming regions.}
    \label{fig:rad_U}
\end{figure}
\begin{figure}
    \centering
    \includegraphics[width=\linewidth]{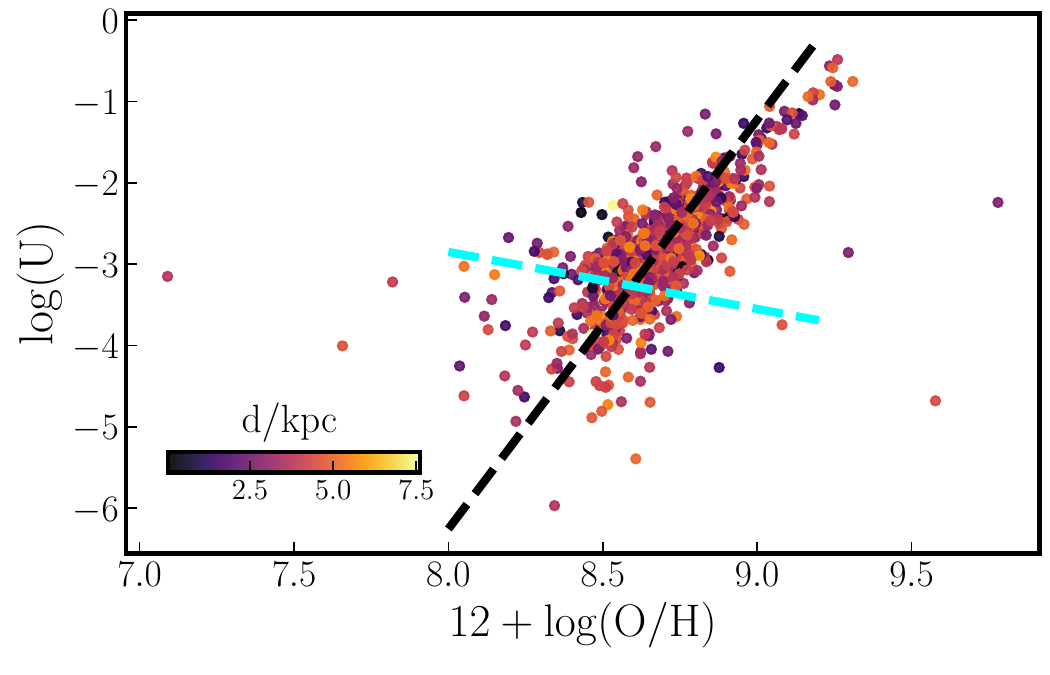}
    \caption{ Comparison between the ionization parameter and the oxygen abundance for our sample of star-forming regions. Data point are color-coded with respect to the galactocentric distance. The black dashed line represent the best linear fit to the data, while the cyan dashed line represents the best fit presented in \citep{Espinosa-Ponce_2022}.}
    \label{fig:UvsO}
\end{figure}

In Fig.\ref{fig:rad_Ne} we show the radial distribution of the electron density, $n_e$, for the sample of \hii\ regions where we can measure this physical property. Despite the large scatter, we find that regions at larger galactocentric distance tend to have lower densities in comparison to those at closer distances. This is somehow expected as regions in the center (where the stellar and gas densities are larger) are prone to be under larger pressure in comparison to regions at the outskirts \citep[e.g.,][]{Barrera-Ballesteros_2021a, Barnes_2021}. This trend is further quantified by the best radial gradient which has a negative slope. As for the other properties we also explore how the residuals of $n_e$ with respect to this gradient are spatially distributed (see Fig.~\ref{fig:NORes_map}). We do not find significant patterns in the spatial distribution of these residuals. 

In Fig.\ref{fig:rad_U} we plot the radial distribution of the ionization parameter for our sample of \hii\ regions. Similar to other properties of the \hii\ regions, we find that this parameters decreases as the galactocentric distance of the regions increases. This is confirmed by the negative slope obtained from the best-fit gradient. Similar radial trends have been observed in other very nearby galaxies \citep[e.g., ][]{Kreckel_2019, Grasha_2022}.  It has been suggested a significant relation between the oxygen abundance (and the N/O ratio) with the ionization parameter for star-forming regions. Initial studies of the properties of \hii\ regions suggested that these two parameters exhibit an anti-correlation, this is that $\log(U)$ decreases as the oxygen abundance increases \citep[e.g., ][]{Dopita_1986}. A similar trend has been found using empirical calibrators of $\log(U)$ as those used in this study for a large sample of star-forming regions included in galaxies observed in the CALIFA \citep[e.g., Fig.~3 in ][]{Sanchez_2015b}. The negative radial gradients from these two parameters derived for our sample of \hii\ regions indicate otherwise. To explore the correlation of these two parameters, we plot in Fig.~\ref{fig:UvsO} these two parameters against each other color coded by the galactocentric distance of the regions. As expected, we find a clear positive trend between these two parameters: metal-rich regions have a large ionization parameter. This is further quantified by the slope of the best linear fit between these two properties (black-dashed line). This result is in contrast to recent characterizations of large samples of \hii\ regions in the nearby universe. Using the same calibrators for both properties, \citet{Espinosa-Ponce_2022} found a negative correlation between these two parameters (see cyan-dashed line). They noted however, that for their sample of regions with low abundances ($12+\log(\mathrm{O/H}) \lesssim$ 8.4) there is indeed a negative trend of $\log(U)$ with $12+\log(\mathrm{O/H})$ whereas that regions with high abundances the trend changes to a positive one. They also mentioned that this relation, contrary to other ones, is highly dependent on the calibrator of the ionization parameter. As we describe above, the average oxygen abundance of the regions from \IC\ is larger than the one from those regions in a larger sample of galaxies such as the one presented by \citet{Espinosa-Ponce_2022}. Thus, our results are in agreement with those presented in that studies, this is, that $\log(U)$ increases with the oxygen abundance for regions with high metallicity. Although there are different possible explanations for the relation between these two parameters \citep[see a detailed discussion in][]{Ji_2022}, it is important to keep in mind that we are using empirical calibrators as tracers of these two parameters. In particular for $\log(U)$ we are considering that these empirical calibrators trace the hardness of the ionizing source. However, as discussed by \citet{Espinosa-Ponce_2022}, those calibrators gauge the excitation level of the ionized gas which in turn depend on both the hardness and the shape of the ionizing source. Finally, we note that similar positive trends have been observed in 20 nearby galaxies included in the PHANGS-MUSE survey \citep{Groves_2023}, suggesting that differences in the correlations could be caused by resolution effects.  

\subsection{Kinematic Properties}
\label{sec:Kin}
\begin{figure}
    \centering
    \includegraphics[width=\linewidth]{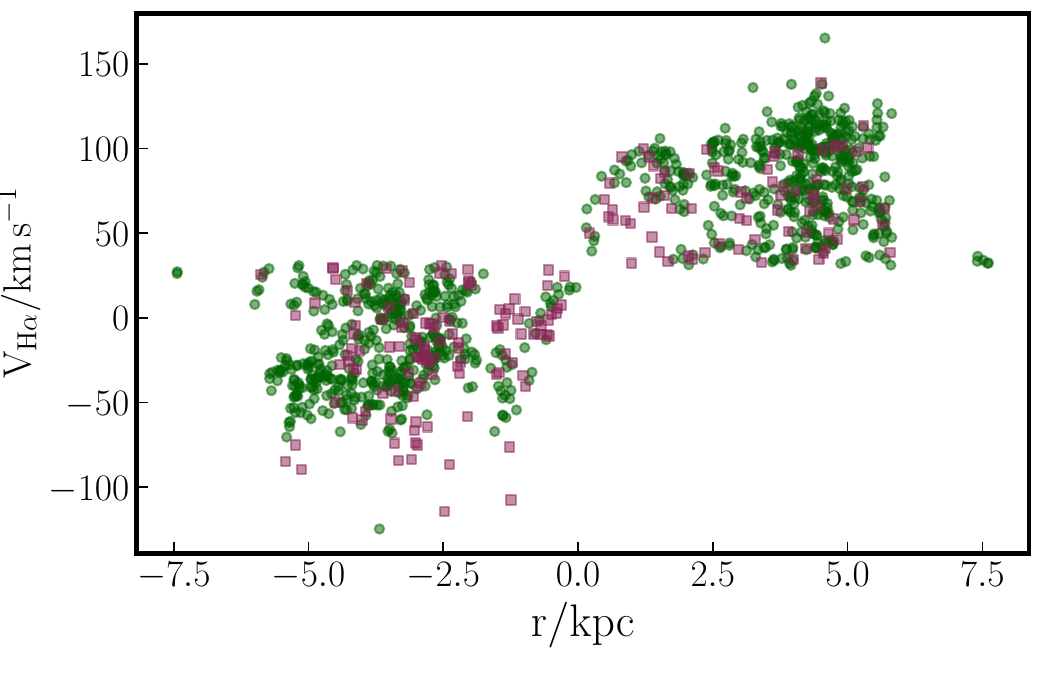}
    \includegraphics[width=\linewidth]{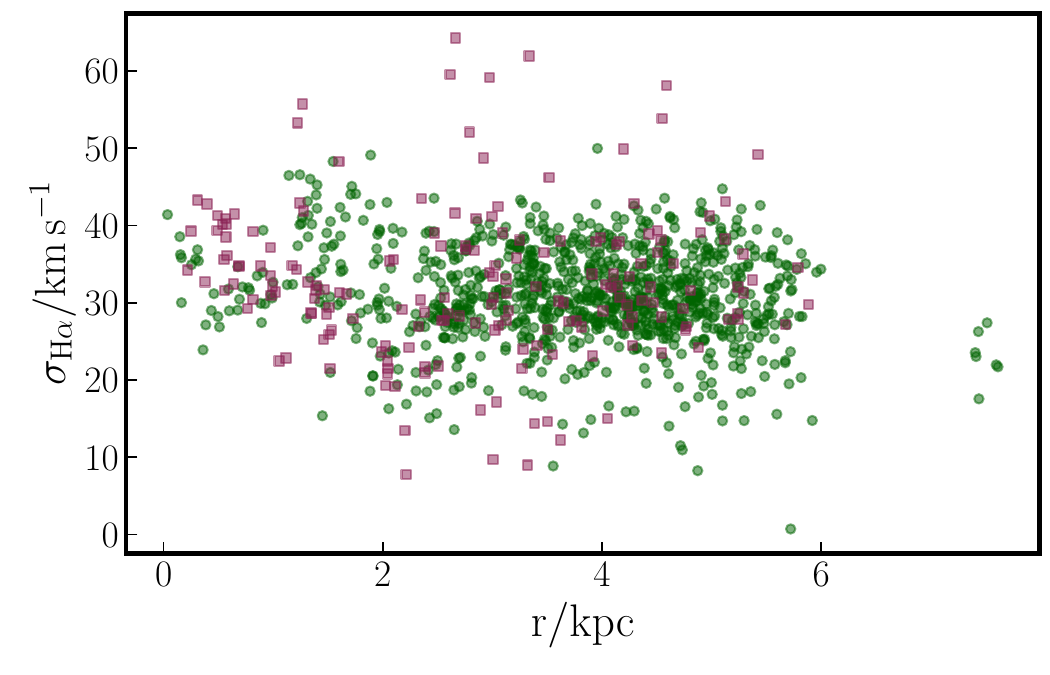}
    \caption{ Radial distribution of the systemic velocity  and the velocity dispersion of the \hii\ regions (top, and bottom panels, respectively). Similar to previous plots, we segragate the regions between star and non-star forming (green circles and pink squares, respectively) .}
    \label{fig:kin}
\end{figure}

In Fig.~\ref{fig:kin} we plot the radial distribution of the kinematic properties of the \hii\ regions detected in \IC. For the systemic velocity (top panel) we use the value reported by \citet[][i.e., 32 km s$^{-1}$]{Tully_1988}. Regarding the line-of-sight velocity of the \hii\ regions we find that the north-west regions presents a receding velocity where as the south-east regions present an approaching velocity. Our results are in agreement as those derived previously for this galaxy using Fabry-Perot observations \citep{Hernandez_2005}. Although it is beyond the scope of this study, it is also necessary to explore how the variations of line-of-sight velocity with respect to the expected circular velocity for a given radius correlates or not with other physical properties of the galaxy. This would be significantly useful in order to understand the impact of, for instance, shocks or other non-circular motions in the physical properties of the ISM. Given the fact that in order to obtain the curve that best describe the radial distribution of the velocity of the \hii\ regions it is necessary invoke a dynamical model, we do not attempt to derive a fit. As mentioned above this is beyond the scope of this article.  

In the bottom panel of Fig.~\ref{fig:kin} we show the radial distribution of the velocity dispersion for our sample of \hii\ regions. As expected for a disk galaxy, the dispersion are small ($\sigma_{H\alpha} \sim 30 \mathrm{\,km\,s^{-1}}$). Furthermore, it is no evident a significant variation of the \ha\ velocity dispersion with respect to the galactocentric distance. We also note that for the outer parts of the galaxy (i.e., r $>$ 2 kpc), the regions with the larger \ha\ velocity dispersion are those that we consider as non-star forming. In Sec.~\ref{sec:residuals} we explore whether the velocity dispersion have any relation with the radial residuals derived for the other physical parameters derived in this study.

\section{\ha\ Luminosity Function}
\label{sec:LF_Ha}
\begin{figure}
    \centering
    \includegraphics[width=\linewidth]{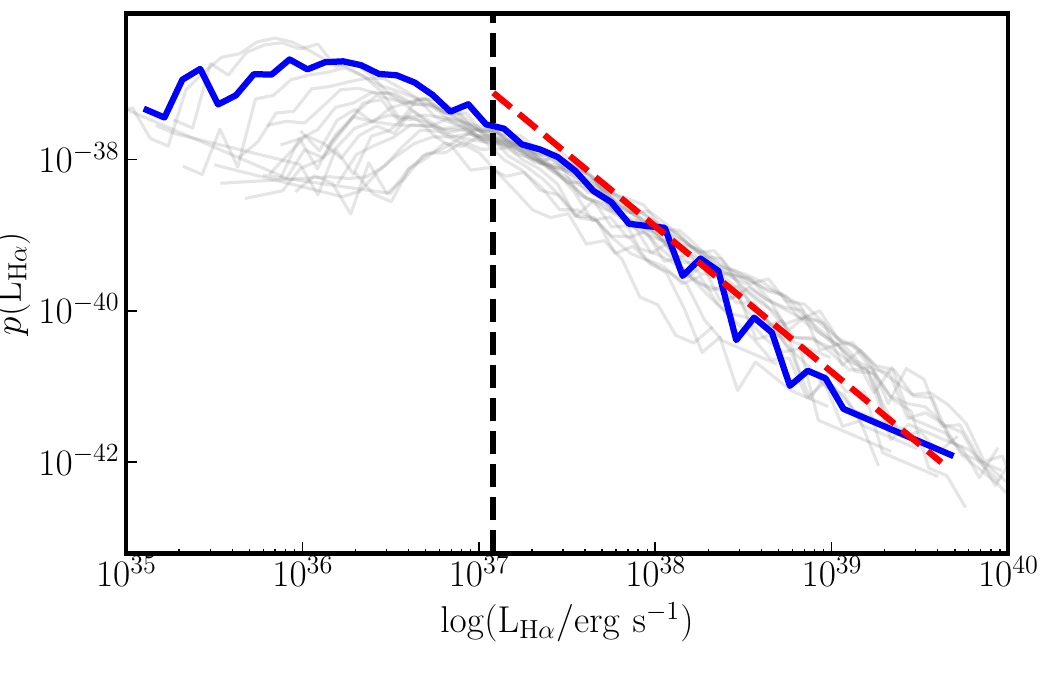}
    \caption{ Probability distribution function of the \ha\ luminosity for the \hii\ regions detected in \IC\ (blue solid line). This distribution function is compared to those derived for galaxies included in the PHANGS-MUSE survey \citep[][gray solid lines]{Santoro_2022}. The red dashed line represents the best fit of a power-law to the dataset. The vertical dashed line represents the minimum luminosity used to derived the fit.}
    \label{fig:LF_comp}
\end{figure}
\begin{figure}
    \centering
    \includegraphics[width=\linewidth]{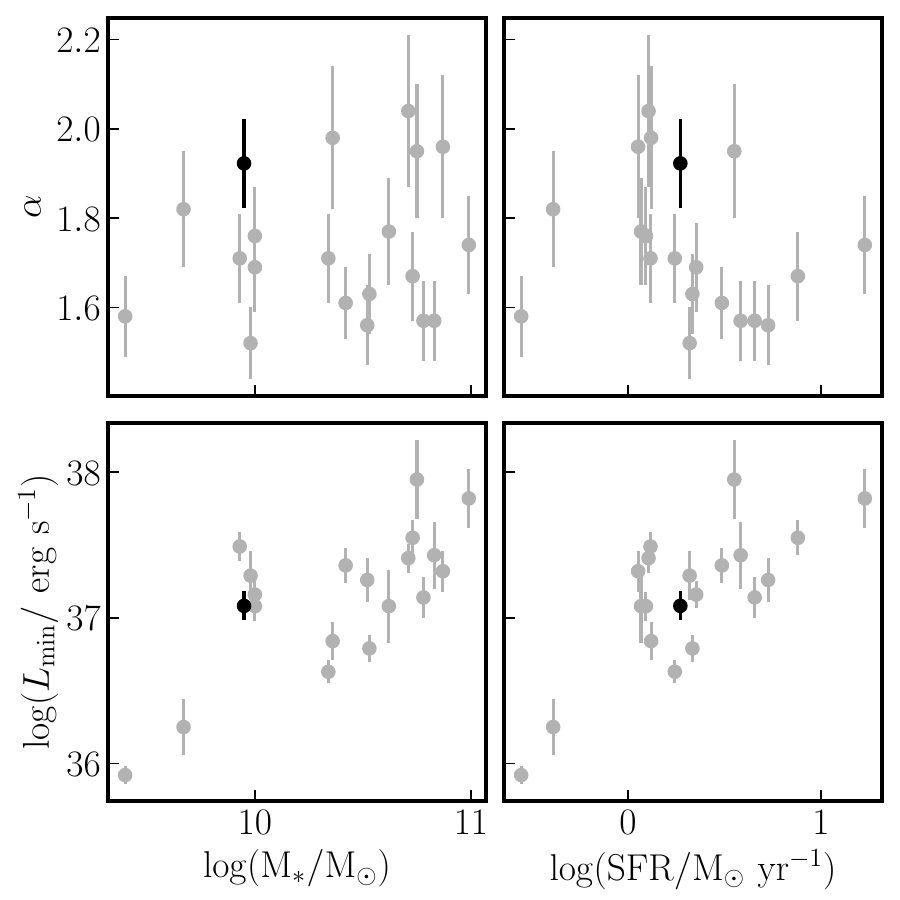}
    \caption{Comparison between the best fit parameters derived for \IC\ ($\alpha$, and $\mathrm{L_{min}}$; black circles) and those derived for the PHANGS-MUSE galaxies (gray circles). Right panels show these two parameters against the total stellar mass of galaxies while left panels show these parameters against the integrated SFR.}
    \label{fig:HaLF_comp}
\end{figure}

Since the \ha\ luminosity in an \hii\ region has been considered as a tracer of the formation of young massive stars, exploring the \ha\ luminosity function in these regions (\haLF) is going to trace the massive end of the mass function in galaxies. Thanks to the spatial resolution of the dataset explored in this study, we are able to derive the \haLF\ for the \hii\ candidates of \IC\ which is one of the closest galaxies where the \haLF\ has been explored so far. Since this function is basically a histogram of the \ha\ luminosity function of \hii\ regions, the estimation of the best slope is heavily depending on how the binning of that histogram is selected. Even more this slope is derived using the brightest \hii\ regions, thus it is required to select a minimum luminosity ($\mathrm{L_{min}}$), which could vary depending on the adopted binning scheme (e.g., same number of regions per bin or fixed luminosity per bin). Instead of using a histogram, we follow a similar procedure as the one presented by \citet{Santoro_2022} to derive the \haLF\ for a sample of 19 galaxies included in the PHANGS-MUSE survey. Following a similar methodology also allows as to provide a direct comparison between nearby galaxies observed with IFS.  In a few words, the method does not depend on any binning scheme as it models the probability distribution function (PDF), $p(L)$, using a power-law of the form:

\begin{equation}
   p(L) =  (\alpha -1)\,L^{\alpha-1}_\mathrm{min}\, L^{-\alpha}\,\,\, \text{with}\,\,\, L \geq L_\mathrm{min}.
\end{equation}
The best fit of the \haLF\ is done using the Python package \textsc{powerlaw} \citep{Alstott_2014}. This algorithm uses a maximum likelihood estimation method in combination with a Kolmogorov-Smirnov statistics to derived the best values of $\mathrm{L_{min}}$ and $\alpha$, providing a robust statistical way to derive these parameters \citep{Clauset_2009, Santoro_2022}. To derive the uncertainties for these parameters, we run a 1000 realizations of the \haLF\ varying the values of the fluxes of the \ha\ emission within their uncertainties. 

In Fig.~\ref{fig:LF_comp}, we plot the PDF of the \hii\ regions detected in \IC\ (blue-solid line), the best fit of the \haLF\ (red-dashed line) along with the PDFs derived for the PHANGS-MUSE galaxies from \citet[][gray lines]{Santoro_2022}. The PDF of \IC\ covers a similar range of luminosities than those derived for the PHANGS-MUSE sample. Its shape is also similar as some of those galaxies. The find that the best-fit values for \IC\ are \mbox{$\alpha = 1.9 \pm 0.1$} and \mbox{$\mathrm{\log(L_{min} / erg\,\,s^{-1}) = 37.1 \pm 0.1}$}. We also compare how these parameters compare with those derived for the PHANGS-MUSE galaxies. In Fig.~\ref{fig:HaLF_comp} we plot $\alpha$ and $\mathrm{L_{min}}$ against the integrated stellar mass, and the SFR (left and right panels, respectively) for the PDFs presented by \citet[][grey circles]{Santoro_2022} and the one derived in this study for \IC. In general, we observe that the slope of the \haLF\ for \IC\ and the value of $\mathrm{L_{min}}$ are within the dynamical range of those derived for nearby galaxies. However, we note that \IC\ exhibits one of the steepest slopes in comparison to those derived for the PHANGS-MUSE galaxies. This is evident for galaxies with similar stellar mass (i.e., $\mathrm{\log(M_{\ast}/M_{\odot}) \sim 10}$) and SFR. On the other hand, $\mathrm{L_{min}}$ is in the average of the values of other galaxies and follows a similar trend as the luminosities derived from the PHANGS-MUSE galaxies, this is that they increases with both the stellar mass and the SFR.
\section{Correlation with residuals from radial trends}
\label{sec:residuals}
\begin{figure*}
    \centering
    \includegraphics[width=\linewidth]{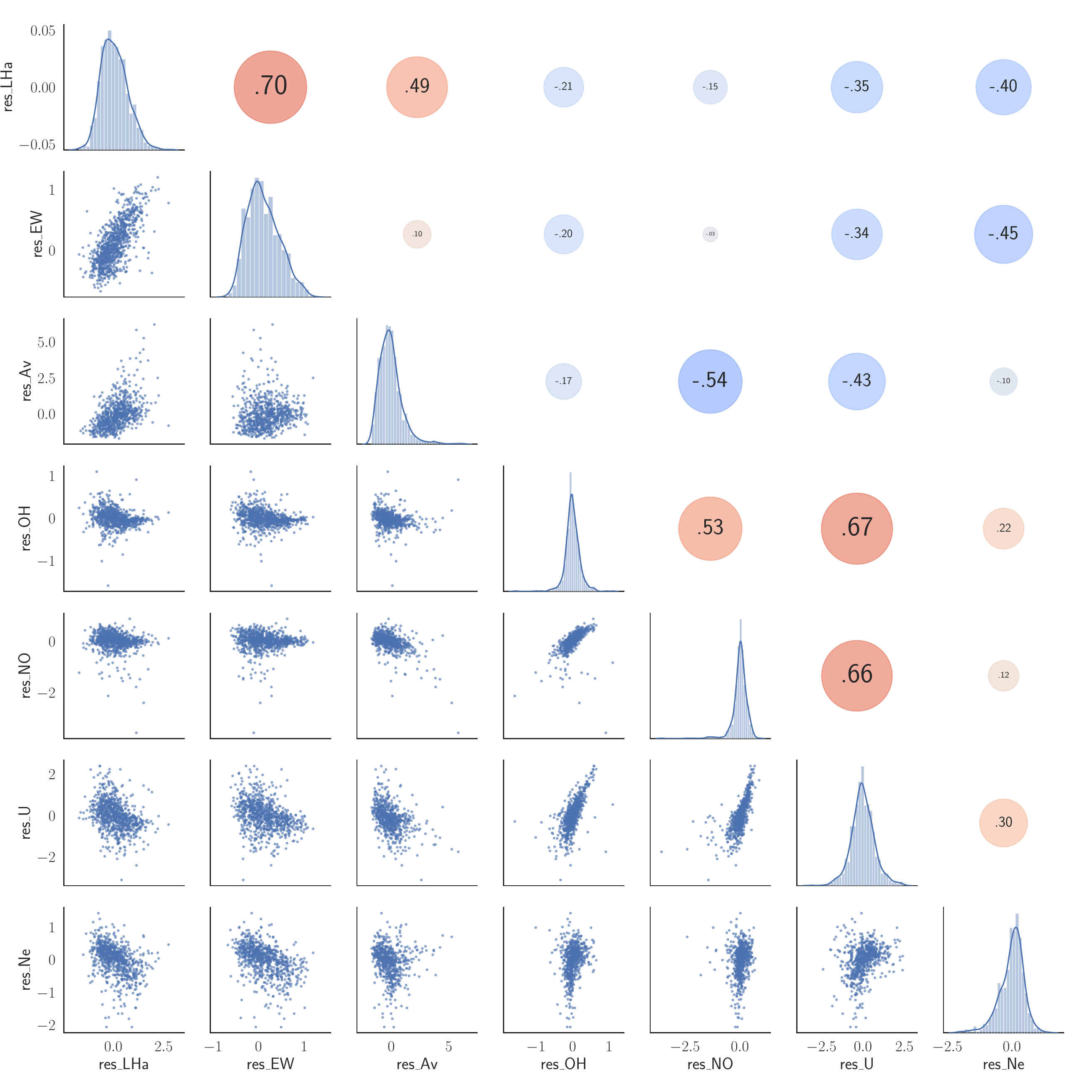}
    \caption{ Correlation matrix comparing the radial residuals of the different physical properties of the regions considered as star-forming. The diagonal panels show the histogram of the residuals whereas the lower off-diagonal panels show the distributions and the upper off-diagonal panels show the Pearson's correlation coefficient for parameters corresponding to the transposed panel. }
    \label{fig:res_plot}
\end{figure*}

In Sec.~\ref{sec:Radial} we describe the radial trend of the derived properties of the \hii\ regions in \IC. As we mention in that section, it is important to explore if once we remove the radial trend of these parameters we still find correlations between the parameters. In Fig.\ref{fig:res_plot} we plot the correlation matrix for the radial residuals of those parameters we describe in Sec.~\ref{sec:Radial}.  This plot gives us a deeper exploration on the possible relations that these residuals could have. 

From left to right, in the first column of Fig.~\ref{fig:res_plot} we plot the radial residuals of the explored properties against the \ha\ luminosity residual, \DeltaL. For \DeltaL, we find that it strongly correlation with the residuals of EW(\ha), $\Delta EW(\ha)$. This is somehow expected, for regions with a large \ha\ luminosity the flux of the \ha\ emission line is significant larger in comparison to the adjacent continuum. Furthermore, the fact that we are finding this strong correlation using the radial residuals suggest that this is a correlation independent of the galactocentric distance. Besides this, the other positive correlation we find for \DeltaL is with the residuals of the optical extinction, $\Delta(Av)$. As we mentioned in \citet{Barrera-Ballesteros_2020}, the optical extinction in nearby galaxies is tracing the amount of cold gas at kpc scales. Thus, since these two parameters are usually associated to the current star-formation and the amount of available cold gas to form new stars, we could think that this is manifestation of the well-known star-formation law \citep[e.g.,][]{Kennicutt_2012}. For the residuals related to the chemical enrichment of the \hii\ regions we do not find significant correlations (i.e., $\Delta O/H$, and $\Delta N/O$, respectively). This is relevant, as it indicates that the luminosity of a given \hii\ region does not strongly correlates with the chemical abundance of their ISM, in turn this suggest that the chemical content of star-forming regions is associated to the chemical enrichment from previous generations of stars \citep[e.g., ][]{Sanchez_2015b}. However, previous studies have suggested significant correlations between $L(\ha)$ and the residuals of the oxygen abundance, suggesting the influence of the star formation with the chemical enrichment  within \hii regions \citep[e.g., ][]{Kreckel_2019}. Finally, we find mild negative correlations between \DeltaL\ and the radial residuals of the ionization parameter and the electronic density. Although we find that these three parameters present a negative radial gradient, in other words they follow the same trend with galactocentric distance, once we remove the contribution from the radial gradient, these two properties decreases as \DeltaL\ increases. At first glance, it is not clear why for a given galactocentric distance brighter regions in \ha\ tend to have lower densities.

Contrary to \DeltaL, for the residuals of EW(\ha) we do not find significant correlations with other radial residuals, except for the one derived for the electronic density. Our analysis indicates that for a given galactocentric distance, regions with large EW(\ha) tend to have lower electronic densities. As for \DeltaL, it is not clear why regions associated with large star-formation activity are less dense than those where star-formation is reduced. 

For the radial residuals of the optical extinction, \DeltaAv, we find that they mainly anti-correlate with the residuals of the N/O ratio, and the ionization parameter (panels in third column from left to right in Fig.~\ref{fig:res_plot}). If N/O traces the ratio between the yield produced by old over young stars into the ISM, thus this result is suggesting that regions with chemical enrichment mostly from young stars have a larger extinction for a given galactocentric radius. This is quantitatively in agreement with the large correlation that we find between \DeltaL\ and \DeltaAv. Regarding the anti-correlation between \DeltaAv\ and the residual of the ionization parameter, it could be that regions with larger amount of dust hampers the strength of the ionization, as we mention above, the empirical calibrator that we are using are actually traced both the hardness and strength and spectral shape of the ionization source.

For the radial residuals of the oxygen abundance, \DeltaOH, we find strong and similar relations to those derived previously between the oxygen abundance, the N/O ratio, the ionization parameter (see Figs.~\ref{fig:NOvsO}, and \ref{fig:UvsO}, respectively). These results suggest that these relations are independent on the galactocentric distance. Regarding the residuals of the N/O ratio, as for the case of \DeltaOH, we also find a significant correlation between this residual and the one derived for the ionization parameter. Finally, we also note that the radial residuals of the ionization parameter have a mild but present correlation with the electronic density.
\begin{figure*}
    \centering
    \includegraphics[width=\linewidth]{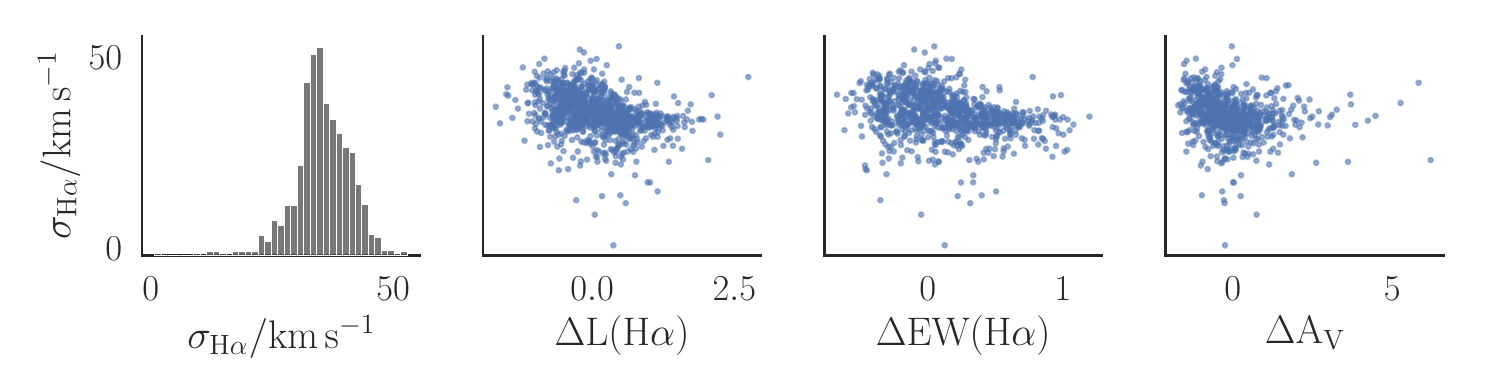}
    \includegraphics[width=\linewidth]{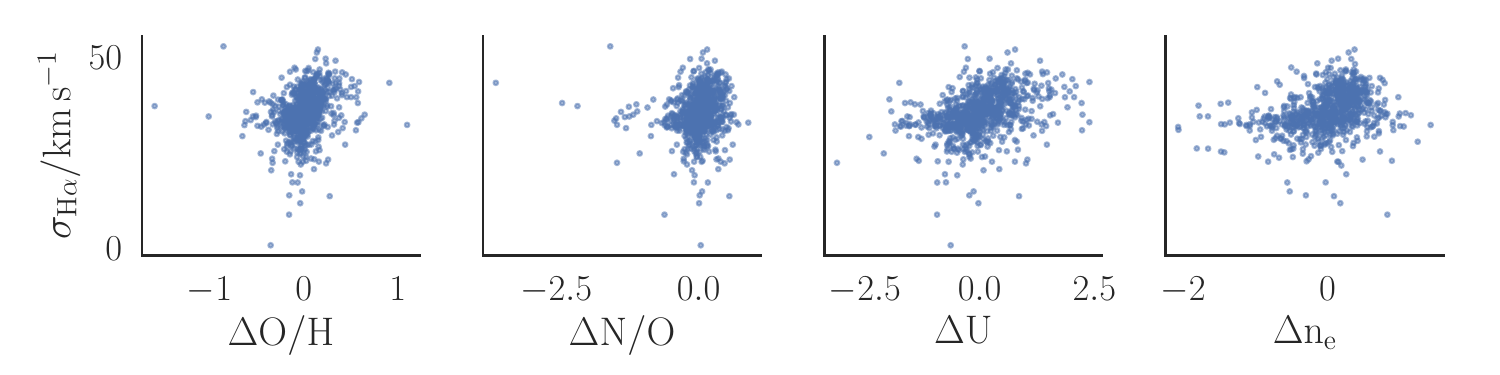}
    \caption{Comparison of the \ha\ velocity dispersion of the \hii\ regions with the residuals from the different physical properties derived from the ionized gas presented in this study.}
    \label{fig:res_dHa}
\end{figure*}

As we mention in Sec.~\ref{sec:Kin}, we can also explore how the residuals of the physical relations correlate or not with the velocity dispersion derived from \ha\ emission line width, \sigmaHa. In Fig.\ref{fig:res_dHa} we compare the radial residuals against \sigmaHa. We find weak but existing anti-correlations with the other parameters derived from the \ha\ emission line (i.e., \DeltaL, and $\Delta \mathrm{EW(\ha)}$, respectively). This is suggesting that for a given region located in a given galactocenctric distance the smaller \sigmaHa, the brighter the luminosity of that region; similary for $\Delta \mathrm{EW(\ha)}$. This suggest, that dynamically cold regions are more prone to form new stars. We also find a mild negative correlation with the radial residual of the extinction, $\Delta \mathrm{A_V}$. This again suggest that regions with a large amount of dust, thus cold gas, are those with the lowest velocity dispersion for a given galactocentric distance. Contrary to these parameters, we find that the parameters related to the chemical abundance and ionization of the ISM mildly correlate with \sigmaHa. Particularly, we find that \sigmaHa\ appears to increase with the radial residual of the oxygen abundance, $\Delta \mathrm{O/H}$ suggesting that those regions with large velocity dispersion are enriched in comparison to dynamically cold ones.  

\section{Discussion and conclusions}
\label{sec:Disc}
\begin{figure*}
    \centering
    \includegraphics[width=\linewidth]{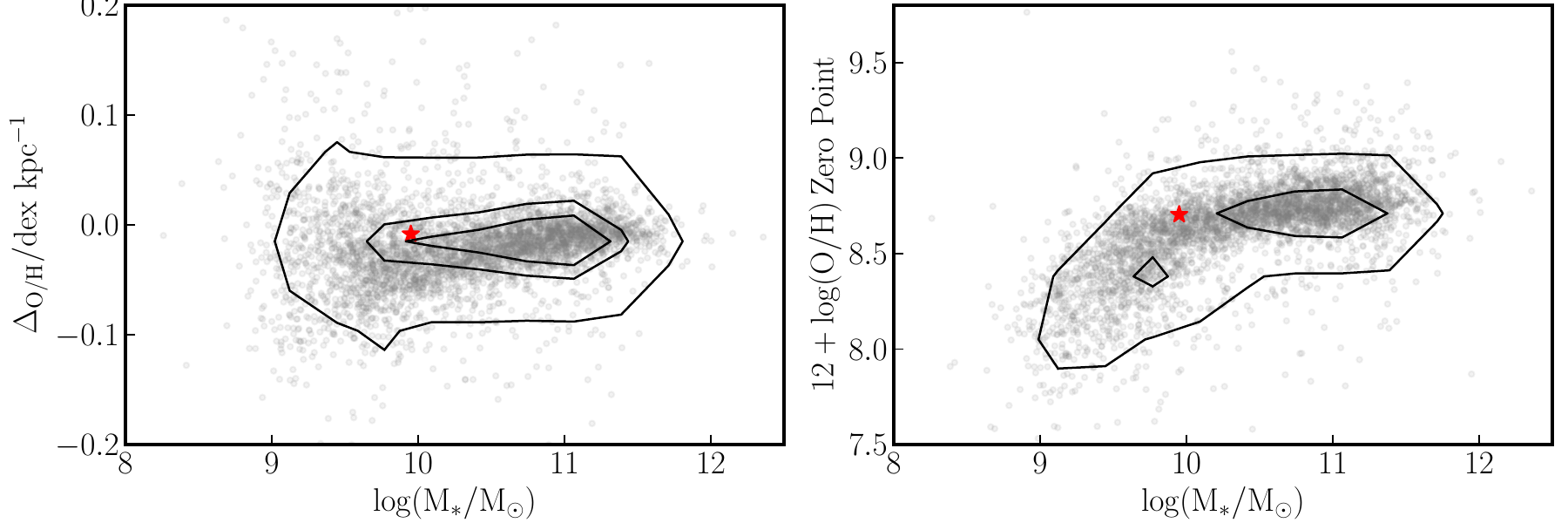}
    \caption{Comparison of the gradient (left-panel) and zero-point (right-panel) of the best of the radial distribution of the oxygen abundance derived in Sec.\ref{sec:rad_OH} (red star) with those derived for 7533 galaxies included in the MaNGA survey (gray data points).}
    \label{fig:manga_comp}
\end{figure*}

In this article we explore the physical properties at sub-kpc scales of \hii\ regions of the nearby galaxy \IC. This study has been possible thanks to the Integral Field Unit observations provided by the MaNGA survey. These observations along with the data analysis pipeline devoted to the SDSS-V Local Volume Mapper (LVM) have allow us to measure properties from the ionized gas in \hii-regions candidates. In comparison to other regions from large IFS surveys (such as AMUSING++), these candidates are similar in a BPT diagram as well as the radial distribution of their properties (see Fig.~\ref{fig:BPT}). Depending on the explored property, \IC\ exhibits positive, flat, or negative gradients (e.g., the $\mathrm{EW(\ha)}$, the \ha\ luminosity, or the N/O ratio, respectively). Using a large heterogeneous set of oxygen abundance calibrators we find that, albeit the large scatter, on average the radial gradient of the oxygen abundance for the star-forming regions in \IC\ is close to zero (see Figs.~\ref{fig:rad_OH} and \ref{fig:bf_OH}). Other properties such as the electron density and ionization parameter have similar trends as those derived for large samples of \hii regions. Regarding the kinematic properties of these \hii\ regions, we find that the velocity dispersion for each \hii\ region is relatively flat across the extension of the galaxy ($\sigma_{H\alpha} \sim 30 \mathrm{\,km\,s^{-1}}$). We also find that the probability distribution funtion of the \haLF\ of \IC\ is similar to the one derive for other nearby galaxies (see Fig.~\ref{fig:LF_comp}). Finally, in Sec.~\ref{sec:residuals} we study the possible relation among the residuals from the radial distribution of the physical properties. Our analysis shows that in general the relations among different physical parameters are independent on the radial distribution of the \hii\ regions. 

Currently there are IFU observations of thousands of galaxies thanks to the MaNGA survey \citep{Bundy_2015}. Thus, it is worth to ask how the radial properties we derive for \IC\ compare to those derived for this sample of galaxies. In particular, one can ask for the stellar mass of \IC\ how its oxygen abundance gradient compares with those derived for the MaNGA survey. In Fig.\ref{fig:manga_comp} we perform such comparison. In the left and right pannel of this figure, we compare the slope and the zero-point derived for \IC\  in Sec.~\ref{sec:rad_OH} using the Ho calibrator (red star) with those derived for 7533 galaxies drawn from the MaNGA survey \citep{Barrera-Ballesteros_2023}\footnote{We note that for the comparison between \IC and the MaNGA survey we derive a single gradient for the oxygen abundance. Furthermore, we quantify this gradient in units of dex/kpc. This is slightly different to our previous study where we use a piece-wise analysis and the slope are measured in units of dex/$\mathrm{R_{eff}}$.}, respectively. We find that for the stellar mass of \IC\ the slope and zero-point of the oxygen abundance gradient is similar to those galaxies with similar stellar mass in the MaNGA survey. In other words, \IC\ shares similar radial properties, at least in its oxygen abundance distribution, with those galaxies of similar mass in the nearby universe. This evidence along with the results from Sec.~\ref{sec:LF_Ha} indicate that \IC\ has similar properties as those reported from other nearby galaxies. Thus this encourage studies in other galaxies in  particular in the local universe (with higher spatial resolution) to quantify the role of the angular resolution in setting the spatially resolved properties of galaxies. 



\section*{Acknowledgments}

J.B-B acknowledges funding from the grant IA-101522 (DGAPA-PAPIIT, UNAM) and support from the DGAPA-PASPA 2025 fellowship (UNAM). L.C. thanks the support from the grant IN103820 (DGAPA-PAPIIT, UNAM). J.B-B thanks Laurent Drissen and Carmelle Robert for useful comments that help to improve the quality of the article. 
KK gratefully acknowledges funding from the Deutsche Forschungsgemeinschaft (DFG, German Research Foundation) in the form of an Emmy Noether Research Group (grant number KR4598/2-1, PI Kreckel) and the European Research Council’s starting grant ERC StG-101077573 (“ISM-METALS"). 
This research made use of Astropy,\footnote{http://www.astropy.org} a community-developed core Python package for Astronomy \citep{astropy:2013, astropy:2018}. 

Funding for the Sloan Digital Sky Survey IV has been provided by the Alfred P. Sloan Foundation, the U.S. Department of Energy Office of  Science, and the Participating  Institutions. 

SDSS-IV acknowledges support and  resources from the Center for High Performance Computing  at the University of Utah. The SDSS 
website is www.sdss.org.

SDSS-IV is managed by the  Astrophysical Research Consortium 
for the Participating Institutions 
of the SDSS Collaboration including 
the Brazilian Participation Group, 
the Carnegie Institution for Science, 
Carnegie Mellon University, Center for 
Astrophysics | Harvard \& 
Smithsonian, the Chilean Participation 
Group, the French Participation Group, 
Instituto de Astrof\'isica de 
Canarias, The Johns Hopkins 
University, Kavli Institute for the 
Physics and Mathematics of the 
Universe (IPMU) / University of 
Tokyo, the Korean Participation Group, 
Lawrence Berkeley National Laboratory, 
Leibniz Institut f\"ur Astrophysik 
Potsdam (AIP),  Max-Planck-Institut 
f\"ur Astronomie (MPIA Heidelberg), 
Max-Planck-Institut f\"ur 
Astrophysik (MPA Garching), 
Max-Planck-Institut f\"ur 
Extraterrestrische Physik (MPE), 
National Astronomical Observatories of 
China, New Mexico State University, 
New York University, University of 
Notre Dame, Observat\'ario 
Nacional / MCTI, The Ohio State 
University, Pennsylvania State 
University, Shanghai 
Astronomical Observatory, United 
Kingdom Participation Group, 
Universidad Nacional Aut\'onoma 
de M\'exico, University of Arizona, 
University of Colorado Boulder, 
University of Oxford, University of 
Portsmouth, University of Utah, 
University of Virginia, University 
of Washington, University of 
Wisconsin, Vanderbilt University, 
and Yale University.

\appendix

\section{Spatial distribution of brightest emission lines}
\label{app:flux}
\begin{figure}
    \centering
    \includegraphics[width=\linewidth]{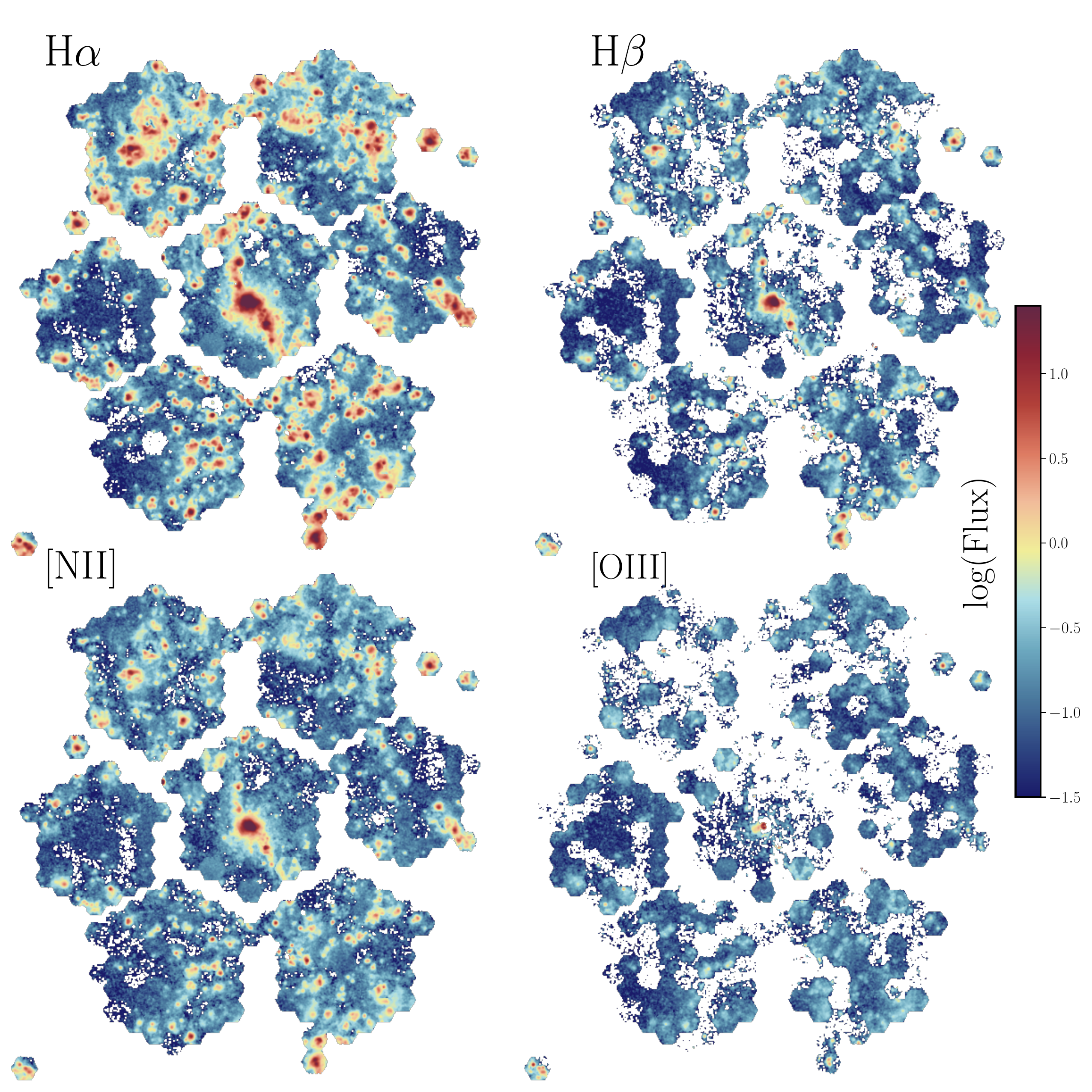}
    \caption{ Maps of the flux from the four brightest emission lines in the optical measured from the MaNGA observations. From top to bottom, left to right, the \ha, \hb, \hii, and \oiii\ emission lines, respectively. The flux units are the same for all the lines: \fluxunits .}
    \label{fig:maps_fluxes}    
\end{figure}

In Sec. \ref{sec:Data} we describe how the LVM-DAP is able to extract the maps of the physical properties in particular of the ionized gas emission. In Fig.~\ref{fig:maps_fluxes} we present the maps of the brightest four emission lines in the optical: \ha, \hb, \nii, \oiii\ (from left to right ,top to bottom). These maps are color coded to the same scale, so it is easy to see that the two most bright lines are \ha, and \nii. The weakest emission among them is \oiii, however it is evident that the nuclear emission is brighter in comparison to the rest of the galaxy. Despite the different in bright, these emission lines follow a similar spatial distribution as the \ha\ emission. We are using the information from these emission lines to derive physical properties of the \hii\ regions.

\section{Spatial distribution of physical properties from candidates}
\label{app:maps}

\begin{figure*}
    \centering
    \includegraphics[width=\linewidth]{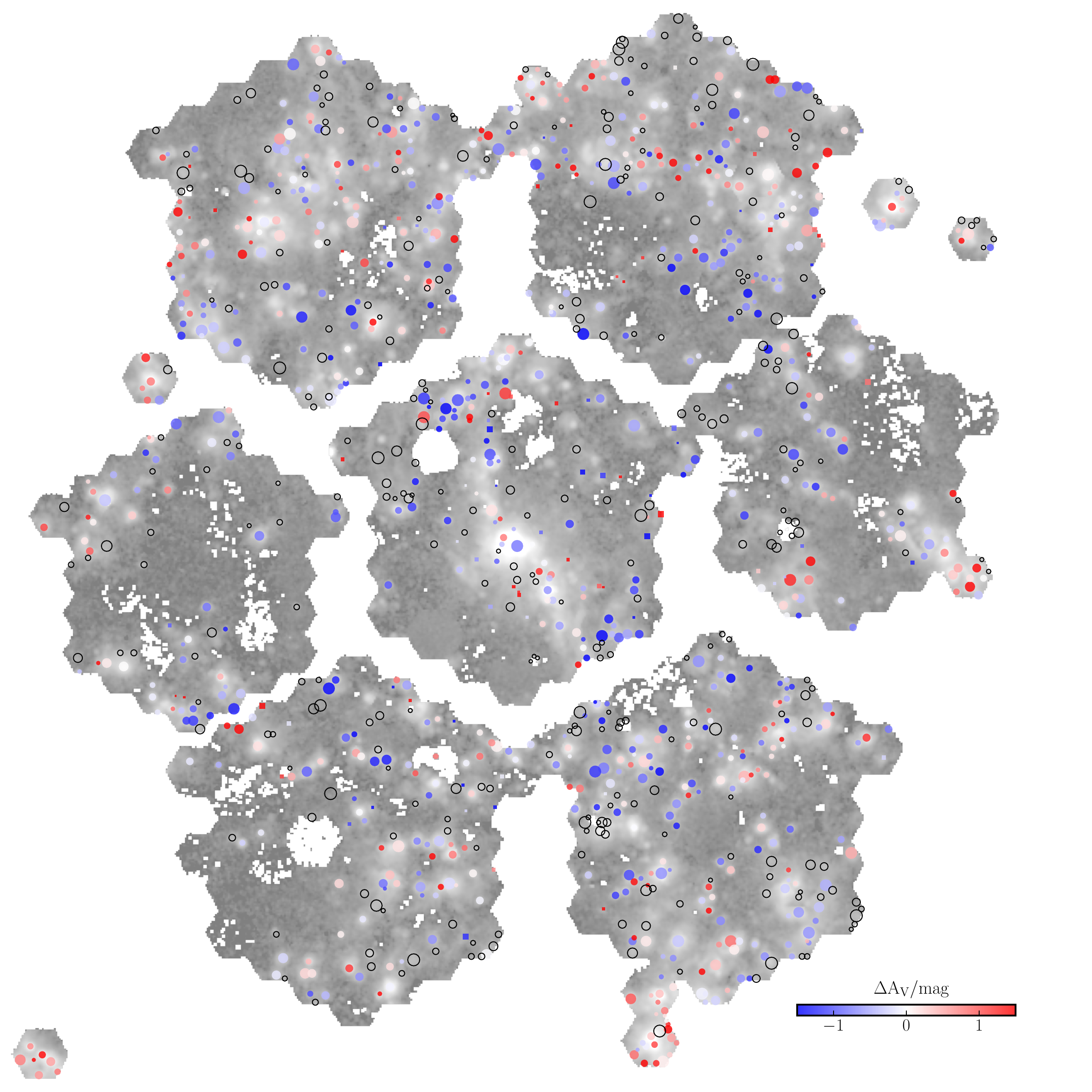}
    \caption{Similar to Fig.~\ref{fig:HIIregions}, with the \hii\ candidates color coded by the residual with respect to the gradient of Av derived in Sec.\ref{sec:rad_Ha}. The empty regions represent those regions where \ha/\hb\ $<$ 2.86. Circles and squares represent those regions selected as star and non-star forming regions according to the selection criteria described in Sec.\ref{sec:BPT}. 
    }
    \label{fig:AvRes_map}
    
\end{figure*}

\begin{figure*}
    \centering
    \includegraphics[width=\linewidth]{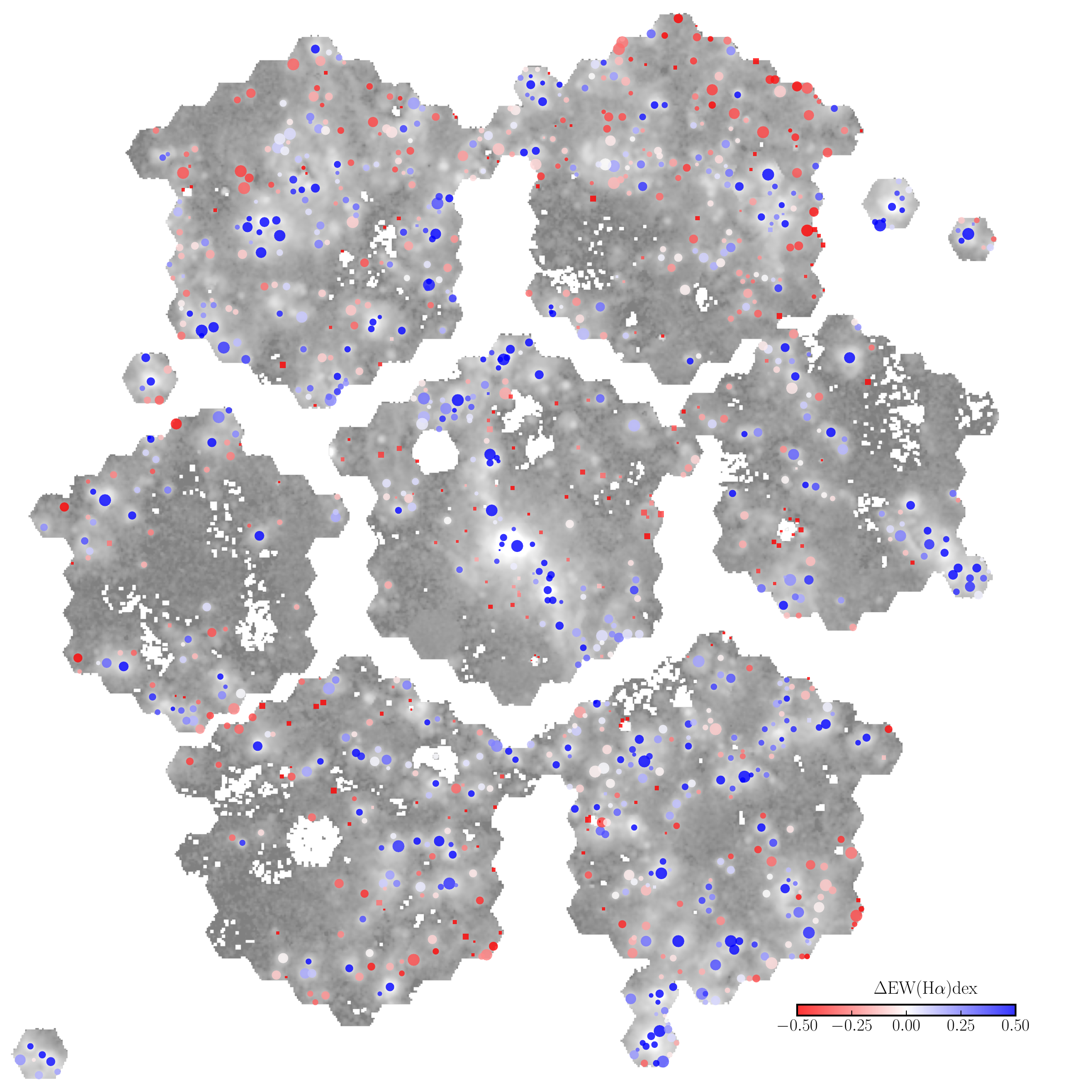}
    \caption{Similar to Fig.~\ref{fig:HIIregions}, with the \hii\ candidates color coded by the residual with respect to the radial gradient of EW(\ha) derived in Sec.\ref{sec:rad_Ha}. 
    Circles and squares represent those regions selected as star and non-star forming regions according to the selection criteria described in Sec.\ref{sec:BPT}. 
    }
    \label{fig:EWRes_map}
\end{figure*}

\begin{figure*}
    \centering
    \includegraphics[width=\linewidth]{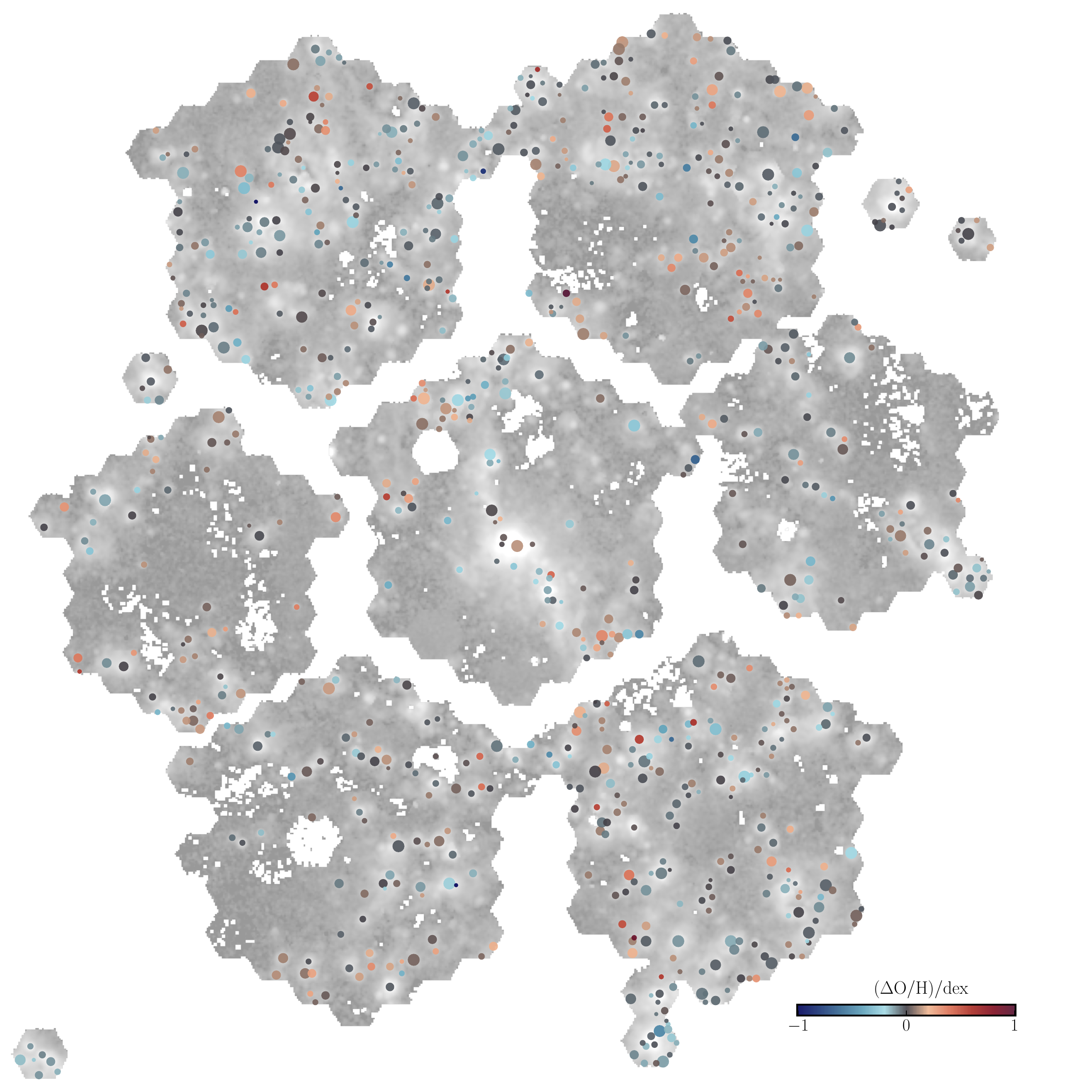}
    \caption{Similar to Fig.~\ref{fig:HIIregions}, with the \hii\ candidates color coded by the residual with respect to the radial gradient of the oxygen abundance derived in Sec.\ref{sec:rad_OH}. 
    }
    \label{fig:OHRes_map}
\end{figure*}

\begin{figure*}
    \centering
    \includegraphics[width=\linewidth]{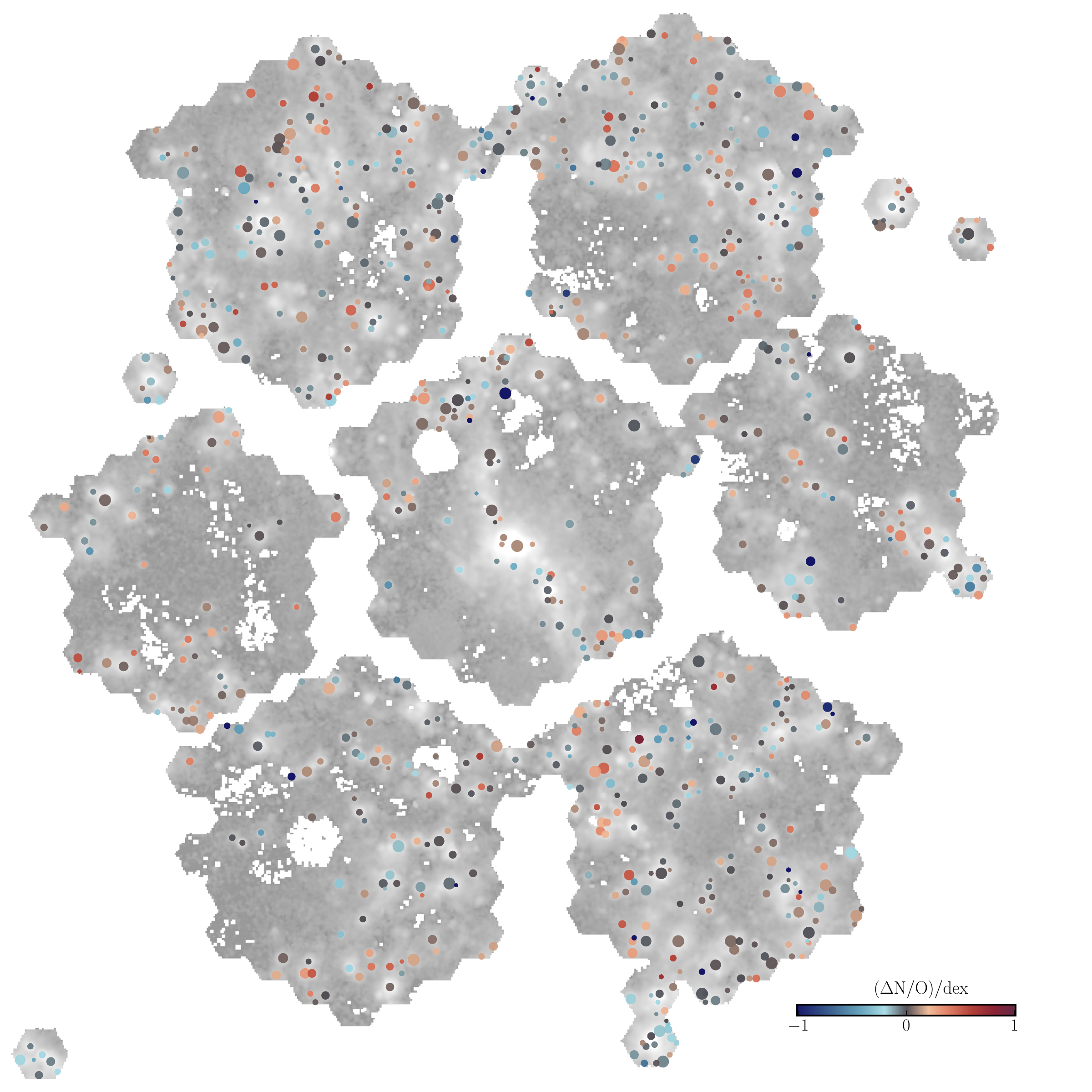}
    \caption{Similar to Fig.~\ref{fig:HIIregions}, with the \hii\ candidates color coded by the residual with respect to the radial gradient of the N/O ratio derived in Sec.\ref{sec:rad_NO}. 
    }
    \label{fig:NORes_map}
\end{figure*}

\bibliography{sample631}{}

\begin{thebibliography}
\expandafter\ifx\csname natexlab\endcsname\relax\def\natexlab#1{#1}\fi
\expandafter\ifx\csname href\endcsname\relax
  \def\href#1#2{}\fi
\expandafter\ifx\csname urllinklabel\endcsname\relax
  \def\urllinklabel{[LINK]}\fi
\expandafter\ifx\csname adsurllinklabel\endcsname\relax
  \def\adsurllinklabel{[ADS]}\fi

\bibitem[{{Alstott} {et~al.}(2014){Alstott}, {Bullmore}, \&
  {Plenz}}]{Alstott_2014}
{Alstott}, J., {Bullmore}, E., \& {Plenz}, D. 2014, PLoS ONE, 9, e85777


\bibitem[{{Anderson}(2014)}]{Anderson_2014}
{Anderson}, L.~D. 2014, in American Astronomical Society Meeting Abstracts,
  Vol. 223, American Astronomical Society Meeting Abstracts \#223, 312.01


\bibitem[{{Astropy Collaboration} {et~al.}(2018){Astropy Collaboration},
  {Price-Whelan}, {Sip{H{o}}cz}, {G{"u}nther}, {Lim}, {Crawford}, {Conseil},
  {Shupe}, {Craig}, {Dencheva}, {Ginsburg}, {Vand erPlas}, {Bradley},
  {P{'e}rez-Su{'a}rez}, {de Val-Borro}, {Aldcroft}, {Cruz}, {Robitaille},
  {Tollerud}, {Ardelean}, {Babej}, {Bach}, {Bachetti}, {Bakanov}, {Bamford},
  {Barentsen}, {Barmby}, {Baumbach}, {Berry}, {Biscani}, {Boquien}, {Bostroem},
  {Bouma}, {Brammer}, {Bray}, {Breytenbach}, {Buddelmeijer}, {Burke},
  {Calderone}, {Cano Rodr{'i}guez}, {Cara}, {Cardoso}, {Cheedella}, {Copin},
  {Corrales}, {Crichton}, {D'Avella}, {Deil}, {Depagne}, {Dietrich}, {Donath},
  {Droettboom}, {Earl}, {Erben}, {Fabbro}, {Ferreira}, {Finethy}, {Fox},
  {Garrison}, {Gibbons}, {Goldstein}, {Gommers}, {Greco}, {Greenfield},
  {Groener}, {Grollier}, {Hagen}, {Hirst}, {Homeier}, {Horton}, {Hosseinzadeh},
  {Hu}, {Hunkeler}, {Ivezi{'c}}, {Jain}, {Jenness}, {Kanarek}, {Kendrew},
  {Kern}, {Kerzendorf}, {Khvalko}, {King}, {Kirkby}, {Kulkarni}, {Kumar},
  {Lee}, {Lenz}, {Littlefair}, {Ma}, {Macleod}, {Mastropietro}, {McCully},
  {Montagnac}, {Morris}, {Mueller}, {Mumford}, {Muna}, {Murphy}, {Nelson},
  {Nguyen}, {Ninan}, {N{"o}the}, {Ogaz}, {Oh}, {Parejko}, {Parley}, {Pascual},
  {Patil}, {Patil}, {Plunkett}, {Prochaska}, {Rastogi}, {Reddy Janga},
  {Sabater}, {Sakurikar}, {Seifert}, {Sherbert}, {Sherwood-Taylor}, {Shih},
  {Sick}, {Silbiger}, {Singanamalla}, {Singer}, {Sladen}, {Sooley},
  {Sornarajah}, {Streicher}, {Teuben}, {Thomas}, {Tremblay}, {Turner},
  {Terr{'o}n}, {van Kerkwijk}, {de la Vega}, {Watkins}, {Weaver}, {Whitmore},
  {Woillez}, {Zabalza}, \& {Astropy Contributors}}]{astropy:2018}
{Astropy Collaboration}, {Price-Whelan}, A.~M., {Sip{H{o}}cz}, B.~M.,
  {G{"u}nther}, H.~M., {Lim}, P.~L., {Crawford}, S.~M., {Conseil}, S., {Shupe},
  D.~L., {Craig}, M.~W., {Dencheva}, N., {Ginsburg}, A., {Vand erPlas}, J.~T.,
  {Bradley}, L.~D., {P{'e}rez-Su{'a}rez}, D., {de Val-Borro}, M., {Aldcroft},
  T.~L., {Cruz}, K.~L., {Robitaille}, T.~P., {Tollerud}, E.~J., {Ardelean}, C.,
  {Babej}, T., {Bach}, Y.~P., {Bachetti}, M., {Bakanov}, A.~V., {Bamford},
  S.~P., {Barentsen}, G., {Barmby}, P., {Baumbach}, A., {Berry}, K.~L.,
  {Biscani}, F., {Boquien}, M., {Bostroem}, K.~A., {Bouma}, L.~G., {Brammer},
  G.~B., {Bray}, E.~M., {Breytenbach}, H., {Buddelmeijer}, H., {Burke}, D.~J.,
  {Calderone}, G., {Cano Rodr{'i}guez}, J.~L., {Cara}, M., {Cardoso}, J.~V.~M.,
  {Cheedella}, S., {Copin}, Y., {Corrales}, L., {Crichton}, D., {D'Avella}, D.,
  {Deil}, C., {Depagne}, E., {Dietrich}, J.~P., {Donath}, A., {Droettboom}, M.,
  {Earl}, N., {Erben}, T., {Fabbro}, S., {Ferreira}, L.~A., {Finethy}, T.,
  {Fox}, R.~T., {Garrison}, L.~H., {Gibbons}, S.~L.~J., {Goldstein}, D.~A.,
  {Gommers}, R., {Greco}, J.~P., {Greenfield}, P., {Groener}, A.~M.,
  {Grollier}, F., {Hagen}, A., {Hirst}, P., {Homeier}, D., {Horton}, A.~J.,
  {Hosseinzadeh}, G., {Hu}, L., {Hunkeler}, J.~S., {Ivezi{'c}}, Z., {Jain}, A.,
  {Jenness}, T., {Kanarek}, G., {Kendrew}, S., {Kern}, N.~S., {Kerzendorf},
  W.~E., {Khvalko}, A., {King}, J., {Kirkby}, D., {Kulkarni}, A.~M., {Kumar},
  A., {Lee}, A., {Lenz}, D., {Littlefair}, S.~P., {Ma}, Z., {Macleod}, D.~M.,
  {Mastropietro}, M., {McCully}, C., {Montagnac}, S., {Morris}, B.~M.,
  {Mueller}, M., {Mumford}, S.~J., {Muna}, D., {Murphy}, N.~A., {Nelson}, S.,
  {Nguyen}, G.~H., {Ninan}, J.~P., {N{"o}the}, M., {Ogaz}, S., {Oh}, S.,
  {Parejko}, J.~K., {Parley}, N., {Pascual}, S., {Patil}, R., {Patil}, A.~A.,
  {Plunkett}, A.~L., {Prochaska}, J.~X., {Rastogi}, T., {Reddy Janga}, V.,
  {Sabater}, J., {Sakurikar}, P., {Seifert}, M., {Sherbert}, L.~E.,
  {Sherwood-Taylor}, H., {Shih}, A.~Y., {Sick}, J., {Silbiger}, M.~T.,
  {Singanamalla}, S., {Singer}, L.~P., {Sladen}, P.~H., {Sooley}, K.~A.,
  {Sornarajah}, S., {Streicher}, O., {Teuben}, P., {Thomas}, S.~W., {Tremblay},
  G.~R., {Turner}, J.~E.~H., {Terr{'o}n}, V., {van Kerkwijk}, M.~H., {de la
  Vega}, A., {Watkins}, L.~L., {Weaver}, B.~A., {Whitmore}, J.~B., {Woillez},
  J., {Zabalza}, V., \& {Astropy Contributors}. 2018, aj, 156, 123


\bibitem[{{Astropy Collaboration} {et~al.}(2013){Astropy Collaboration},
  {Robitaille}, {Tollerud}, {Greenfield}, {Droettboom}, {Bray}, {Aldcroft},
  {Davis}, {Ginsburg}, {Price-Whelan}, {Kerzendorf}, {Conley}, {Crighton},
  {Barbary}, {Muna}, {Ferguson}, {Grollier}, {Parikh}, {Nair}, {Unther},
  {Deil}, {Woillez}, {Conseil}, {Kramer}, {Turner}, {Singer}, {Fox}, {Weaver},
  {Zabalza}, {Edwards}, {Azalee Bostroem}, {Burke}, {Casey}, {Crawford},
  {Dencheva}, {Ely}, {Jenness}, {Labrie}, {Lim}, {Pierfederici}, {Pontzen},
  {Ptak}, {Refsdal}, {Servillat}, \& {Streicher}}]{astropy:2013}
{Astropy Collaboration}, {Robitaille}, T.~P., {Tollerud}, E.~J., {Greenfield},
  P., {Droettboom}, M., {Bray}, E., {Aldcroft}, T., {Davis}, M., {Ginsburg},
  A., {Price-Whelan}, A.~M., {Kerzendorf}, W.~E., {Conley}, A., {Crighton}, N.,
  {Barbary}, K., {Muna}, D., {Ferguson}, H., {Grollier}, F., {Parikh}, M.~M.,
  {Nair}, P.~H., {Unther}, H.~M., {Deil}, C., {Woillez}, J., {Conseil}, S.,
  {Kramer}, R., {Turner}, J.~E.~H., {Singer}, L., {Fox}, R., {Weaver}, B.~A.,
  {Zabalza}, V., {Edwards}, Z.~I., {Azalee Bostroem}, K., {Burke}, D.~J.,
  {Casey}, A.~R., {Crawford}, S.~M., {Dencheva}, N., {Ely}, J., {Jenness}, T.,
  {Labrie}, K., {Lim}, P.~L., {Pierfederici}, F., {Pontzen}, A., {Ptak}, A.,
  {Refsdal}, B., {Servillat}, M., \& {Streicher}, O. 2013, \aap, 558, A33


\bibitem[{{Baldwin} {et~al.}(1981){Baldwin}, {Phillips}, \& {Terlevich}}]{BPT}
{Baldwin}, J.~A., {Phillips}, M.~M., \& {Terlevich}, R. 1981, \pasp, 93, 5


\bibitem[{{Barnes} {et~al.}(2021){Barnes}, {Glover}, {Kreckel}, {Ostriker},
  {Bigiel}, {Belfiore}, {Be{\v{s}}li{\'c}}, {Blanc}, {Chevance}, {Dale},
  {Egorov}, {Eibensteiner}, {Emsellem}, {Grasha}, {Groves}, {Klessen},
  {Kruijssen}, {Leroy}, {Longmore}, {Lopez}, {McElroy}, {Meidt}, {Murphy},
  {Rosolowsky}, {Saito}, {Santoro}, {Schinnerer}, {Schruba}, {Sun}, {Watkins},
  \& {Williams}}]{Barnes_2021}
{Barnes}, A.~T., {Glover}, S.~C.~O., {Kreckel}, K., {Ostriker}, E.~C.,
  {Bigiel}, F., {Belfiore}, F., {Be{\v{s}}li{\'c}}, I., {Blanc}, G.~A.,
  {Chevance}, M., {Dale}, D.~A., {Egorov}, O., {Eibensteiner}, C., {Emsellem},
  E., {Grasha}, K., {Groves}, B.~A., {Klessen}, R.~S., {Kruijssen}, J.~M.~D.,
  {Leroy}, A.~K., {Longmore}, S.~N., {Lopez}, L., {McElroy}, R., {Meidt},
  S.~E., {Murphy}, E.~J., {Rosolowsky}, E., {Saito}, T., {Santoro}, F.,
  {Schinnerer}, E., {Schruba}, A., {Sun}, J., {Watkins}, E.~J., \& {Williams},
  T.~G. 2021, \mnras, 508, 5362


\bibitem[{{Barrera-Ballesteros} {et~al.}(2023){Barrera-Ballesteros},
  {S{\'a}nchez}, {Espinosa-Ponce}, {L{\'o}pez-Cob{\'a}}, {Carigi},
  {Lugo-Aranda}, {Lacerda}, {Bruzual}, {Hernandez-Toledo}, {Boardman}, {Drory},
  {Lane}, \& {Brownstein}}]{Barrera-Ballesteros_2023}
{Barrera-Ballesteros}, J.~K., {S{\'a}nchez}, S.~F., {Espinosa-Ponce}, C.,
  {L{\'o}pez-Cob{\'a}}, C., {Carigi}, L., {Lugo-Aranda}, A.~Z., {Lacerda}, E.,
  {Bruzual}, G., {Hernandez-Toledo}, H., {Boardman}, N., {Drory}, N., {Lane},
  R.~R., \& {Brownstein}, J.~R. 2023, \rmxaa, 59, 213


\bibitem[{{Barrera-Ballesteros} {et~al.}(2021){Barrera-Ballesteros},
  {S{\'a}nchez}, {Heckman}, {Wong}, {Bolatto}, {Ostriker}, {Rosolowsky},
  {Carigi}, {Vogel}, {Levy}, {Colombo}, {Luo}, \&
  {Cao}}]{Barrera-Ballesteros_2021a}
{Barrera-Ballesteros}, J.~K., {S{\'a}nchez}, S.~F., {Heckman}, T., {Wong}, T.,
  {Bolatto}, A., {Ostriker}, E., {Rosolowsky}, E., {Carigi}, L., {Vogel}, S.,
  {Levy}, R.~C., {Colombo}, D., {Luo}, Y., \& {Cao}, Y. 2021, \mnras, 503, 3643


\bibitem[{{Barrera-Ballesteros} {et~al.}(2020){Barrera-Ballesteros}, {Utomo},
  {Bolatto}, {S{\'a}nchez}, {Vogel}, {Wong}, {Levy}, {Colombo}, {Kalinova},
  {Teuben}, {Garc{\'\i}a-Benito}, {Husemann}, {Mast}, \&
  {Blitz}}]{Barrera-Ballesteros_2020}
{Barrera-Ballesteros}, J.~K., {Utomo}, D., {Bolatto}, A.~D., {S{\'a}nchez},
  S.~F., {Vogel}, S.~N., {Wong}, T., {Levy}, R.~C., {Colombo}, D., {Kalinova},
  V., {Teuben}, P., {Garc{\'\i}a-Benito}, R., {Husemann}, B., {Mast}, D., \&
  {Blitz}, L. 2020, \mnras, 492, 2651


\bibitem[{{Belfiore} {et~al.}(2017){Belfiore}, {Maiolino}, {Tremonti},
  {S{\'a}nchez}, {Bundy}, {Bershady}, {Westfall}, {Lin}, {Drory}, {Boquien},
  {Thomas}, \& {Brinkmann}}]{Belfiore_2017}
{Belfiore}, F., {Maiolino}, R., {Tremonti}, C., {S{\'a}nchez}, S.~F., {Bundy},
  K., {Bershady}, M., {Westfall}, K., {Lin}, L., {Drory}, N., {Boquien}, M.,
  {Thomas}, D., \& {Brinkmann}, J. 2017, \mnras, 469, 151


\bibitem[{{Bigiel} {et~al.}(2008){Bigiel}, {Leroy}, {Walter}, {Brinks}, {de
  Blok}, {Madore}, \& {Thornley}}]{Bigiel_2008}
{Bigiel}, F., {Leroy}, A., {Walter}, F., {Brinks}, E., {de Blok}, W.~J.~G.,
  {Madore}, B., \& {Thornley}, M.~D. 2008, \aj, 136, 2846


\bibitem[{{Blanton} {et~al.}(2017){Blanton}, {Bershady}, {Abolfathi},
  {Albareti}, {Allende Prieto}, {Almeida}, {Alonso-Garc{\'\i}a}, {Anders},
  {Anderson}, {Andrews}, {Aquino-Ort{\'\i}z}, {Arag{\'o}n-Salamanca},
  {Argudo-Fern{\'a}ndez}, {Armengaud}, {Aubourg}, {Avila-Reese}, {Badenes},
  {Bailey}, {Barger}, {Barrera-Ballesteros}, {Bartosz}, {Bates}, {Baumgarten},
  {Bautista}, {Beaton}, {Beers}, {Belfiore}, {Bender}, {Berlind}, {Bernardi},
  {Beutler}, {Bird}, {Bizyaev}, {Blanc}, {Blomqvist}, {Bolton}, {Boquien},
  {Borissova}, {van den Bosch}, {Bovy}, {Brandt}, {Brinkmann}, {Brownstein},
  {Bundy}, {Burgasser}, {Burtin}, {Busca}, {Cappellari}, {Delgado Carigi},
  {Carlberg}, {Carnero Rosell}, {Carrera}, {Chanover}, {Cherinka}, {Cheung},
  {G{\'o}mez Maqueo Chew}, {Chiappini}, {Choi}, {Chojnowski}, {Chuang},
  {Chung}, {Cirolini}, {Clerc}, {Cohen}, {Comparat}, {da Costa}, {Cousinou},
  {Covey}, {Crane}, {Croft}, {Cruz-Gonzalez}, {Garrido Cuadra}, {Cunha},
  {Damke}, {Darling}, {Davies}, {Dawson}, {de la Macorra}, {Dell'Agli}, {De
  Lee}, {Delubac}, {Di Mille}, {Diamond-Stanic}, {Cano-D{\'\i}az}, {Donor},
  {Downes}, {Drory}, {du Mas des Bourboux}, {Duckworth}, {Dwelly}, {Dyer},
  {Ebelke}, {Eigenbrot}, {Eisenstein}, {Emsellem}, {Eracleous}, {Escoffier},
  {Evans}, {Fan}, {Fern{\'a}ndez-Alvar}, {Fernandez-Trincado}, {Feuillet},
  {Finoguenov}, {Fleming}, {Font-Ribera}, {Fredrickson}, {Freischlad},
  {Frinchaboy}, {Fuentes}, {Galbany}, {Garcia-Dias},
  {Garc{\'\i}a-Hern{\'a}ndez}, {Gaulme}, {Geisler}, {Gelfand},
  {Gil-Mar{\'\i}n}, {Gillespie}, {Goddard}, {Gonzalez-Perez}, {Grabowski},
  {Green}, {Grier}, {Gunn}, {Guo}, {Guy}, {Hagen}, {Hahn}, {Hall}, {Harding},
  {Hasselquist}, {Hawley}, {Hearty}, {Gonzalez Hern{\'a}ndez}, {Ho}, {Hogg},
  {Holley-Bockelmann}, {Holtzman}, {Holzer}, {Huehnerhoff}, {Hutchinson},
  {Hwang}, {Ibarra-Medel}, {da Silva Ilha}, {Ivans}, {Ivory}, {Jackson},
  {Jensen}, {Johnson}, {Jones}, {J{\"o}nsson}, {Jullo}, {Kamble}, {Kinemuchi},
  {Kirkby}, {Kitaura}, {Klaene}, {Knapp}, {Kneib}, {Kollmeier}, {Lacerna},
  {Lane}, {Lang}, {Law}, {Lazarz}, {Lee}, {Le Goff}, {Liang}, {Li}, {Li},
  {Lian}, {Lima}, {Lin}, {Lin}, {Bertran de Lis}, {Liu}, {de Icaza Lizaola},
  {Long}, {Lucatello}, {Lundgren}, {MacDonald}, {Deconto Machado}, {MacLeod},
  {Mahadevan}, {Geimba Maia}, {Maiolino}, {Majewski}, {Malanushenko},
  {Malanushenko}, {Manchado}, {Mao}, {Maraston}, {Marques-Chaves}, {Masseron},
  {Masters}, {McBride}, {McDermid}, {McGrath}, {McGreer}, {Medina Pe{\~n}a},
  {Melendez}, {Merloni}, {Merrifield}, {Meszaros}, {Meza}, {Minchev},
  {Minniti}, {Miyaji}, {More}, {Mulchaey}, {M{\"u}ller-S{\'a}nchez}, {Muna},
  {Munoz}, {Myers}, {Nair}, {Nandra}, {Correa do Nascimento}, {Negrete},
  {Ness}, {Newman}, {Nichol}, {Nidever}, {Nitschelm}, {Ntelis}, {O'Connell},
  {Oelkers}, {Oravetz}, {Oravetz}, {Pace}, {Padilla}, {Palanque-Delabrouille},
  {Alonso Palicio}, {Pan}, {Parejko}, {Parikh}, {P{\^a}ris}, {Park}, {Patten},
  {Peirani}, {Pellejero-Ibanez}, {Penny}, {Percival}, {Perez-Fournon},
  {Petitjean}, {Pieri}, {Pinsonneault}, {Pisani}, {Poleski}, {Prada},
  {Prakash}, {Queiroz}, {Raddick}, {Raichoor}, {Barboza Rembold}, {Richstein},
  {Riffel}, {Riffel}, {Rix}, {Robin}, {Rockosi}, {Rodr{\'\i}guez-Torres},
  {Roman-Lopes}, {Rom{\'a}n-Z{\'u}{\~n}iga}, {Rosado}, {Ross}, {Rossi}, {Ruan},
  {Ruggeri}, {Rykoff}, {Salazar-Albornoz}, {Salvato}, {S{\'a}nchez}, {Aguado},
  {S{\'a}nchez-Gallego}, {Santana}, {Santiago}, {Sayres}, {Schiavon}, {da Silva
  Schimoia}, {Schlafly}, {Schlegel}, {Schneider}, {Schultheis}, {Schuster},
  {Schwope}, {Seo}, {Shao}, {Shen}, {Shetrone}, {Shull}, {Simon}, {Skinner},
  {Skrutskie}, {Slosar}, {Smith}, {Sobeck}, {Sobreira}, {Somers}, {Souto},
  {Stark}, {Stassun}, {Stauffer}, {Steinmetz}, {Storchi-Bergmann},
  {Streblyanska}, {Stringfellow}, {Su{\'a}rez}, {Sun}, {Suzuki}, {Szigeti},
  {Taghizadeh-Popp}, {Tang}, {Tao}, {Tayar}, {Tembe}, {Teske}, {Thakar},
  {Thomas}, {Thompson}, {Tinker}, {Tissera}, {Tojeiro}, {Hernandez Toledo}, {de
  la Torre}, {Tremonti}, {Troup}, {Valenzuela}, {Martinez Valpuesta},
  {Vargas-Gonz{\'a}lez}, {Vargas-Maga{\~n}a}, {Vazquez}, {Villanova}, {Vivek},
  {Vogt}, {Wake}, {Walterbos}, {Wang}, {Weaver}, {Weijmans}, {Weinberg},
  {Westfall}, {Whelan}, {Wild}, {Wilson}, {Wood-Vasey}, {Wylezalek}, {Xiao},
  {Yan}, {Yang}, {Ybarra}, {Y{\`e}che}, {Zakamska}, {Zamora}, {Zarrouk},
  {Zasowski}, {Zhang}, {Zhao}, {Zheng}, {Zheng}, {Zhou}, {Zhou}, {Zhu},
  {Zoccali}, \& {Zou}}]{Blanton_2017}
{Blanton}, M.~R., {Bershady}, M.~A., {Abolfathi}, B., {Albareti}, F.~D.,
  {Allende Prieto}, C., {Almeida}, A., {Alonso-Garc{\'\i}a}, J., {Anders}, F.,
  {Anderson}, S.~F., {Andrews}, B., {Aquino-Ort{\'\i}z}, E.,
  {Arag{\'o}n-Salamanca}, A., {Argudo-Fern{\'a}ndez}, M., {Armengaud}, E.,
  {Aubourg}, E., {Avila-Reese}, V., {Badenes}, C., {Bailey}, S., {Barger},
  K.~A., {Barrera-Ballesteros}, J., {Bartosz}, C., {Bates}, D., {Baumgarten},
  F., {Bautista}, J., {Beaton}, R., {Beers}, T.~C., {Belfiore}, F., {Bender},
  C.~F., {Berlind}, A.~A., {Bernardi}, M., {Beutler}, F., {Bird}, J.~C.,
  {Bizyaev}, D., {Blanc}, G.~A., {Blomqvist}, M., {Bolton}, A.~S., {Boquien},
  M., {Borissova}, J., {van den Bosch}, R., {Bovy}, J., {Brandt}, W.~N.,
  {Brinkmann}, J., {Brownstein}, J.~R., {Bundy}, K., {Burgasser}, A.~J.,
  {Burtin}, E., {Busca}, N.~G., {Cappellari}, M., {Delgado Carigi}, M.~L.,
  {Carlberg}, J.~K., {Carnero Rosell}, A., {Carrera}, R., {Chanover}, N.~J.,
  {Cherinka}, B., {Cheung}, E., {G{\'o}mez Maqueo Chew}, Y., {Chiappini}, C.,
  {Choi}, P.~D., {Chojnowski}, D., {Chuang}, C.-H., {Chung}, H., {Cirolini},
  R.~F., {Clerc}, N., {Cohen}, R.~E., {Comparat}, J., {da Costa}, L.,
  {Cousinou}, M.-C., {Covey}, K., {Crane}, J.~D., {Croft}, R. A.~C.,
  {Cruz-Gonzalez}, I., {Garrido Cuadra}, D., {Cunha}, K., {Damke}, G.~J.,
  {Darling}, J., {Davies}, R., {Dawson}, K., {de la Macorra}, A., {Dell'Agli},
  F., {De Lee}, N., {Delubac}, T., {Di Mille}, F., {Diamond-Stanic}, A.,
  {Cano-D{\'\i}az}, M., {Donor}, J., {Downes}, J.~J., {Drory}, N., {du Mas des
  Bourboux}, H., {Duckworth}, C.~J., {Dwelly}, T., {Dyer}, J., {Ebelke}, G.,
  {Eigenbrot}, A.~D., {Eisenstein}, D.~J., {Emsellem}, E., {Eracleous}, M.,
  {Escoffier}, S., {Evans}, M.~L., {Fan}, X., {Fern{\'a}ndez-Alvar}, E.,
  {Fernandez-Trincado}, J.~G., {Feuillet}, D.~K., {Finoguenov}, A., {Fleming},
  S.~W., {Font-Ribera}, A., {Fredrickson}, A., {Freischlad}, G., {Frinchaboy},
  P.~M., {Fuentes}, C.~E., {Galbany}, L., {Garcia-Dias}, R.,
  {Garc{\'\i}a-Hern{\'a}ndez}, D.~A., {Gaulme}, P., {Geisler}, D., {Gelfand},
  J.~D., {Gil-Mar{\'\i}n}, H., {Gillespie}, B.~A., {Goddard}, D.,
  {Gonzalez-Perez}, V., {Grabowski}, K., {Green}, P.~J., {Grier}, C.~J.,
  {Gunn}, J.~E., {Guo}, H., {Guy}, J., {Hagen}, A., {Hahn}, C., {Hall}, M.,
  {Harding}, P., {Hasselquist}, S., {Hawley}, S.~L., {Hearty}, F., {Gonzalez
  Hern{\'a}ndez}, J.~I., {Ho}, S., {Hogg}, D.~W., {Holley-Bockelmann}, K.,
  {Holtzman}, J.~A., {Holzer}, P.~H., {Huehnerhoff}, J., {Hutchinson}, T.~A.,
  {Hwang}, H.~S., {Ibarra-Medel}, H.~J., {da Silva Ilha}, G., {Ivans}, I.~I.,
  {Ivory}, K., {Jackson}, K., {Jensen}, T.~W., {Johnson}, J.~A., {Jones}, A.,
  {J{\"o}nsson}, H., {Jullo}, E., {Kamble}, V., {Kinemuchi}, K., {Kirkby}, D.,
  {Kitaura}, F.-S., {Klaene}, M., {Knapp}, G.~R., {Kneib}, J.-P., {Kollmeier},
  J.~A., {Lacerna}, I., {Lane}, R.~R., {Lang}, D., {Law}, D.~R., {Lazarz}, D.,
  {Lee}, Y., {Le Goff}, J.-M., {Liang}, F.-H., {Li}, C., {Li}, H., {Lian}, J.,
  {Lima}, M., {Lin}, L., {Lin}, Y.-T., {Bertran de Lis}, S., {Liu}, C., {de
  Icaza Lizaola}, M. A.~C., {Long}, D., {Lucatello}, S., {Lundgren}, B.,
  {MacDonald}, N.~K., {Deconto Machado}, A., {MacLeod}, C.~L., {Mahadevan}, S.,
  {Geimba Maia}, M.~A., {Maiolino}, R., {Majewski}, S.~R., {Malanushenko}, E.,
  {Malanushenko}, V., {Manchado}, A., {Mao}, S., {Maraston}, C.,
  {Marques-Chaves}, R., {Masseron}, T., {Masters}, K.~L., {McBride}, C.~K.,
  {McDermid}, R.~M., {McGrath}, B., {McGreer}, I.~D., {Medina Pe{\~n}a}, N.,
  {Melendez}, M., {Merloni}, A., {Merrifield}, M.~R., {Meszaros}, S., {Meza},
  A., {Minchev}, I., {Minniti}, D., {Miyaji}, T., {More}, S., {Mulchaey}, J.,
  {M{\"u}ller-S{\'a}nchez}, F., {Muna}, D., {Munoz}, R.~R., {Myers}, A.~D.,
  {Nair}, P., {Nandra}, K., {Correa do Nascimento}, J., {Negrete}, A., {Ness},
  M., {Newman}, J.~A., {Nichol}, R.~C., {Nidever}, D.~L., {Nitschelm}, C.,
  {Ntelis}, P., {O'Connell}, J.~E., {Oelkers}, R.~J., {Oravetz}, A., {Oravetz},
  D., {Pace}, Z., {Padilla}, N., {Palanque-Delabrouille}, N., {Alonso Palicio},
  P., {Pan}, K., {Parejko}, J.~K., {Parikh}, T., {P{\^a}ris}, I., {Park}, C.,
  {Patten}, A.~Y., {Peirani}, S., {Pellejero-Ibanez}, M., {Penny}, S.,
  {Percival}, W.~J., {Perez-Fournon}, I., {Petitjean}, P., {Pieri}, M.~M.,
  {Pinsonneault}, M., {Pisani}, A., {Poleski}, R., {Prada}, F., {Prakash}, A.,
  {Queiroz}, A. B. d.~A., {Raddick}, M.~J., {Raichoor}, A., {Barboza Rembold},
  S., {Richstein}, H., {Riffel}, R.~A., {Riffel}, R., {Rix}, H.-W., {Robin},
  A.~C., {Rockosi}, C.~M., {Rodr{\'\i}guez-Torres}, S., {Roman-Lopes}, A.,
  {Rom{\'a}n-Z{\'u}{\~n}iga}, C., {Rosado}, M., {Ross}, A.~J., {Rossi}, G.,
  {Ruan}, J., {Ruggeri}, R., {Rykoff}, E.~S., {Salazar-Albornoz}, S.,
  {Salvato}, M., {S{\'a}nchez}, A.~G., {Aguado}, D.~S., {S{\'a}nchez-Gallego},
  J.~R., {Santana}, F.~A., {Santiago}, B.~X., {Sayres}, C., {Schiavon}, R.~P.,
  {da Silva Schimoia}, J., {Schlafly}, E.~F., {Schlegel}, D.~J., {Schneider},
  D.~P., {Schultheis}, M., {Schuster}, W.~J., {Schwope}, A., {Seo}, H.-J.,
  {Shao}, Z., {Shen}, S., {Shetrone}, M., {Shull}, M., {Simon}, J.~D.,
  {Skinner}, D., {Skrutskie}, M.~F., {Slosar}, A., {Smith}, V.~V., {Sobeck},
  J.~S., {Sobreira}, F., {Somers}, G., {Souto}, D., {Stark}, D.~V., {Stassun},
  K., {Stauffer}, F., {Steinmetz}, M., {Storchi-Bergmann}, T., {Streblyanska},
  A., {Stringfellow}, G.~S., {Su{\'a}rez}, G., {Sun}, J., {Suzuki}, N.,
  {Szigeti}, L., {Taghizadeh-Popp}, M., {Tang}, B., {Tao}, C., {Tayar}, J.,
  {Tembe}, M., {Teske}, J., {Thakar}, A.~R., {Thomas}, D., {Thompson}, B.~A.,
  {Tinker}, J.~L., {Tissera}, P., {Tojeiro}, R., {Hernandez Toledo}, H., {de la
  Torre}, S., {Tremonti}, C., {Troup}, N.~W., {Valenzuela}, O., {Martinez
  Valpuesta}, I., {Vargas-Gonz{\'a}lez}, J., {Vargas-Maga{\~n}a}, M.,
  {Vazquez}, J.~A., {Villanova}, S., {Vivek}, M., {Vogt}, N., {Wake}, D.,
  {Walterbos}, R., {Wang}, Y., {Weaver}, B.~A., {Weijmans}, A.-M., {Weinberg},
  D.~H., {Westfall}, K.~B., {Whelan}, D.~G., {Wild}, V., {Wilson}, J.,
  {Wood-Vasey}, W.~M., {Wylezalek}, D., {Xiao}, T., {Yan}, R., {Yang}, M.,
  {Ybarra}, J.~E., {Y{\`e}che}, C., {Zakamska}, N., {Zamora}, O., {Zarrouk},
  P., {Zasowski}, G., {Zhang}, K., {Zhao}, G.-B., {Zheng}, Z., {Zheng}, Z.,
  {Zhou}, X., {Zhou}, Z.-M., {Zhu}, G.~B., {Zoccali}, M., \& {Zou}, H. 2017,
  \aj, 154, 28


\bibitem[{{Bradley} {et~al.}(2006){Bradley}, {Knapen}, {Beckman}, \&
  {Folkes}}]{Bradley_2006}
{Bradley}, T.~R., {Knapen}, J.~H., {Beckman}, J.~E., \& {Folkes}, S.~L. 2006,
  \aap, 459, L13


\bibitem[{{Bundy} {et~al.}(2015){Bundy}, {Bershady}, {Law}, {Yan}, {Drory},
  {MacDonald}, {Wake}, {Cherinka}, {S{\'a}nchez-Gallego}, {Weijmans}, {Thomas},
  {Tremonti}, {Masters}, {Coccato}, {Diamond-Stanic}, {Arag{\'o}n-Salamanca},
  {Avila-Reese}, {Badenes}, {Falc{\'o}n-Barroso}, {Belfiore}, {Bizyaev},
  {Blanc}, {Bland-Hawthorn}, {Blanton}, {Brownstein}, {Byler}, {Cappellari},
  {Conroy}, {Dutton}, {Emsellem}, {Etherington}, {Frinchaboy}, {Fu}, {Gunn},
  {Harding}, {Johnston}, {Kauffmann}, {Kinemuchi}, {Klaene}, {Knapen},
  {Leauthaud}, {Li}, {Lin}, {Maiolino}, {Malanushenko}, {Malanushenko}, {Mao},
  {Maraston}, {McDermid}, {Merrifield}, {Nichol}, {Oravetz}, {Pan}, {Parejko},
  {Sanchez}, {Schlegel}, {Simmons}, {Steele}, {Steinmetz}, {Thanjavur},
  {Thompson}, {Tinker}, {van den Bosch}, {Westfall}, {Wilkinson}, {Wright},
  {Xiao}, \& {Zhang}}]{Bundy_2015}
{Bundy}, K., {Bershady}, M.~A., {Law}, D.~R., {Yan}, R., {Drory}, N.,
  {MacDonald}, N., {Wake}, D.~A., {Cherinka}, B., {S{\'a}nchez-Gallego}, J.~R.,
  {Weijmans}, A.-M., {Thomas}, D., {Tremonti}, C., {Masters}, K., {Coccato},
  L., {Diamond-Stanic}, A.~M., {Arag{\'o}n-Salamanca}, A., {Avila-Reese}, V.,
  {Badenes}, C., {Falc{\'o}n-Barroso}, J., {Belfiore}, F., {Bizyaev}, D.,
  {Blanc}, G.~A., {Bland-Hawthorn}, J., {Blanton}, M.~R., {Brownstein}, J.~R.,
  {Byler}, N., {Cappellari}, M., {Conroy}, C., {Dutton}, A.~A., {Emsellem}, E.,
  {Etherington}, J., {Frinchaboy}, P.~M., {Fu}, H., {Gunn}, J.~E., {Harding},
  P., {Johnston}, E.~J., {Kauffmann}, G., {Kinemuchi}, K., {Klaene}, M.~A.,
  {Knapen}, J.~H., {Leauthaud}, A., {Li}, C., {Lin}, L., {Maiolino}, R.,
  {Malanushenko}, V., {Malanushenko}, E., {Mao}, S., {Maraston}, C.,
  {McDermid}, R.~M., {Merrifield}, M.~R., {Nichol}, R.~C., {Oravetz}, D.,
  {Pan}, K., {Parejko}, J.~K., {Sanchez}, S.~F., {Schlegel}, D., {Simmons}, A.,
  {Steele}, O., {Steinmetz}, M., {Thanjavur}, K., {Thompson}, B.~A., {Tinker},
  J.~L., {van den Bosch}, R.~C.~E., {Westfall}, K.~B., {Wilkinson}, D.,
  {Wright}, S., {Xiao}, T., \& {Zhang}, K. 2015, \apj, 798, 7


\bibitem[{{Catal{\'a}n-Torrecilla} {et~al.}(2015){Catal{\'a}n-Torrecilla}, {Gil
  de Paz}, {Castillo-Morales}, {Iglesias-P{\'a}ramo}, {S{\'a}nchez},
  {Kennicutt}, {P{\'e}rez-Gonz{\'a}lez}, {Marino}, {Walcher}, {Husemann},
  {Garc{\'\i}a-Benito}, {Mast}, {Gonz{\'a}lez Delgado}, {Mu{\~n}oz-Mateos},
  {Bland-Hawthorn}, {Bomans}, {Del Olmo}, {Galbany}, {Gomes}, {Kehrig},
  {L{\'o}pez-S{\'a}nchez}, {Mendoza}, {Monreal-Ibero}, {P{\'e}rez-Torres},
  {S{\'a}nchez-Bl{\'a}zquez}, {Vilchez}, \& {Califa
  Collaboration}}]{Catalan-Torrecilla_2015}
{Catal{\'a}n-Torrecilla}, C., {Gil de Paz}, A., {Castillo-Morales}, A.,
  {Iglesias-P{\'a}ramo}, J., {S{\'a}nchez}, S.~F., {Kennicutt}, R.~C.,
  {P{\'e}rez-Gonz{\'a}lez}, P.~G., {Marino}, R.~A., {Walcher}, C.~J.,
  {Husemann}, B., {Garc{\'\i}a-Benito}, R., {Mast}, D., {Gonz{\'a}lez Delgado},
  R.~M., {Mu{\~n}oz-Mateos}, J.~C., {Bland-Hawthorn}, J., {Bomans}, D.~J., {Del
  Olmo}, A., {Galbany}, L., {Gomes}, J.~M., {Kehrig}, C.,
  {L{\'o}pez-S{\'a}nchez}, {\'A}.~R., {Mendoza}, M.~A., {Monreal-Ibero}, A.,
  {P{\'e}rez-Torres}, M., {S{\'a}nchez-Bl{\'a}zquez}, P., {Vilchez}, J.~M., \&
  {Califa Collaboration}. 2015, \aap, 584, A87


\bibitem[{{Cid Fernandes} {et~al.}(2011){Cid Fernandes}, {Stasi{\'n}ska},
  {Mateus}, \& {Vale Asari}}]{Cid-Fernandes_2011}
{Cid Fernandes}, R., {Stasi{\'n}ska}, G., {Mateus}, A., \& {Vale Asari}, N.
  2011, \mnras, 413, 1687


\bibitem[{{Cid Fernandes} {et~al.}(2010){Cid Fernandes}, {Stasi{\'n}ska},
  {Schlickmann}, {Mateus}, {Vale Asari}, {Schoenell}, \&
  {Sodr{\'e}}}]{Cid-Fernandes2010}
{Cid Fernandes}, R., {Stasi{\'n}ska}, G., {Schlickmann}, M.~S., {Mateus}, A.,
  {Vale Asari}, N., {Schoenell}, W., \& {Sodr{\'e}}, L. 2010, \mnras, 403, 1036


\bibitem[{{Clauset} {et~al.}(2009){Clauset}, {Shalizi}, \&
  {Newman}}]{Clauset_2009}
{Clauset}, A., {Shalizi}, C.~R., \& {Newman}, M.~E.~J. 2009, SIAM Review, 51,
  661


\bibitem[{{Croom} {et~al.}(2012){Croom}, {Lawrence}, {Bland-Hawthorn},
  {Bryant}, {Fogarty}, {Richards}, {Goodwin}, {Farrell}, {Miziarski}, {Heald},
  {Jones}, {Lee}, {Colless}, {Brough}, {Hopkins}, {Bauer}, {Birchall}, {Ellis},
  {Horton}, {Leon-Saval}, {Lewis}, {L{\'o}pez-S{\'a}nchez}, {Min}, {Trinh}, \&
  {Trowland}}]{Croom_2012}
{Croom}, S.~M., {Lawrence}, J.~S., {Bland-Hawthorn}, J., {Bryant}, J.~J.,
  {Fogarty}, L., {Richards}, S., {Goodwin}, M., {Farrell}, T., {Miziarski}, S.,
  {Heald}, R., {Jones}, D.~H., {Lee}, S., {Colless}, M., {Brough}, S.,
  {Hopkins}, A.~M., {Bauer}, A.~E., {Birchall}, M.~N., {Ellis}, S., {Horton},
  A., {Leon-Saval}, S., {Lewis}, G., {L{\'o}pez-S{\'a}nchez}, {\'A}.~R., {Min},
  S.-S., {Trinh}, C., \& {Trowland}, H. 2012, \mnras, 421, 872


\bibitem[{{Crosthwaite} {et~al.}(2000){Crosthwaite}, {Turner}, \&
  {Ho}}]{Crosthwaite_2000}
{Crosthwaite}, L.~P., {Turner}, J.~L., \& {Ho}, P. T.~P. 2000, \aj, 119, 1720


\bibitem[{{D'Agostino} {et~al.}(2018){D'Agostino}, {Poetrodjojo}, {Ho},
  {Groves}, {Kewley}, {Madore}, {Rich}, \& {Seibert}}]{DAgostino_2018}
{D'Agostino}, J.~J., {Poetrodjojo}, H., {Ho}, I.~T., {Groves}, B., {Kewley},
  L., {Madore}, B.~F., {Rich}, J., \& {Seibert}, M. 2018, \mnras, 479, 4907


\bibitem[{{Della Bruna} {et~al.}(2022){Della Bruna}, {Adamo}, {Amram},
  {Rosolowsky}, {Usher}, {Sirressi}, {Schruba}, {Emsellem}, {Leroy}, {Bik},
  {Blair}, {McLeod}, {{\"O}stlin}, {Renaud}, {Robert}, {Rousseau-Nepton}, \&
  {Smith}}]{Della_Bruna_2022}
{Della Bruna}, L., {Adamo}, A., {Amram}, P., {Rosolowsky}, E., {Usher}, C.,
  {Sirressi}, M., {Schruba}, A., {Emsellem}, E., {Leroy}, A., {Bik}, A.,
  {Blair}, W.~P., {McLeod}, A.~F., {{\"O}stlin}, G., {Renaud}, F., {Robert},
  C., {Rousseau-Nepton}, L., \& {Smith}, L.~J. 2022, \aap, 660, A77


\bibitem[{{Della Bruna} {et~al.}(2020){Della Bruna}, {Adamo}, {Bik},
  {Fumagalli}, {Walterbos}, {{\"O}stlin}, {Bruzual}, {Calzetti}, {Charlot},
  {Grasha}, {Smith}, {Thilker}, \& {Wofford}}]{Della_Bruna_2020}
{Della Bruna}, L., {Adamo}, A., {Bik}, A., {Fumagalli}, M., {Walterbos}, R.,
  {{\"O}stlin}, G., {Bruzual}, G., {Calzetti}, D., {Charlot}, S., {Grasha}, K.,
  {Smith}, L.~J., {Thilker}, D., \& {Wofford}, A. 2020, \aap, 635, A134


\bibitem[{{Della Bruna} {et~al.}(2021){Della Bruna}, {Adamo}, {Lee}, {Smith},
  {Krumholz}, {Bik}, {Calzetti}, {Fox}, {Fumagalli}, {Grasha}, {Messa},
  {{\"O}stlin}, {Walterbos}, \& {Wofford}}]{Della_Bruna_2021}
{Della Bruna}, L., {Adamo}, A., {Lee}, J.~C., {Smith}, L.~J., {Krumholz}, M.,
  {Bik}, A., {Calzetti}, D., {Fox}, A., {Fumagalli}, M., {Grasha}, K., {Messa},
  M., {{\"O}stlin}, G., {Walterbos}, R., \& {Wofford}, A. 2021, \aap, 650, A103


\bibitem[{{Dopita} \& {Evans}(1986)}]{Dopita_1986}
{Dopita}, M.~A. \& {Evans}, I.~N. 1986, \apj, 307, 431


\bibitem[{{Dottori} \& {Copetti}(1989)}]{Dottori-Copetti_1989}
{Dottori}, H.~A. \& {Copetti}, M. V.~F. 1989, \rmxaa, 18, 115


\bibitem[{{Drory} {et~al.}(2024){Drory}, {Blanc}, {Kreckel}, {Sanchez},
  {Mejia-Narvaez}, {Johnston}, {Jones}, {Pellegrini}, {Konidaris}, {Herbst},
  {Sanchez-Gallego}, {Kollmeier}, {de Almeida}, {Barrera-Ballesteros},
  {Bizyaev}, {Brownstein}, {Saguer}, {Cherinka}, {Cioni}, {Congiu}, {Cosens},
  {Dias}, {Donor}, {Egorov}, {Egorova}, {Froning}, {Garcia}, {Glover}, {Greve},
  {Haeberle}, {Hoy}, {Ibarra}, {Li}, {Klessen}, {Krishnarao}, {Kumari}, {Long},
  {Mendez-Delgado}, {Popa}, {Ramirez}, {Rix}, {Mata Sanchez}, {Sankrit},
  {Sattler}, {Sayres}, {Singh}, {Stringfellow}, {Wachter}, {Watkins}, {Wong},
  \& {Wofford}}]{Drory_2024}
{Drory}, N., {Blanc}, G.~A., {Kreckel}, K., {Sanchez}, S.~F., {Mejia-Narvaez},
  A., {Johnston}, E.~J., {Jones}, A.~M., {Pellegrini}, E.~W., {Konidaris},
  N.~P., {Herbst}, T., {Sanchez-Gallego}, J., {Kollmeier}, J.~A., {de Almeida},
  F., {Barrera-Ballesteros}, J.~K., {Bizyaev}, D., {Brownstein}, J.~R.,
  {Saguer}, M. C.~i., {Cherinka}, B., {Cioni}, M.-R.~L., {Congiu}, E.,
  {Cosens}, M., {Dias}, B., {Donor}, J., {Egorov}, O., {Egorova}, E.,
  {Froning}, C.~S., {Garcia}, P., {Glover}, S. C.~O., {Greve}, H., {Haeberle},
  M., {Hoy}, K., {Ibarra}, H., {Li}, J., {Klessen}, R.~S., {Krishnarao}, D.,
  {Kumari}, N., {Long}, K.~S., {Mendez-Delgado}, J.~E., {Popa}, S.~A.,
  {Ramirez}, S., {Rix}, H.-W., {Mata Sanchez}, A., {Sankrit}, R., {Sattler},
  N., {Sayres}, C., {Singh}, A., {Stringfellow}, G., {Wachter}, S., {Watkins},
  E.~J., {Wong}, T., \& {Wofford}, A. 2024, arXiv e-prints, arXiv:2405.01637


\bibitem[{{Drory} {et~al.}(2015){Drory}, {MacDonald}, {Bershady}, {Bundy},
  {Gunn}, {Law}, {Smith}, {Stoll}, {Tremonti}, {Wake}, {Yan}, {Weijmans},
  {Byler}, {Cherinka}, {Cope}, {Eigenbrot}, {Harding}, {Holder}, {Huehnerhoff},
  {Jaehnig}, {Jansen}, {Klaene}, {Paat}, {Percival}, \& {Sayres}}]{Drory_2015}
{Drory}, N., {MacDonald}, N., {Bershady}, M.~A., {Bundy}, K., {Gunn}, J.,
  {Law}, D.~R., {Smith}, M., {Stoll}, R., {Tremonti}, C.~A., {Wake}, D.~A.,
  {Yan}, R., {Weijmans}, A.~M., {Byler}, N., {Cherinka}, B., {Cope}, F.,
  {Eigenbrot}, A., {Harding}, P., {Holder}, D., {Huehnerhoff}, J., {Jaehnig},
  K., {Jansen}, T.~C., {Klaene}, M., {Paat}, A.~M., {Percival}, J., \&
  {Sayres}, C. 2015, \aj, 149, 77


\bibitem[{{Emsellem} {et~al.}(2022){Emsellem}, {Schinnerer}, {Santoro},
  {Belfiore}, {Pessa}, {McElroy}, {Blanc}, {Congiu}, {Groves}, {Ho}, {Kreckel},
  {Razza}, {Sanchez-Blazquez}, {Egorov}, {Faesi}, {Klessen}, {Leroy}, {Meidt},
  {Querejeta}, {Rosolowsky}, {Scheuermann}, {Anand}, {Barnes},
  {Be{\v{s}}li{\'c}}, {Bigiel}, {Boquien}, {Cao}, {Chevance}, {Dale},
  {Eibensteiner}, {Glover}, {Grasha}, {Henshaw}, {Hughes}, {Koch}, {Kruijssen},
  {Lee}, {Liu}, {Pan}, {Pety}, {Saito}, {Sandstrom}, {Schruba}, {Sun},
  {Thilker}, {Usero}, {Watkins}, \& {Williams}}]{Emsellem_2022}
{Emsellem}, E., {Schinnerer}, E., {Santoro}, F., {Belfiore}, F., {Pessa}, I.,
  {McElroy}, R., {Blanc}, G.~A., {Congiu}, E., {Groves}, B., {Ho}, I.~T.,
  {Kreckel}, K., {Razza}, A., {Sanchez-Blazquez}, P., {Egorov}, O., {Faesi},
  C., {Klessen}, R.~S., {Leroy}, A.~K., {Meidt}, S., {Querejeta}, M.,
  {Rosolowsky}, E., {Scheuermann}, F., {Anand}, G.~S., {Barnes}, A.~T.,
  {Be{\v{s}}li{\'c}}, I., {Bigiel}, F., {Boquien}, M., {Cao}, Y., {Chevance},
  M., {Dale}, D.~A., {Eibensteiner}, C., {Glover}, S. C.~O., {Grasha}, K.,
  {Henshaw}, J.~D., {Hughes}, A., {Koch}, E.~W., {Kruijssen}, J.~M.~D., {Lee},
  J., {Liu}, D., {Pan}, H.-A., {Pety}, J., {Saito}, T., {Sandstrom}, K.~M.,
  {Schruba}, A., {Sun}, J., {Thilker}, D.~A., {Usero}, A., {Watkins}, E.~J., \&
  {Williams}, T.~G. 2022, \aap, 659, A191


\bibitem[{{Espinosa-Ponce} {et~al.}(2022){Espinosa-Ponce}, {S{\'a}nchez},
  {Morisset}, {Barrera-Ballesteros}, {Galbany}, {Garc{\'\i}a-Benito},
  {Lacerda}, \& {Mast}}]{Espinosa-Ponce_2022}
{Espinosa-Ponce}, C., {S{\'a}nchez}, S.~F., {Morisset}, C.,
  {Barrera-Ballesteros}, J.~K., {Galbany}, L., {Garc{\'\i}a-Benito}, R.,
  {Lacerda}, E.~A.~D., \& {Mast}, D. 2022, \mnras


\bibitem[{{Galbany} {et~al.}(2016){Galbany}, {Anderson}, {Rosales-Ortega},
  {Kuncarayakti}, {Kr{\"u}hler}, {S{\'a}nchez}, {Falc{\'o}n-Barroso},
  {P{\'e}rez}, {Maureira}, {Hamuy}, {Gonz{\'a}lez-Gait{\'a}n}, {F{\"o}rster},
  \& {Moral}}]{Galbany_2016}
{Galbany}, L., {Anderson}, J.~P., {Rosales-Ortega}, F.~F., {Kuncarayakti}, H.,
  {Kr{\"u}hler}, T., {S{\'a}nchez}, S.~F., {Falc{\'o}n-Barroso}, J.,
  {P{\'e}rez}, E., {Maureira}, J.~C., {Hamuy}, M., {Gonz{\'a}lez-Gait{\'a}n},
  S., {F{\"o}rster}, F., \& {Moral}, V. 2016, \mnras, 455, 4087


\bibitem[{{Galbany} {et~al.}(2020){Galbany}, {Anderson}, {S{\'a}nchez},
  {Kunkarayakti}, {Lyman}, {Kruehler}, {Aquino}, {Afonso}, {Ascasibar},
  {Ashall}, {Badenes}, {Burns}, {Dessart}, {Dominguez}, {Dong},
  {Falc{\'o}n-Barroso}, {F{\"o}rster}, {Gonz{\'a}lez-Gait{\'a}n},
  {Guti{\'e}rrez}, {Hamuy}, {Hoeflich}, {Holoien}, {Hsiao}, {James}, {Kangas},
  {Kankare}, {Kochanek}, {Mattila}, {Mour{\~a}o}, {Ortega-Minakata},
  {P{\'e}rez}, {P{\'e}rez}, {Phillips}, {Paulina-Afonso}, {Prieto}, {Razza},
  {Rosales-Ortega}, {Ruiz-Lara}, {S{\'a}nchez-Bl{\'a}zquez},
  {S{\'a}nchez-Menguiano}, {Schady}, {Shapee}, {Smith}, {Stanek},
  {Stritzinger}, {Sullivan}, \& {Wang}}]{Galbany_2020}
{Galbany}, L., {Anderson}, J.~P., {S{\'a}nchez}, S.~F., {Kunkarayakti}, H.,
  {Lyman}, J., {Kruehler}, T., {Aquino}, E., {Afonso}, A.~S., {Ascasibar}, Y.,
  {Ashall}, C., {Badenes}, C., {Burns}, C., {Dessart}, L., {Dominguez}, I.,
  {Dong}, S., {Falc{\'o}n-Barroso}, J., {F{\"o}rster}, F.,
  {Gonz{\'a}lez-Gait{\'a}n}, S., {Guti{\'e}rrez}, C., {Hamuy}, M., {Hoeflich},
  P., {Holoien}, T., {Hsiao}, E., {James}, P., {Kangas}, T., {Kankare}, E.,
  {Kochanek}, C., {Mattila}, S., {Mour{\~a}o}, A., {Ortega-Minakata}, R.,
  {P{\'e}rez}, E., {P{\'e}rez}, I., {Phillips}, M., {Paulina-Afonso}, A.,
  {Prieto}, J.~L., {Razza}, A., {Rosales-Ortega}, F., {Ruiz-Lara}, T.,
  {S{\'a}nchez-Bl{\'a}zquez}, P., {S{\'a}nchez-Menguiano}, L., {Schady}, P.,
  {Shapee}, B., {Smith}, M., {Stanek}, K., {Stritzinger}, M., {Sullivan}, M.,
  \& {Wang}, L. 2020, in XIV.0 Scientific Meeting (virtual) of the Spanish
  Astronomical Society, 38


\bibitem[{{Garc{\'\i}a-Benito} {et~al.}(2011){Garc{\'\i}a-Benito}, {P{\'e}rez},
  {D{\'\i}az}, {Ma{\'\i}z Apell{\'a}niz}, \&
  {Cervi{\~n}o}}]{Garcia-Benito_2011}
{Garc{\'\i}a-Benito}, R., {P{\'e}rez}, E., {D{\'\i}az}, {\'A}.~I., {Ma{\'\i}z
  Apell{\'a}niz}, J., \& {Cervi{\~n}o}, M. 2011, \aj, 141, 126


\bibitem[{{Gonz{\'a}lez Delgado} \& {P{\'e}rez}(1997)}]{Gonzalez-Delgado_1997}
{Gonz{\'a}lez Delgado}, R.~M. \& {P{\'e}rez}, E. 1997, \apjs, 108, 199


\bibitem[{{Grasha} {et~al.}(2022){Grasha}, {Chen}, {Battisti}, {Acharyya},
  {Ridolfo}, {Poehler}, {Mably}, {Verma}, {Hayward}, {Kharbanda},
  {Poetrodjojo}, {Seibert}, {Rich}, {Madore}, \& {Kewley}}]{Grasha_2022}
{Grasha}, K., {Chen}, Q.~H., {Battisti}, A.~J., {Acharyya}, A., {Ridolfo}, S.,
  {Poehler}, E., {Mably}, S., {Verma}, A.~A., {Hayward}, K.~L., {Kharbanda},
  A., {Poetrodjojo}, H., {Seibert}, M., {Rich}, J.~A., {Madore}, B.~F., \&
  {Kewley}, L.~J. 2022, \apj, 929, 118


\bibitem[{{Groves} {et~al.}(2023){Groves}, {Kreckel}, {Santoro}, {Belfiore},
  {Zavodnik}, {Congiu}, {Egorov}, {Emsellem}, {Grasha}, {Leroy}, {Scheuermann},
  {Schinnerer}, {Watkins}, {Barnes}, {Bigiel}, {Dale}, {Glover}, {Pessa},
  {Sanchez-Blazquez}, \& {Williams}}]{Groves_2023}
{Groves}, B., {Kreckel}, K., {Santoro}, F., {Belfiore}, F., {Zavodnik}, E.,
  {Congiu}, E., {Egorov}, O.~V., {Emsellem}, E., {Grasha}, K., {Leroy}, A.,
  {Scheuermann}, F., {Schinnerer}, E., {Watkins}, E.~J., {Barnes}, A.~T.,
  {Bigiel}, F., {Dale}, D.~A., {Glover}, S.~C.~O., {Pessa}, I.,
  {Sanchez-Blazquez}, P., \& {Williams}, T.~G. 2023, \mnras, 520, 4902


\bibitem[{{Gunn} {et~al.}(2006){Gunn}, {Siegmund}, {Mannery}, {Owen}, {Hull},
  {Leger}, {Carey}, {Knapp}, {York}, {Boroski}, {Kent}, {Lupton}, {Rockosi},
  {Evans}, {Waddell}, {Anderson}, {Annis}, {Barentine}, {Bartoszek}, {Bastian},
  {Bracker}, {Brewington}, {Briegel}, {Brinkmann}, {Brown}, {Carr},
  {Czarapata}, {Drennan}, {Dombeck}, {Federwitz}, {Gillespie}, {Gonzales},
  {Hansen}, {Harvanek}, {Hayes}, {Jordan}, {Kinney}, {Klaene}, {Kleinman},
  {Kron}, {Kresinski}, {Lee}, {Limmongkol}, {Lindenmeyer}, {Long}, {Loomis},
  {McGehee}, {Mantsch}, {Neilsen}, {Neswold}, {Newman}, {Nitta}, {Peoples},
  {Pier}, {Prieto}, {Prosapio}, {Rivetta}, {Schneider}, {Snedden}, \&
  {Wang}}]{Gunn_2006}
{Gunn}, J.~E., {Siegmund}, W.~A., {Mannery}, E.~J., {Owen}, R.~E., {Hull},
  C.~L., {Leger}, R.~F., {Carey}, L.~N., {Knapp}, G.~R., {York}, D.~G.,
  {Boroski}, W.~N., {Kent}, S.~M., {Lupton}, R.~H., {Rockosi}, C.~M., {Evans},
  M.~L., {Waddell}, P., {Anderson}, J.~E., {Annis}, J., {Barentine}, J.~C.,
  {Bartoszek}, L.~M., {Bastian}, S., {Bracker}, S.~B., {Brewington}, H.~J.,
  {Briegel}, C.~I., {Brinkmann}, J., {Brown}, Y.~J., {Carr}, M.~A.,
  {Czarapata}, P.~C., {Drennan}, C.~C., {Dombeck}, T., {Federwitz}, G.~R.,
  {Gillespie}, B.~A., {Gonzales}, C., {Hansen}, S.~U., {Harvanek}, M., {Hayes},
  J., {Jordan}, W., {Kinney}, E., {Klaene}, M., {Kleinman}, S.~J., {Kron},
  R.~G., {Kresinski}, J., {Lee}, G., {Limmongkol}, S., {Lindenmeyer}, C.~W.,
  {Long}, D.~C., {Loomis}, C.~L., {McGehee}, P.~M., {Mantsch}, P.~M.,
  {Neilsen}, Jr., E.~H., {Neswold}, R.~M., {Newman}, P.~R., {Nitta}, A.,
  {Peoples}, Jr., J., {Pier}, J.~R., {Prieto}, P.~S., {Prosapio}, A.,
  {Rivetta}, C., {Schneider}, D.~P., {Snedden}, S., \& {Wang}, S.-i. 2006, \aj,
  131, 2332


\bibitem[{{Hernandez} {et~al.}(2005){Hernandez}, {Carignan}, {Amram}, {Chemin},
  \& {Daigle}}]{Hernandez_2005}
{Hernandez}, O., {Carignan}, C., {Amram}, P., {Chemin}, L., \& {Daigle}, O.
  2005, \mnras, 360, 1201


\bibitem[{{Ho}(2019)}]{Ho_2019}
{Ho}, I.~T. 2019, \mnras, 485, 3569


\bibitem[{{Hodge} \& {Kennicutt}(1983)}]{Hodge-Kennicutt_1983}
{Hodge}, P.~W. \& {Kennicutt}, R.~C., J. 1983, \apj, 267, 563


\bibitem[{{Ishizuki} {et~al.}(1990){Ishizuki}, {Kawabe}, {Ishiguro}, {Okumura},
  \& {Morita}}]{Ishizuki_1990}
{Ishizuki}, S., {Kawabe}, R., {Ishiguro}, M., {Okumura}, S.~K., \& {Morita},
  K.-I. 1990, \nat, 344, 224


\bibitem[{{Jarrett} {et~al.}(2013){Jarrett}, {Masci}, {Tsai}, {Petty},
  {Cluver}, {Assef}, {Benford}, {Blain}, {Bridge}, {Donoso}, {Eisenhardt},
  {Koribalski}, {Lake}, {Neill}, {Seibert}, {Sheth}, {Stanford}, \&
  {Wright}}]{Jarrett_2013}
{Jarrett}, T.~H., {Masci}, F., {Tsai}, C.~W., {Petty}, S., {Cluver}, M.~E.,
  {Assef}, R.~J., {Benford}, D., {Blain}, A., {Bridge}, C., {Donoso}, E.,
  {Eisenhardt}, P., {Koribalski}, B., {Lake}, S., {Neill}, J.~D., {Seibert},
  M., {Sheth}, K., {Stanford}, S., \& {Wright}, E. 2013, \aj, 145, 6


\bibitem[{{Ji} \& {Yan}(2022)}]{Ji_2022}
{Ji}, X. \& {Yan}, R. 2022, \aap, 659, A112


\bibitem[{{Kauffmann} {et~al.}(2003){Kauffmann}, {Heckman}, {White}, {Charlot},
  {Tremonti}, {Peng}, {Seibert}, {Brinkmann}, {Nichol}, {SubbaRao}, \&
  {York}}]{Kauffmann_2003}
{Kauffmann}, G., {Heckman}, T.~M., {White}, S. D.~M., {Charlot}, S.,
  {Tremonti}, C., {Peng}, E.~W., {Seibert}, M., {Brinkmann}, J., {Nichol},
  R.~C., {SubbaRao}, M., \& {York}, D. 2003, \mnras, 341, 54


\bibitem[{{Kennicutt} {et~al.}(2011){Kennicutt}, {Calzetti}, {Aniano},
  {Appleton}, {Armus}, {Beir{\~a}o}, {Bolatto}, {Brandl}, {Crocker}, {Croxall},
  {Dale}, {Donovan Meyer}, {Draine}, {Engelbracht}, {Galametz}, {Gordon},
  {Groves}, {Hao}, {Helou}, {Hinz}, {Hunt}, {Johnson}, {Koda}, {Krause},
  {Leroy}, {Li}, {Meidt}, {Montiel}, {Murphy}, {Rahman}, {Rix}, {Roussel},
  {Sandstrom}, {Sauvage}, {Schinnerer}, {Skibba}, {Smith}, {Srinivasan},
  {Vigroux}, {Walter}, {Wilson}, {Wolfire}, \& {Zibetti}}]{Kennicutt_2011}
{Kennicutt}, R.~C., {Calzetti}, D., {Aniano}, G., {Appleton}, P., {Armus}, L.,
  {Beir{\~a}o}, P., {Bolatto}, A.~D., {Brandl}, B., {Crocker}, A., {Croxall},
  K., {Dale}, D.~A., {Donovan Meyer}, J., {Draine}, B.~T., {Engelbracht},
  C.~W., {Galametz}, M., {Gordon}, K.~D., {Groves}, B., {Hao}, C.~N., {Helou},
  G., {Hinz}, J., {Hunt}, L.~K., {Johnson}, B., {Koda}, J., {Krause}, O.,
  {Leroy}, A.~K., {Li}, Y., {Meidt}, S., {Montiel}, E., {Murphy}, E.~J.,
  {Rahman}, N., {Rix}, H.~W., {Roussel}, H., {Sandstrom}, K., {Sauvage}, M.,
  {Schinnerer}, E., {Skibba}, R., {Smith}, J.~D.~T., {Srinivasan}, S.,
  {Vigroux}, L., {Walter}, F., {Wilson}, C.~D., {Wolfire}, M., \& {Zibetti}, S.
  2011, \pasp, 123, 1347


\bibitem[{{Kennicutt} \& {Evans}(2012)}]{Kennicutt_2012}
{Kennicutt}, R.~C. \& {Evans}, N.~J. 2012, \araa, 50, 531


\bibitem[{{Kewley} {et~al.}(2001){Kewley}, {Dopita}, {Sutherland}, {Heisler},
  \& {Trevena}}]{Kewley_2001}
{Kewley}, L.~J., {Dopita}, M.~A., {Sutherland}, R.~S., {Heisler}, C.~A., \&
  {Trevena}, J. 2001, \apj, 556, 121


\bibitem[{{Knapen}(1998)}]{Knapen_1998}
{Knapen}, J.~H. 1998, \mnras, 297, 255


\bibitem[{{Knapen} {et~al.}(2004){Knapen}, {Stedman}, {Bramich}, {Folkes}, \&
  {Bradley}}]{Knapen_2004}
{Knapen}, J.~H., {Stedman}, S., {Bramich}, D.~M., {Folkes}, S.~L., \&
  {Bradley}, T.~R. 2004, \aap, 426, 1135


\bibitem[{{Kollmeier} {et~al.}(2017){Kollmeier}, {Zasowski}, {Rix}, {Johns},
  {Anderson}, {Drory}, {Johnson}, {Pogge}, {Bird}, {Blanc}, {Brownstein},
  {Crane}, {De Lee}, {Klaene}, {Kreckel}, {MacDonald}, {Merloni}, {Ness},
  {O'Brien}, {Sanchez-Gallego}, {Sayres}, {Shen}, {Thakar}, {Tkachenko},
  {Aerts}, {Blanton}, {Eisenstein}, {Holtzman}, {Maoz}, {Nandra}, {Rockosi},
  {Weinberg}, {Bovy}, {Casey}, {Chaname}, {Clerc}, {Conroy}, {Eracleous},
  {G{\"a}nsicke}, {Hekker}, {Horne}, {Kauffmann}, {McQuinn}, {Pellegrini},
  {Schinnerer}, {Schlafly}, {Schwope}, {Seibert}, {Teske}, \& {van
  Saders}}]{Kollmeier_2017}
{Kollmeier}, J.~A., {Zasowski}, G., {Rix}, H.-W., {Johns}, M., {Anderson},
  S.~F., {Drory}, N., {Johnson}, J.~A., {Pogge}, R.~W., {Bird}, J.~C., {Blanc},
  G.~A., {Brownstein}, J.~R., {Crane}, J.~D., {De Lee}, N.~M., {Klaene}, M.~A.,
  {Kreckel}, K., {MacDonald}, N., {Merloni}, A., {Ness}, M.~K., {O'Brien}, T.,
  {Sanchez-Gallego}, J.~R., {Sayres}, C.~C., {Shen}, Y., {Thakar}, A.~R.,
  {Tkachenko}, A., {Aerts}, C., {Blanton}, M.~R., {Eisenstein}, D.~J.,
  {Holtzman}, J.~A., {Maoz}, D., {Nandra}, K., {Rockosi}, C., {Weinberg},
  D.~H., {Bovy}, J., {Casey}, A.~R., {Chaname}, J., {Clerc}, N., {Conroy}, C.,
  {Eracleous}, M., {G{\"a}nsicke}, B.~T., {Hekker}, S., {Horne}, K.,
  {Kauffmann}, J., {McQuinn}, K. B.~W., {Pellegrini}, E.~W., {Schinnerer}, E.,
  {Schlafly}, E.~F., {Schwope}, A.~D., {Seibert}, M., {Teske}, J.~K., \& {van
  Saders}, J.~L. 2017, arXiv e-prints, arXiv:1711.03234


\bibitem[{{Konidaris} {et~al.}(2020){Konidaris}, {Drory}, {Froning}, {Hebert},
  {Bilgi}, {Blanc}, {Lanz}, {Hull}, {Kollmeier}, {Ramirez}, {Wachter},
  {Kreckel}, {Pak}, {Pellegrini}, {Almeida}, {Case}, {Zhelem}, {Feger},
  {Lawrence}, {Lesser}, {Herbst}, {Sanchez-Gallego}, {Bershady},
  {Chattopadhyay}, {Hauser}, {Smith}, {Wolf}, \& {Yan}}]{Konidaris_2020}
{Konidaris}, N.~P., {Drory}, N., {Froning}, C.~S., {Hebert}, A., {Bilgi}, P.,
  {Blanc}, G.~A., {Lanz}, A.~E., {Hull}, C.~L., {Kollmeier}, J.~A., {Ramirez},
  S., {Wachter}, S., {Kreckel}, K., {Pak}, S., {Pellegrini}, E., {Almeida}, A.,
  {Case}, S., {Zhelem}, R., {Feger}, T., {Lawrence}, J., {Lesser}, M.,
  {Herbst}, T., {Sanchez-Gallego}, J., {Bershady}, M.~A., {Chattopadhyay}, S.,
  {Hauser}, A., {Smith}, M., {Wolf}, M.~J., \& {Yan}, R. Society of
  Photo-Optical Instrumentation Engineers (SPIE) Conference Series, Vol. 11447,
  , Ground-based and Airborne Instrumentation for Astronomy VIII, ed. C.~J.
  {Evans}J.~J. {Bryant} \& K.~{Motohara}, 1144718


\bibitem[{{Kreckel} {et~al.}(2019){Kreckel}, {Ho}, {Blanc}, {Groves},
  {Santoro}, {Schinnerer}, {Bigiel}, {Chevance}, {Congiu}, {Emsellem}, {Faesi},
  {Glover}, {Grasha}, {Kruijssen}, {Lang}, {Leroy}, {Meidt}, {McElroy}, {Pety},
  {Rosolowsky}, {Saito}, {Sandstrom}, {Sanchez-Blazquez}, \&
  {Schruba}}]{Kreckel_2019}
{Kreckel}, K., {Ho}, I.~T., {Blanc}, G.~A., {Groves}, B., {Santoro}, F.,
  {Schinnerer}, E., {Bigiel}, F., {Chevance}, M., {Congiu}, E., {Emsellem}, E.,
  {Faesi}, C., {Glover}, S.~C.~O., {Grasha}, K., {Kruijssen}, J.~M.~D., {Lang},
  P., {Leroy}, A.~K., {Meidt}, S.~E., {McElroy}, R., {Pety}, J., {Rosolowsky},
  E., {Saito}, T., {Sandstrom}, K., {Sanchez-Blazquez}, P., \& {Schruba}, A.
  2019, \apj, 887, 80


\bibitem[{{Kuno} {et~al.}(2007){Kuno}, {Sato}, {Nakanishi}, {Hirota}, {Tosaki},
  {Shioya}, {Sorai}, {Nakai}, {Nishiyama}, \& {Vila-Vilar{\'o}}}]{Kuno_2007}
{Kuno}, N., {Sato}, N., {Nakanishi}, H., {Hirota}, A., {Tosaki}, T., {Shioya},
  Y., {Sorai}, K., {Nakai}, N., {Nishiyama}, K., \& {Vila-Vilar{\'o}}, B. 2007,
  \pasj, 59, 117


\bibitem[{{Lacerda} {et~al.}(2018){Lacerda}, {Cid Fernandes}, {Couto},
  {Stasi{\'n}ska}, {Garc{\'\i}a-Benito}, {Vale Asari}, {P{\'e}rez},
  {Gonz{\'a}lez Delgado}, {S{\'a}nchez}, \& {de Amorim}}]{Lacerda_2018}
{Lacerda}, E.~A.~D., {Cid Fernandes}, R., {Couto}, G.~S., {Stasi{\'n}ska}, G.,
  {Garc{\'\i}a-Benito}, R., {Vale Asari}, N., {P{\'e}rez}, E., {Gonz{\'a}lez
  Delgado}, R.~M., {S{\'a}nchez}, S.~F., \& {de Amorim}, A.~L. 2018, \mnras,
  474, 3727


\bibitem[{{Lacerda} {et~al.}(2020){Lacerda}, {S{\'a}nchez}, {Cid Fernandes},
  {L{\'o}pez-Cob{\'a}}, {Espinosa-Ponce}, \& {Galbany}}]{Lacerda_2020}
{Lacerda}, E. A.~D., {S{\'a}nchez}, S.~F., {Cid Fernandes}, R.,
  {L{\'o}pez-Cob{\'a}}, C., {Espinosa-Ponce}, C., \& {Galbany}, L. 2020,
  \mnras, 492, 3073


\bibitem[{{Lacerda} {et~al.}(2022){Lacerda}, {S{\'a}nchez},
  {Mej{\'\i}a-Narv{\'a}ez}, {Camps-Fari{\~n}a}, {Espinosa-Ponce},
  {Barrera-Ballesteros}, {Ibarra-Medel}, \& {Lugo-Aranda}}]{Lacerda_2022}
{Lacerda}, E. A.~D., {S{\'a}nchez}, S.~F., {Mej{\'\i}a-Narv{\'a}ez}, A.,
  {Camps-Fari{\~n}a}, A., {Espinosa-Ponce}, C., {Barrera-Ballesteros}, J.~K.,
  {Ibarra-Medel}, H., \& {Lugo-Aranda}, A.~Z. 2022, \na, 97, 101895


\bibitem[{{Law} {et~al.}(2016){Law}, {Cherinka}, {Yan}, {Andrews}, {Bershady},
  {Bizyaev}, {Blanc}, {Blanton}, {Bolton}, {Brownstein}, {Bundy}, {Chen},
  {Drory}, {D'Souza}, {Fu}, {Jones}, {Kauffmann}, {MacDonald}, {Masters},
  {Newman}, {Parejko}, {S{\'a}nchez-Gallego}, {S{\'a}nchez}, {Schlegel},
  {Thomas}, {Wake}, {Weijmans}, {Westfall}, \& {Zhang}}]{Law_2016}
{Law}, D.~R., {Cherinka}, B., {Yan}, R., {Andrews}, B.~H., {Bershady}, M.~A.,
  {Bizyaev}, D., {Blanc}, G.~A., {Blanton}, M.~R., {Bolton}, A.~S.,
  {Brownstein}, J.~R., {Bundy}, K., {Chen}, Y., {Drory}, N., {D'Souza}, R.,
  {Fu}, H., {Jones}, A., {Kauffmann}, G., {MacDonald}, N., {Masters}, K.~L.,
  {Newman}, J.~A., {Parejko}, J.~K., {S{\'a}nchez-Gallego}, J.~R.,
  {S{\'a}nchez}, S.~F., {Schlegel}, D.~J., {Thomas}, D., {Wake}, D.~A.,
  {Weijmans}, A.-M., {Westfall}, K.~B., \& {Zhang}, K. 2016, \aj, 152, 83


\bibitem[{{L{\'o}pez-Cob{\'a}} {et~al.}(2020){L{\'o}pez-Cob{\'a}},
  {S{\'a}nchez}, {Anderson}, {Cruz-Gonz{\'a}lez}, {Galbany}, {Ruiz-Lara},
  {Barrera-Ballesteros}, {Prieto}, \& {Kuncarayakti}}]{Lopez-Coba_2020}
{L{\'o}pez-Cob{\'a}}, C., {S{\'a}nchez}, S.~F., {Anderson}, J.~P.,
  {Cruz-Gonz{\'a}lez}, I., {Galbany}, L., {Ruiz-Lara}, T.,
  {Barrera-Ballesteros}, J.~K., {Prieto}, J.~L., \& {Kuncarayakti}, H. 2020,
  \aj, 159, 167


\bibitem[{{L{\'o}pez-Cob{\'a}} {et~al.}(2017){L{\'o}pez-Cob{\'a}},
  {S{\'a}nchez}, {Moiseev}, {Oparin}, {Bitsakis}, {Cruz-Gonz{\'a}lez},
  {Morisset}, {Galbany}, {Bland-Hawthorn}, {Roth}, {Dettmar}, {Bomans},
  {Gonz{\'a}lez Delgado}, {Cano-D{\'\i}az}, {Marino}, {Kehrig}, {Monreal
  Ibero}, \& {Abril-Melgarejo}}]{Lopez-Coba_2017}
{L{\'o}pez-Cob{\'a}}, C., {S{\'a}nchez}, S.~F., {Moiseev}, A.~V., {Oparin},
  D.~V., {Bitsakis}, T., {Cruz-Gonz{\'a}lez}, I., {Morisset}, C., {Galbany},
  L., {Bland-Hawthorn}, J., {Roth}, M.~M., {Dettmar}, R.~J., {Bomans}, D.~J.,
  {Gonz{\'a}lez Delgado}, R.~M., {Cano-D{\'\i}az}, M., {Marino}, R.~A.,
  {Kehrig}, C., {Monreal Ibero}, A., \& {Abril-Melgarejo}, V. 2017, \mnras,
  467, 4951


\bibitem[{{Lugo-Aranda} {et~al.}(2024){Lugo-Aranda}, {S{\'a}nchez},
  {Barrera-Ballesteros}, {L{\'o}pez-Cob{\'a}}, {Espinosa-Ponce}, {Galbany}, \&
  {Anderson}}]{Lugo-Aranda_2024}
{Lugo-Aranda}, A.~Z., {S{\'a}nchez}, S.~F., {Barrera-Ballesteros}, J.~K.,
  {L{\'o}pez-Cob{\'a}}, C., {Espinosa-Ponce}, C., {Galbany}, L., \& {Anderson},
  J.~P. 2024, \mnras, 528, 6099


\bibitem[{{Lugo-Aranda} {et~al.}(2022){Lugo-Aranda}, {S{\'a}nchez},
  {Espinosa-Ponce}, {L{\'o}pez-Cob{\'a}}, {Galbany}, {Barrera-Ballesteros},
  {S{\'a}nchez-Menguiano}, \& {Anderson}}]{LugoAranda_2022}
{Lugo-Aranda}, A.~Z., {S{\'a}nchez}, S.~F., {Espinosa-Ponce}, C.,
  {L{\'o}pez-Cob{\'a}}, C., {Galbany}, L., {Barrera-Ballesteros}, J.~K.,
  {S{\'a}nchez-Menguiano}, L., \& {Anderson}, J.~P. 2022, RAS Techniques and
  Instruments, 1, 3


\bibitem[{{McCall} {et~al.}(1985){McCall}, {Rybski}, \&
  {Shields}}]{McCall_1985}
{McCall}, M.~L., {Rybski}, P.~M., \& {Shields}, G.~A. 1985, \apjs, 57, 1


\bibitem[{{McLeod} {et~al.}(2021){McLeod}, {Ali}, {Chevance}, {Della Bruna},
  {Schruba}, {Stevance}, {Adamo}, {Kruijssen}, {Longmore}, {Weisz}, \&
  {Zeidler}}]{McLeod_2021}
{McLeod}, A.~F., {Ali}, A.~A., {Chevance}, M., {Della Bruna}, L., {Schruba},
  A., {Stevance}, H.~F., {Adamo}, A., {Kruijssen}, J.~M.~D., {Longmore}, S.~N.,
  {Weisz}, D.~R., \& {Zeidler}, P. 2021, \mnras, 508, 5425


\bibitem[{{McLeod} {et~al.}(2019){McLeod}, {Dale}, {Evans}, {Ginsburg},
  {Kruijssen}, {Pellegrini}, {Ramsay}, \& {Testi}}]{McLeod_2019}
{McLeod}, A.~F., {Dale}, J.~E., {Evans}, C.~J., {Ginsburg}, A., {Kruijssen},
  J.~M.~D., {Pellegrini}, E.~W., {Ramsay}, S.~K., \& {Testi}, L. 2019, \mnras,
  486, 5263


\bibitem[{{McLeod} {et~al.}(2020){McLeod}, {Kruijssen}, {Weisz}, {Zeidler},
  {Schruba}, {Dalcanton}, {Longmore}, {Chevance}, {Faesi}, \&
  {Byler}}]{McLeod_2020}
{McLeod}, A.~F., {Kruijssen}, J.~M.~D., {Weisz}, D.~R., {Zeidler}, P.,
  {Schruba}, A., {Dalcanton}, J.~J., {Longmore}, S.~N., {Chevance}, M.,
  {Faesi}, C.~M., \& {Byler}, N. 2020, \apj, 891, 25


\bibitem[{{Pilyugin} \& {Grebel}(2016)}]{Pilyugin_2016}
{Pilyugin}, L.~S. \& {Grebel}, E.~K. 2016, \mnras, 457, 3678


\bibitem[{{Pilyugin} {et~al.}(2004){Pilyugin}, {V{\'\i}lchez}, \&
  {Contini}}]{Pilyugin_2004}
{Pilyugin}, L.~S., {V{\'\i}lchez}, J.~M., \& {Contini}, T. 2004, \aap, 425, 849


\bibitem[{{Rickard} \& {Palmer}(1981)}]{Rickard_1981}
{Rickard}, L.~J. \& {Palmer}, P. 1981, \aap, 102, L13


\bibitem[{{Rousseau-Nepton} {et~al.}(2019){Rousseau-Nepton}, {Martin},
  {Robert}, {Drissen}, {Amram}, {Prunet}, {Martin}, {Moumen}, {Adamo},
  {Alarie}, {Barmby}, {Boselli}, {Bresolin}, {Bureau}, {Chemin}, {Fernandes},
  {Combes}, {Crowder}, {Della Bruna}, {Duarte Puertas}, {Egusa}, {Epinat},
  {Ksoll}, {Girard}, {G{\'o}mez Llanos}, {Gouliermis}, {Grasha}, {Higgs},
  {Hlavacek-Larrondo}, {Ho}, {Iglesias-P{\'a}ramo}, {Joncas}, {Kam}, {Karera},
  {Kennicutt}, {Klessen}, {Lianou}, {Liu}, {Liu}, {de Amorim}, {Lyman},
  {Martel}, {Mazzilli-Ciraulo}, {McLeod}, {Melchior}, {Millan}, {Moll{\'a}},
  {Momose}, {Morisset}, {Pan}, {Pati}, {Pellerin}, {Pellegrini}, {P{\'e}rez},
  {Petric}, {Plana}, {Rahner}, {Ruiz Lara}, {S{\'a}nchez-Menguiano},
  {Spekkens}, {Stasi{\'n}ska}, {Takamiya}, {Vale Asari}, \&
  {V{\'\i}lchez}}]{Rousseau-Nepton_2019}
{Rousseau-Nepton}, L., {Martin}, R.~P., {Robert}, C., {Drissen}, L., {Amram},
  P., {Prunet}, S., {Martin}, T., {Moumen}, I., {Adamo}, A., {Alarie}, A.,
  {Barmby}, P., {Boselli}, A., {Bresolin}, F., {Bureau}, M., {Chemin}, L.,
  {Fernandes}, R.~C., {Combes}, F., {Crowder}, C., {Della Bruna}, L., {Duarte
  Puertas}, S., {Egusa}, F., {Epinat}, B., {Ksoll}, V.~F., {Girard}, M.,
  {G{\'o}mez Llanos}, V., {Gouliermis}, D., {Grasha}, K., {Higgs}, C.,
  {Hlavacek-Larrondo}, J., {Ho}, I.~T., {Iglesias-P{\'a}ramo}, J., {Joncas},
  G., {Kam}, Z.~S., {Karera}, P., {Kennicutt}, R.~C., {Klessen}, R.~S.,
  {Lianou}, S., {Liu}, L., {Liu}, Q., {de Amorim}, A.~L., {Lyman}, J.~D.,
  {Martel}, H., {Mazzilli-Ciraulo}, B., {McLeod}, A.~F., {Melchior}, A.~L.,
  {Millan}, I., {Moll{\'a}}, M., {Momose}, R., {Morisset}, C., {Pan}, H.~A.,
  {Pati}, A.~K., {Pellerin}, A., {Pellegrini}, E., {P{\'e}rez}, I., {Petric},
  A., {Plana}, H., {Rahner}, D., {Ruiz Lara}, T., {S{\'a}nchez-Menguiano}, L.,
  {Spekkens}, K., {Stasi{\'n}ska}, G., {Takamiya}, M., {Vale Asari}, N., \&
  {V{\'\i}lchez}, J.~M. 2019, \mnras, 489, 5530


\bibitem[{{Saha} {et~al.}(2002){Saha}, {Claver}, \& {Hoessel}}]{Saha_2002}
{Saha}, A., {Claver}, J., \& {Hoessel}, J.~G. 2002, \aj, 124, 839


\bibitem[{{Sanchez}(in prep)}]{sanchez24}
{Sanchez}, S. in prep, {LVM Data Analysis Pipeline}


\bibitem[{{S{\'a}nchez} {et~al.}(2012){S{\'a}nchez}, {Kennicutt}, {Gil de Paz},
  {van de Ven}, {V{\'{\i}}lchez}, {Wisotzki}, {Walcher}, {Mast}, {Aguerri},
  {Albiol-P{\'e}rez}, {Alonso-Herrero}, {Alves}, {Bakos}, {Bart{\'a}kov{\'a}},
  {Bland-Hawthorn}, {Boselli}, {Bomans}, {Castillo-Morales}, {Cortijo-Ferrero},
  {de Lorenzo-C{\'a}ceres}, {Del Olmo}, {Dettmar}, {D{\'{\i}}az}, {Ellis},
  {Falc{\'o}n-Barroso}, {Flores}, {Gallazzi}, {Garc{\'{\i}}a-Lorenzo},
  {Gonz{\'a}lez Delgado}, {Gruel}, {Haines}, {Hao}, {Husemann},
  {Igl{\'e}sias-P{\'a}ramo}, {Jahnke}, {Johnson}, {Jungwiert}, {Kalinova},
  {Kehrig}, {Kupko}, {L{\'o}pez-S{\'a}nchez}, {Lyubenova}, {Marino},
  {M{\'a}rmol-Queralt{\'o}}, {M{\'a}rquez}, {Masegosa}, {Meidt},
  {Mendez-Abreu}, {Monreal-Ibero}, {Montijo}, {Mour{\~a}o}, {Palacios-Navarro},
  {Papaderos}, {Pasquali}, {Peletier}, {P{\'e}rez}, {P{\'e}rez}, {Quirrenbach},
  {Rela{\~n}o}, {Rosales-Ortega}, {Roth}, {Ruiz-Lara},
  {S{\'a}nchez-Bl{\'a}zquez}, {Sengupta}, {Singh}, {Stanishev}, {Trager},
  {Vazdekis}, {Viironen}, {Wild}, {Zibetti}, \& {Ziegler}}]{Sanchez_2012}
{S{\'a}nchez}, S.~F., {Kennicutt}, R.~C., {Gil de Paz}, A., {van de Ven}, G.,
  {V{\'{\i}}lchez}, J.~M., {Wisotzki}, L., {Walcher}, C.~J., {Mast}, D.,
  {Aguerri}, J.~A.~L., {Albiol-P{\'e}rez}, S., {Alonso-Herrero}, A., {Alves},
  J., {Bakos}, J., {Bart{\'a}kov{\'a}}, T., {Bland-Hawthorn}, J., {Boselli},
  A., {Bomans}, D.~J., {Castillo-Morales}, A., {Cortijo-Ferrero}, C., {de
  Lorenzo-C{\'a}ceres}, A., {Del Olmo}, A., {Dettmar}, R.-J., {D{\'{\i}}az},
  A., {Ellis}, S., {Falc{\'o}n-Barroso}, J., {Flores}, H., {Gallazzi}, A.,
  {Garc{\'{\i}}a-Lorenzo}, B., {Gonz{\'a}lez Delgado}, R., {Gruel}, N.,
  {Haines}, T., {Hao}, C., {Husemann}, B., {Igl{\'e}sias-P{\'a}ramo}, J.,
  {Jahnke}, K., {Johnson}, B., {Jungwiert}, B., {Kalinova}, V., {Kehrig}, C.,
  {Kupko}, D., {L{\'o}pez-S{\'a}nchez}, {\'A}.~R., {Lyubenova}, M., {Marino},
  R.~A., {M{\'a}rmol-Queralt{\'o}}, E., {M{\'a}rquez}, I., {Masegosa}, J.,
  {Meidt}, S., {Mendez-Abreu}, J., {Monreal-Ibero}, A., {Montijo}, C.,
  {Mour{\~a}o}, A.~M., {Palacios-Navarro}, G., {Papaderos}, P., {Pasquali}, A.,
  {Peletier}, R., {P{\'e}rez}, E., {P{\'e}rez}, I., {Quirrenbach}, A.,
  {Rela{\~n}o}, M., {Rosales-Ortega}, F.~F., {Roth}, M.~M., {Ruiz-Lara}, T.,
  {S{\'a}nchez-Bl{\'a}zquez}, P., {Sengupta}, C., {Singh}, R., {Stanishev}, V.,
  {Trager}, S.~C., {Vazdekis}, A., {Viironen}, K., {Wild}, V., {Zibetti}, S.,
  \& {Ziegler}, B. 2012, \aap, 538, A8


\bibitem[{{S{\'a}nchez} {et~al.}(2025){S{\'a}nchez}, {Mej{\'\i}a-Narv{\'a}ez},
  {Egorov}, {Kreckel}, {Drory}, {Blanc}, {M{\'e}ndez-Delgado},
  {Barrera-Ballesteros}, {Ibarra-Medel}, {Bizyaev}, {Garc{\'\i}a}, {Wofford},
  \& {Lugo-Aranda}}]{Sanchez_2025}
{S{\'a}nchez}, S.~F., {Mej{\'\i}a-Narv{\'a}ez}, A., {Egorov}, O.~V., {Kreckel},
  K., {Drory}, N., {Blanc}, G.~A., {M{\'e}ndez-Delgado}, J.~E.,
  {Barrera-Ballesteros}, J.~K., {Ibarra-Medel}, H., {Bizyaev}, D.,
  {Garc{\'\i}a}, P., {Wofford}, A., \& {Lugo-Aranda}, A.~Z. 2025, \aj, 169, 52


\bibitem[{{S{\'a}nchez} {et~al.}(2015{\natexlab{a}}){S{\'a}nchez}, {P{\'e}rez},
  {Rosales-Ortega}, {Miralles-Caballero}, {L{\'o}pez-S{\'a}nchez},
  {Iglesias-P{\'a}ramo}, {Marino}, {S{\'a}nchez-Menguiano},
  {Garc{\'\i}a-Benito}, {Mast}, {Mendoza}, {Papaderos}, {Ellis}, {Galbany},
  {Kehrig}, {Monreal-Ibero}, {Gonz{\'a}lez Delgado}, {Moll{\'a}}, {Ziegler},
  {de Lorenzo-C{\'a}ceres}, {Mendez-Abreu}, {Bland-Hawthorn},
  {Bekerait{\.{e}}}, {Roth}, {Pasquali}, {D{\'\i}az}, {Bomans}, {van de Ven},
  \& {Wisotzki}}]{Sanchez_2015b}
{S{\'a}nchez}, S.~F., {P{\'e}rez}, E., {Rosales-Ortega}, F.~F.,
  {Miralles-Caballero}, D., {L{\'o}pez-S{\'a}nchez}, A.~R.,
  {Iglesias-P{\'a}ramo}, J., {Marino}, R.~A., {S{\'a}nchez-Menguiano}, L.,
  {Garc{\'\i}a-Benito}, R., {Mast}, D., {Mendoza}, M.~A., {Papaderos}, P.,
  {Ellis}, S., {Galbany}, L., {Kehrig}, C., {Monreal-Ibero}, A., {Gonz{\'a}lez
  Delgado}, R., {Moll{\'a}}, M., {Ziegler}, B., {de Lorenzo-C{\'a}ceres}, A.,
  {Mendez-Abreu}, J., {Bland-Hawthorn}, J., {Bekerait{\.{e}}}, S., {Roth},
  M.~M., {Pasquali}, A., {D{\'\i}az}, A., {Bomans}, D., {van de Ven}, G., \&
  {Wisotzki}, L. 2015{\natexlab{a}}, \aap, 574, A47


\bibitem[{{S{\'a}nchez} {et~al.}(2016){S{\'a}nchez}, {P{\'e}rez},
  {S{\'a}nchez-Bl{\'a}zquez}, {Garc{\'\i}a-Benito}, {Ibarra-Mede},
  {Gonz{\'a}lez}, {Rosales-Ortega}, {S{\'a}nchez-Menguiano}, {Ascasibar},
  {Bitsakis}, {Law}, {Cano-D{\'\i}az}, {L{\'o}pez-Cob{\'a}}, {Marino}, {Gil de
  Paz}, {L{\'o}pez-S{\'a}nchez}, {Barrera-Ballesteros}, {Galbany}, {Mast},
  {Abril-Melgarejo}, \& {Roman-Lopes}}]{Sanchez_2016}
{S{\'a}nchez}, S.~F., {P{\'e}rez}, E., {S{\'a}nchez-Bl{\'a}zquez}, P.,
  {Garc{\'\i}a-Benito}, R., {Ibarra-Mede}, H.~J., {Gonz{\'a}lez}, J.~J.,
  {Rosales-Ortega}, F.~F., {S{\'a}nchez-Menguiano}, L., {Ascasibar}, Y.,
  {Bitsakis}, T., {Law}, D., {Cano-D{\'\i}az}, M., {L{\'o}pez-Cob{\'a}}, C.,
  {Marino}, R.~A., {Gil de Paz}, A., {L{\'o}pez-S{\'a}nchez}, A.~R.,
  {Barrera-Ballesteros}, J., {Galbany}, L., {Mast}, D., {Abril-Melgarejo}, V.,
  \& {Roman-Lopes}, A. 2016, \rmxaa, 52, 171


\bibitem[{{S{\'a}nchez} {et~al.}(2015{\natexlab{b}}){S{\'a}nchez}, {P{\'e}rez},
  {S{\'a}nchez-Bl{\'a}zquez}, {Gonz{\'a}lez}, {Rosales-Ortega},
  {Cano-D{\'{\i}}az}, {L{\'o}pez-Cob{\'a}}, {Marino}, {Gil de Paz},
  {Moll{\'a}}, {L{\'o}pez-S{\'a}nchez}, {Ascasibar}, \&
  {Barrera-Ballesteros}}]{Sanchez_2015}
{S{\'a}nchez}, S.~F., {P{\'e}rez}, E., {S{\'a}nchez-Bl{\'a}zquez}, P.,
  {Gonz{\'a}lez}, J.~J., {Rosales-Ortega}, F.~F., {Cano-D{\'{\i}}az}, M.,
  {L{\'o}pez-Cob{\'a}}, C., {Marino}, R.~A., {Gil de Paz}, A., {Moll{\'a}}, M.,
  {L{\'o}pez-S{\'a}nchez}, A.~R., {Ascasibar}, Y., \& {Barrera-Ballesteros}, J.
  2015{\natexlab{b}}, ArXiv e-prints


\bibitem[{{S{\'a}nchez-Menguiano} {et~al.}(2018){S{\'a}nchez-Menguiano},
  {S{\'a}nchez}, {P{\'e}rez}, {Ruiz-Lara}, {Galbany}, {Anderson},
  {Kr{\"u}hler}, {Kuncarayakti}, \& {Lyman}}]{Sanchez-Menguiano_2018}
{S{\'a}nchez-Menguiano}, L., {S{\'a}nchez}, S.~F., {P{\'e}rez}, I.,
  {Ruiz-Lara}, T., {Galbany}, L., {Anderson}, J.~P., {Kr{\"u}hler}, T.,
  {Kuncarayakti}, H., \& {Lyman}, J.~D. 2018, \aap, 609, A119


\bibitem[{{Santoro} {et~al.}(2022){Santoro}, {Kreckel}, {Belfiore}, {Groves},
  {Congiu}, {Thilker}, {Blanc}, {Schinnerer}, {Ho}, {Kruijssen}, {Meidt},
  {Klessen}, {Schruba}, {Querejeta}, {Pessa}, {Chevance}, {Kim}, {Emsellem},
  {McElroy}, {Barnes}, {Bigiel}, {Boquien}, {Dale}, {Glover}, {Grasha}, {Lee},
  {Leroy}, {Pan}, {Rosolowsky}, {Saito}, {Sanchez-Blazquez}, {Watkins}, \&
  {Williams}}]{Santoro_2022}
{Santoro}, F., {Kreckel}, K., {Belfiore}, F., {Groves}, B., {Congiu}, E.,
  {Thilker}, D.~A., {Blanc}, G.~A., {Schinnerer}, E., {Ho}, I.~T., {Kruijssen},
  J.~M.~D., {Meidt}, S., {Klessen}, R.~S., {Schruba}, A., {Querejeta}, M.,
  {Pessa}, I., {Chevance}, M., {Kim}, J., {Emsellem}, E., {McElroy}, R.,
  {Barnes}, A.~T., {Bigiel}, F., {Boquien}, M., {Dale}, D.~A., {Glover}, S.
  C.~O., {Grasha}, K., {Lee}, J., {Leroy}, A.~K., {Pan}, H.-A., {Rosolowsky},
  E., {Saito}, T., {Sanchez-Blazquez}, P., {Watkins}, E.~J., \& {Williams},
  T.~G. 2022, \aap, 658, A188


\bibitem[{{Schlafly} \& {Finkbeiner}(2011)}]{Schlafly_2011}
{Schlafly}, E.~F. \& {Finkbeiner}, D.~P. 2011, \apj, 737, 103


\bibitem[{{Smee} {et~al.}(2013){Smee}, {Gunn}, {Uomoto}, {Roe}, {Schlegel},
  {Rockosi}, {Carr}, {Leger}, {Dawson}, {Olmstead}, {Brinkmann}, {Owen},
  {Barkhouser}, {Honscheid}, {Harding}, {Long}, {Lupton}, {Loomis}, {Anderson},
  {Annis}, {Bernardi}, {Bhardwaj}, {Bizyaev}, {Bolton}, {Brewington}, {Briggs},
  {Burles}, {Burns}, {Castander}, {Connolly}, {Davenport}, {Ebelke}, {Epps},
  {Feldman}, {Friedman}, {Frieman}, {Heckman}, {Hull}, {Knapp}, {Lawrence},
  {Loveday}, {Mannery}, {Malanushenko}, {Malanushenko}, {Merrelli}, {Muna},
  {Newman}, {Nichol}, {Oravetz}, {Pan}, {Pope}, {Ricketts}, {Shelden},
  {Sandford}, {Siegmund}, {Simmons}, {Smith}, {Snedden}, {Schneider},
  {SubbaRao}, {Tremonti}, {Waddell}, \& {York}}]{Smee_2013}
{Smee}, S.~A., {Gunn}, J.~E., {Uomoto}, A., {Roe}, N., {Schlegel}, D.,
  {Rockosi}, C.~M., {Carr}, M.~A., {Leger}, F., {Dawson}, K.~S., {Olmstead},
  M.~D., {Brinkmann}, J., {Owen}, R., {Barkhouser}, R.~H., {Honscheid}, K.,
  {Harding}, P., {Long}, D., {Lupton}, R.~H., {Loomis}, C., {Anderson}, L.,
  {Annis}, J., {Bernardi}, M., {Bhardwaj}, V., {Bizyaev}, D., {Bolton}, A.~S.,
  {Brewington}, H., {Briggs}, J.~W., {Burles}, S., {Burns}, J.~G., {Castander},
  F.~J., {Connolly}, A., {Davenport}, J.~R.~A., {Ebelke}, G., {Epps}, H.,
  {Feldman}, P.~D., {Friedman}, S.~D., {Frieman}, J., {Heckman}, T., {Hull},
  C.~L., {Knapp}, G.~R., {Lawrence}, D.~M., {Loveday}, J., {Mannery}, E.~J.,
  {Malanushenko}, E., {Malanushenko}, V., {Merrelli}, A.~J., {Muna}, D.,
  {Newman}, P.~R., {Nichol}, R.~C., {Oravetz}, D., {Pan}, K., {Pope}, A.~C.,
  {Ricketts}, P.~G., {Shelden}, A., {Sandford}, D., {Siegmund}, W., {Simmons},
  A., {Smith}, D.~S., {Snedden}, S., {Schneider}, D.~P., {SubbaRao}, M.,
  {Tremonti}, C., {Waddell}, P., \& {York}, D.~G. 2013, \aj, 146, 32


\bibitem[{{Tully}(1988)}]{Tully_1988}
{Tully}, R.~B. 1988, Journal of the British Astronomical Association, 98, 316


\bibitem[{{Vicens-Mouret} {et~al.}(2023){Vicens-Mouret}, {Drissen}, {Robert},
  {Rousseau-Nepton}, {Martin}, \& {Amram}}]{Vicens-Mouret_2023}
{Vicens-Mouret}, S., {Drissen}, L., {Robert}, C., {Rousseau-Nepton}, L.,
  {Martin}, R.~P., \& {Amram}, P. 2023, \mnras, 524, 3623


\bibitem[{{Yan} {et~al.}(2016){Yan}, {Tremonti}, {Bershady}, {Law}, {Schlegel},
  {Bundy}, {Drory}, {MacDonald}, {Bizyaev}, {Blanc}, {Blanton}, {Cherinka},
  {Eigenbrot}, {Gunn}, {Harding}, {Hogg}, {S{\'a}nchez-Gallego}, {S{\'a}nchez},
  {Wake}, {Weijmans}, {Xiao}, \& {Zhang}}]{Yan_2016}
{Yan}, R., {Tremonti}, C., {Bershady}, M.~A., {Law}, D.~R., {Schlegel}, D.~J.,
  {Bundy}, K., {Drory}, N., {MacDonald}, N., {Bizyaev}, D., {Blanc}, G.~A.,
  {Blanton}, M.~R., {Cherinka}, B., {Eigenbrot}, A., {Gunn}, J.~E., {Harding},
  P., {Hogg}, D.~W., {S{\'a}nchez-Gallego}, J.~R., {S{\'a}nchez}, S.~F.,
  {Wake}, D.~A., {Weijmans}, A.-M., {Xiao}, T., \& {Zhang}, K. 2016, \aj, 151,
  8


\bibitem[{{Zhu} {et~al.}(2010){Zhu}, {Wu}, {Li}, \& {Cao}}]{Zhu_2010}
{Zhu}, Y.-N., {Wu}, H., {Li}, H.-N., \& {Cao}, C. 2010, Research in Astronomy
  and Astrophysics, 10, 329


\bibitem[{{Zibetti} {et~al.}(2009){Zibetti}, {Charlot}, \&
  {Rix}}]{Zibetti_2009}
{Zibetti}, S., {Charlot}, S., \& {Rix}, H.-W. 2009, \mnras, 400, 1181


\end{thebibliography}
\bibliographystyle{aasjournal}

\label{RMAALastPage}
\end{document}